\documentclass[twocolumn]{aastex631}

\usepackage{commath}
\usepackage{floatrow}
\usepackage{soul}
\usepackage{CJK}
\usepackage{threeparttable}
\usepackage{soul}
\usepackage{ulem}
\usepackage[caption=false]{subfig}
\usepackage{appendix}
\usepackage{xspace}

\newcommand{\msasd}{{\sc MSA-3D}\xspace}

\newcommand{\oiii}{\hbox{[O\,{\scriptsize III}]}}
\newcommand{\nii}{\hbox{[N\,{\scriptsize II}]}}

\defcitealias{Ju2025}{Ju25}

\shorttitle{\msasd: gas kinematics and metallicity gradients in $z\sim1$ galaxies}
\shortauthors{Ju et al.}
\graphicspath{{./}{figures/}}
\begin{document}
\begin{CJK*}{UTF8}{gbsn}

\title{MSA-3D: Connecting the Chemical and Kinematic Structures of Galaxies at $z \sim 1$}

\correspondingauthor{Mengting Ju, Xin Wang}
\email{jumengting@ucas.ac.cn, xwang@ucas.ac.cn}


\author[0000-0002-5815-2387]{Mengting Ju}
\affiliation{School of Astronomy and Space Science, University of Chinese Academy of Sciences (UCAS), Beijing 100049, China}

\author[0000-0002-9373-3865]{Xin Wang}
\affiliation{School of Astronomy and Space Science, University of Chinese Academy of Sciences (UCAS), Beijing 100049, China}
\affiliation{National Astronomical Observatories, Chinese Academy of Sciences, Beijing 100101, China}
\affiliation{Institute for Frontiers in Astronomy and Astrophysics, Beijing Normal University, Beijing 102206, China}

\author[0000-0001-5860-3419]{Tucker Jones}
\affiliation{Department of Physics and Astronomy, University of California, Davis, 1 Shields Avenue, Davis, CA 95616, USA}

\author[0000-0001-6371-6274]{Ivana Bari\v{s}i\'{c}}
\affiliation{Department of Physics and Astronomy, University of California, Davis, 1 Shields Avenue, Davis, CA 95616, USA}

\author[0000-0001-6703-4676]{Juan M. Espejo Salcedo}
\affiliation{Max-Planck-Institut f\"{u}r Extraterrestrische Physik (MPE), Giessenbachstr. 1, D-85748 Garching, Germany}

\author[0000-0002-3254-9044]{Karl Glazebrook}
\affiliation{Centre for Astrophysics and Supercomputing, Swinburne University of Technology, Hawthorn, VIC 3122, Australia}

\author[0000-0002-1527-0762]{Danail Obreschkow}
\affiliation{International Centre for Radio Astronomy Research (ICRAR), M468, University of Western Australia, Perth, WA 6009, Australia}
\affiliation{Australian Research Council, ARC Centre of Excellence for All Sky Astrophysics in 3 Dimensions (ASTRO 3D), Australia}

\author[0000-0002-1499-6377]{Takafumi Tsukui}
\affiliation{Astronomical Institute, Tohoku University, 6-3, Aramaki, Aoba-ku, Sendai, Miyagi, 980-8578, Japan}

\author[0009-0006-1255-9567]{Qianqiao Zhou}
\affiliation{School of Astronomy and Space Science, University of Chinese Academy of Sciences (UCAS), Beijing 100049, China}

\author[0000-0001-9742-3138]{Kevin Bundy}
\affiliation{UCO/Lick Observatory, University of California, Santa Cruz, 1156 High Street, Santa Cruz, CA 95064, USA}

\author[0000-0002-6586-4446]{Alaina Henry}
\affiliation{Space Telescope Science Institute, 3700 San Martin Drive, Baltimore, MD 21218, USA}

\author[0000-0001-6919-1237]{Matthew A. Malkan}
\affiliation{Department of Physics and Astronomy, University of California, Los Angeles, CA 90095-1547, USA}

\author[0000-0003-2804-0648]{Themiya Nanayakkara}
\affiliation{Centre for Astrophysics and Supercomputing, Swinburne University of Technology, Hawthorn, VIC 3122, Australia}

\author[0000-0002-4430-8846]{Namrata Roy}
\affiliation{Center for Astrophysical Sciences, Department of Physics and Astronomy, Johns Hopkins University, Baltimore, MD, 21218}

\author[0009-0005-8170-5153]{Xunda Sun}
\affiliation{School of Astronomy and Space Science, University of Chinese Academy of Sciences (UCAS), Beijing 100049, China}


\begin{abstract}

We investigate the connection between ionized gas kinematics and gas-phase metallicity gradients in 21 star-forming galaxies at $0.5 < z < 1.7$ from the MSA-3D survey, using spatially resolved JWST/NIRSpec slit-stepping observations. 
Galaxy kinematics are characterized by the ratio of rotational velocity to intrinsic velocity dispersion, $v/\sigma$, measured at $1.5\,R_e$, where $R_e$ is the effective radius. 
We find that dynamically hotter disks exhibit systematically flatter metallicity gradients, with a moderate anti-correlation between $\nabla \mathrm{O/H}$ and $v/\sigma$ ($\rho = -0.30^{+0.16}_{-0.15}$). A linear fit yields a slope of $\sim 0.005$ dex per dex in $v/\sigma$, weaker than the dependence on stellar mass. Similarly, the N2 gradients show a consistent trend, with $\rho(v/\sigma, \nabla \mathrm{N2}) = -0.23^{+0.07}_{-0.08}$, indicating that the correlation is robust to the choice of metallicity indicator.
A significantly  stronger anti-correlation is observed with $R_{\rm e}/\sigma$, interpreted as a proxy for the radial mixing timescale (Spearman rank correlation coefficient $\rho = -0.43^{+0.14}_{-0.13}$), suggesting that cumulative radial mixing more directly regulates chemical stratification.
The metallicity gradients in our sample are uniformly shallow, indicating that efficient turbulent mixing in kinematically settled disks regulates the chemical structure of typical star-forming galaxies at $z\sim1$.

\end{abstract}


\section{Introduction} \label{sec:intro}

Gas-phase metallicity gradients, which describe the radial distribution of heavy elements within galaxies, serve as crucial probes into the chemical and structural evolution of galaxies. These gradients are intimately linked to fundamental processes such as star formation, gas accretion, galactic outflows, and radial mixing, all of which collectively influence the properties of galaxies over cosmic time \citep[e.g.,][]{Thielemann2017, Maiolino2019, Tremonti2004, Mannucci2010, Cresci2010, Queyrel2012, Jones2010_gradient, Jones2013, WangX2017, WangX2019, WangX2020, WangX2022a, Ju2022, Ju2025, Sun2025, Venturi2024, Ratcliffe2025, Garcia2025, Wangenci2023, Li2025}. 
Metallicity gradients provide key insights into the formation and evolution of galactic disks.
For example, an inside-out growth scenario typically leads to more rapid star formation and chemical enrichment in the central regions compared to the outskirts, leading to negative radial gradients in metallicity. In contrast, flat gradients—where central and outer regions exhibit similar metallicities—or even central metal deficiencies may indicate more complex evolutionary processes, such as galaxy mergers, inward gas accretion, or strong feedback mechanisms that redistribute metals throughout the galaxy. Many of these phenomena have been extensively explored in simulations \citep[e.g.,][]{Gibson2013, Ma2017,Bellardini2021,Bellardini2022, Sun2025, Simons2021, Hemler2021, Tissera2022, Wangenci2024}.

The dynamical state of star-forming galaxies provides a window into disk assembly and the regulation of baryons \citep[e.g.,][]{Gillman2019, Girard2020, Sharda2021b,Tsukui2025,Romeo2016,Romeo2011,Romeo2013}. \cite{Bosch2002} has emphasized the tight coupling between angular-momentum acquisition, disk settling, and feedback-driven turbulence. \cite{Kassin2007} introduced the modified kinematic parameter, designed to incorporate the elevated turbulence characteristic of high-redshift systems and to recover a tighter Tully-Fisher relation. \cite{Miller2011,Miller2012} found relatively low scatter in the Tully-Fisher relation at $z\sim1$, suggesting that sample selection, particularly whether galaxies are morphologically ``pure'' disks, plays a crucial role in interpreting rotational support and dynamical evolution. Integral field unit (IFU) surveys such as SINS/zC-SINF AO \citep{Schreiber2018} and KMOS$^{\rm 3D}$ \citep{Wisnioski2015,Tiley2021} have revealed the diversity of kinematic structures at $z\sim1$--3, ranging from rotation-dominated disks to dispersion-supported systems. A central diagnostic of this diversity is the balance between ordered rotation ($v_{rot}$) and turbulent motions ($\sigma_0$), often quantified by their ratio $v/\sigma$. 

Recent studies have linked these dynamical measurements to internal processes such as gas inflows, star-formation-driven outflows, and metal transport, which influence the structural evolution of galaxies \citep[e.g.,][]{Furlanetto2021, Yang2024, Danhaive2025, Graaff2024, Wang2024, Amvrosiadis2025, Esparza2025,Romeo2014}. 
However, observational studies connecting gas kinematics and metallicity gradients at high redshift remain limited to modest sample sizes \citep[e.g.,][]{Queyrel2012, Gillman2021, Yuan2011, Jones2013, Leethochawalit2016}. 
Simulations have also explored the joint evolution of gradients and kinematics \citep[e.g.,][]{Ma2017, Hemler2021, Sun2025}. Based on ground-based IFU surveys spanning $0<z<2.5$, \cite{Sharda2021b} found that galaxies with high velocity dispersion typically exhibit flat metallicity gradients, whereas rotation-dominated disks show steeper negative gradients.
High gas velocity dispersion traces strong turbulence and non-circular motions that enhance radial mixing and tend to flatten any radial metallicity gradients \citep[e.g.,][]{Yang2012,Forbes2014}.
In contrast, rotation-dominated systems reside in deeper gravitational potentials, where the competing effects of gas accretion, star formation, and feedback allow metallicity gradients to persist but with a broader intrinsic scatter \citep[e.g.,][]{Tully1977,Faucher2011,Ostriker2011,Romeo2017}, motivating a systematic investigation of the connection between $v/\sigma$ and metallicity gradients.

The launch of the James Webb Space Telescope (\textit{JWST}) has revolutionized our ability to probe the distant universe with unprecedented sensitivity and spatial resolution.
Its Near-Infrared Spectrograph (NIRSpec) with the Multi-Object Spectroscopy (MOS) mode, employing the Micro-Shutter Assembly (MSA) and a slit-stepping strategy, enables efficient acquisition of 3D spectroscopic data cubes \citep[e.g., MSA-3D:][]{Ivana2025, Ju2025, Roy2025}. This capability is crucial for resolving gas kinematics and measuring metallicity gradients within large samples of distant galaxies, marking a leap forward in studying galaxy assembly and chemical enrichment in the early universe. 
In an earlier paper we analyzed spatially resolved gas-phase metallicity gradients
from a sample of 25 galaxies in the MSA-3D project \citep[][hereafter \citetalias{Ju2025}]{Ju2025}. This study investigates the connection between the dynamical state of these galaxies, characterized by their $v/\sigma$ ratio, and the 
slope of their metallicity profiles.

This paper is organized as follows. Section~\ref{sec:data} describes the data and the derivation of the $v/\sigma$ ratios. Section~\ref{sec:method} examines the connection between galaxy dynamical state and gas-phase metallicity gradients, with the goal of characterizing how metallicity gradients vary across different dynamical state. The discussion and summary are presented in Section~\ref{sec:summary}. Throughout this work, we adopt a $\Lambda$ cold dark matter cosmology with $H_0 = 69.32\ \mathrm{km\,s^{-1}\,Mpc^{-1}}$ and $\Omega_{\rm M} = 0.2865$.

\section{Data} \label{sec:data}

\begin{table*}[]
  \scriptsize
  \caption{Morphologies of 43 galaxies in the MSA-3D project.}
  \tabcolsep=0.1cm
  \label{table:1}
  \centering
\begin{tabular}{lcccc|cccc|cccc}
\hline\hline
ID	&	 RA	&	 Dec		&	redshift	&	M$_*^a$ &	\multicolumn{4}{c|}{F160W} & \multicolumn{4}{c}{F444W} 	\\
-	&	 Degree	&	 Degrees		&	-	& $\rm log(\frac{M_{*}}{M_{\odot}})$&   n & R$_e$(arcsec) & q &  PA &   n & R$_e$(arcsec) & q  & PA \\
\hline
2111 	&	215.0628 	&	52.9071 	&	0.58	&	9.97	&	2.39 	$\pm$	0.06 	& 	0.95 	$\pm$	0.03 	& 	0.50 	$\pm$	0.00 	& 	86.97 	$\pm$	0.41 	& 	2.05 	$\pm$	0.02 	& 	0.66 	$\pm$	0.01 	& 	0.56 	$\pm$	0.00 	& 	87.30 	$\pm$	0.24 	\\
2145 	&	215.0695 	&	52.9109 	&	1.17	&	9.19	&	2.06 	$\pm$	0.21 	& 	0.20 	$\pm$	0.01 	& 	0.35 	$\pm$	0.02 	& 	8.01 	$\pm$	1.24 	& 	2.04 	$\pm$	0.06 	& 	0.20 	$\pm$	0.00 	& 	0.42 	$\pm$	0.01 	& 	2.67 	$\pm$	0.55 	\\
2465 	&	215.0704 	&	52.9137 	&	1.25	&	9.3	&	0.55 	$\pm$	0.04 	& 	0.47 	$\pm$	0.01 	& 	0.28 	$\pm$	0.01 	& 	4.69 	$\pm$	0.54 	& 	0.88 	$\pm$	0.02 	& 	0.43 	$\pm$	0.00 	& 	0.28 	$\pm$	0.00 	& 	4.42 	$\pm$	0.23 	\\
2824 	&	215.0685 	&	52.9143 	&	0.98	&	9.49	&		-		& 		-		& 		-		& 		-		& 	1.90 	$\pm$	0.04 	& 	0.32 	$\pm$	0.01 	& 	0.46 	$\pm$	0.01 	& 	-84.72 	$\pm$	0.50 	\\
3399 	&	215.0425 	&	52.8996 	&	1.34	&	9.81	&	0.52 	$\pm$	0.02 	& 	0.46 	$\pm$	0.00 	& 	0.30 	$\pm$	0.00 	& 	68.29 	$\pm$	0.34 	& 	0.80 	$\pm$	0.01 	& 	0.40 	$\pm$	0.00 	& 	0.30 	$\pm$	0.00 	& 	67.80 	$\pm$	0.16 	\\
4391 	&	215.0676 	&	52.9232 	&	1.08	&	9.48	&	2.78 	$\pm$	0.15 	& 	0.20 	$\pm$	0.00 	& 	0.92 	$\pm$	0.02 	& 	-67.99 	$\pm$	8.04 	& 	1.88 	$\pm$	0.02 	& 	0.18 	$\pm$	0.00 	& 	0.90 	$\pm$	0.00 	& 	-74.27 	$\pm$	1.75 	\\
6199 	&	215.0450 	&	52.9195 	&	1.59	&	10	&	1.29 	$\pm$	0.08 	& 	0.89 	$\pm$	0.05 	& 	0.49 	$\pm$	0.01 	& 	-27.39 	$\pm$	1.11 	& 	2.30 	$\pm$	0.12 	& 	1.21 	$\pm$	0.09 	& 	0.49 	$\pm$	0.01 	& 	-33.79 	$\pm$	0.86 	\\
6430 	&	215.0131 	&	52.8980 	&	1.17	&	9.79	&	1.43 	$\pm$	0.04 	& 	0.32 	$\pm$	0.00 	& 	0.34 	$\pm$	0.01 	& 	30.00 	$\pm$	0.39 	& 		-		& 		-		& 		-		& 		-		\\
6848 	&	215.0356 	&	52.9167 	&	1.57	&	10.64	&	3.45 	$\pm$	0.30 	& 	0.82 	$\pm$	0.10 	& 	0.71 	$\pm$	0.02 	& 	-74.31 	$\pm$	2.52 	& 	1.57 	$\pm$	0.01 	& 	0.18 	$\pm$	0.00 	& 	0.66 	$\pm$	0.00 	& 	-85.49 	$\pm$	0.33 	\\
7314 	&	214.9989 	&	52.8925 	&	1.28	&	9.55	&		-		& 		-		& 		-		& 		-		& 		-		& 		-		& 		-		& 		-		\\
7561 	&	215.0609 	&	52.9384 	&	1.03	&	9.21	&	1.92 	$\pm$	0.12 	& 	0.50 	$\pm$	0.02 	& 	0.32 	$\pm$	0.01 	& 	-31.89 	$\pm$	0.67 	& 	1.28 	$\pm$	0.03 	& 	0.40 	$\pm$	0.00 	& 	0.32 	$\pm$	0.00 	& 	-33.14 	$\pm$	0.23 	\\
8365 	&	215.0600 	&	52.9422 	&	1.68	&	9.56	&	6.35 	$\pm$	1.20 	& 	0.92 	$\pm$	0.30 	& 	0.50 	$\pm$	0.03 	& 	-42.65 	$\pm$	2.10 	& 	1.06 	$\pm$	0.03 	& 	0.26 	$\pm$	0.00 	& 	0.51 	$\pm$	0.01 	& 	-38.88 	$\pm$	0.53 	\\
8512 	&	215.0498 	&	52.9381 	&	1.1	&	10.32	&	1.84 	$\pm$	0.08 	& 	1.62 	$\pm$	0.11 	& 	0.69 	$\pm$	0.01 	& 	89.64 	$\pm$	1.19 	& 	2.39 	$\pm$	0.06 	& 	1.81 	$\pm$	0.09 	& 	0.69 	$\pm$	0.00 	& 	-86.31 	$\pm$	0.70 	\\
8576 	&	215.0596 	&	52.9434 	&	1.57	&	9.6	&	1.01 	$\pm$	0.03 	& 	0.33 	$\pm$	0.01 	& 	0.82 	$\pm$	0.01 	& 	0.16 	$\pm$	2.54 	& 	1.11 	$\pm$	0.02 	& 	0.30 	$\pm$	0.00 	& 	0.85 	$\pm$	0.00 	& 	-2.16 	$\pm$	1.42 	\\
8942 	&	215.0094 	&	52.9101 	&	1.18	&	9.86	&	0.95 	$\pm$	0.02 	& 	0.22 	$\pm$	0.00 	& 	0.74 	$\pm$	0.01 	& 	5.40 	$\pm$	1.02 	& 	0.80 	$\pm$	0.01 	& 	0.22 	$\pm$	0.00 	& 	0.74 	$\pm$	0.00 	& 	2.65 	$\pm$	0.41 	\\
9337 	&	214.9957 	&	52.9019 	&	1.17	&	9.27	&		-		& 		-		& 		-		& 		-		& 		-		& 		-		& 		-		& 		-		\\
9424 	&	214.9927 	&	52.9009 	&	0.98	&	9.76	&	1.35 	$\pm$	0.05 	& 	0.58 	$\pm$	0.01 	& 	0.40 	$\pm$	0.01 	& 	89.58 	$\pm$	0.48 	& 		-		& 		-		& 		-		& 		-		\\
9482 	&	215.0530 	&	52.9442 	&	1.21	&	9.77	&		-		& 		-		& 		-		& 		-		& 		-		& 		-		& 		-		& 		-		\\
9527 	&	215.0085 	&	52.9124 	&	1.42	&	10.02	&	6.29 	$\pm$	0.59 	& 	0.42 	$\pm$	0.04 	& 	0.62 	$\pm$	0.01 	& 	-57.17 	$\pm$	1.66 	& 	2.70 	$\pm$	0.06 	& 	0.16 	$\pm$	0.00 	& 	0.65 	$\pm$	0.01 	& 	-57.04 	$\pm$	0.82 	\\
9636 	&	215.0366 	&	52.9329 	&	0.74	&	9.35	&	1.29 	$\pm$	0.04 	& 	0.70 	$\pm$	0.01 	& 	0.25 	$\pm$	0.00 	& 	70.96 	$\pm$	0.20 	& 		-		& 		-		& 		-		& 		-		\\
9812 	&	215.0403 	&	52.9376 	&	0.74	&	9.97	&	1.76 	$\pm$	0.05 	& 	0.88 	$\pm$	0.03 	& 	0.73 	$\pm$	0.01 	& 	80.71 	$\pm$	1.19 	& 		-		& 		-		& 		-		& 		-		\\
9960 	&	215.0319 	&	52.9332 	&	1.51	&	11.06	&	6.23 	$\pm$	0.31 	& 	1.93 	$\pm$	0.22 	& 	0.45 	$\pm$	0.01 	& 	6.40 	$\pm$	0.44 	& 		-		& 		-		& 		-		& 		-		\\
10107 	&	214.9818 	&	52.8976 	&	1.01	&	10.01	&	2.13 	$\pm$	0.07 	& 	0.36 	$\pm$	0.01 	& 	0.73 	$\pm$	0.01 	& 	29.47 	$\pm$	1.46 	& 	1.81 	$\pm$	0.02 	& 	0.31 	$\pm$	0.00 	& 	0.83 	$\pm$	0.01 	& 	27.65 	$\pm$	1.34 	\\
10502 	&	214.9858 	&	52.9033 	&	1.23	&	10.13	&	4.51 	$\pm$	0.43 	& 	6.02 	$\pm$	1.68 	& 	0.59 	$\pm$	0.01 	& 	47.13 	$\pm$	1.20 	& 	2.81 	$\pm$	0.09 	& 	1.50 	$\pm$	0.08 	& 	0.59 	$\pm$	0.01 	& 	45.33 	$\pm$	0.58 	\\
10752 	&	215.0404 	&	52.9414 	&	1.73	&	9.62	&		-		& 		-		& 		-		& 		-		& 		-		& 		-		& 		-		& 		-		\\
10863 	&	215.0551 	&	52.9530 	&	1.03	&	9.24	&		-		& 		-		& 		-		& 		-		& 		-		& 		-		& 		-		& 		-		\\
10910 	&	215.0562 	&	52.9554 	&	0.74	&	9.57	&	2.19 	$\pm$	0.12 	& 	1.30 	$\pm$	0.10 	& 	0.77 	$\pm$	0.01 	& 	-14.33 	$\pm$	2.06 	& 		-		& 		-		& 		-		& 		-		\\
11225 	&	215.0416 	&	52.9455 	&	1.05	&	9.67	&	2.15 	$\pm$	0.07 	& 	0.19 	$\pm$	0.00 	& 	0.40 	$\pm$	0.01 	& 	-25.91 	$\pm$	0.49 	& 		-		& 		-		& 		-		& 		-		\\
11539 	&	214.9820 	&	52.9051 	&	1.61	&	10.39	&	4.09 	$\pm$	0.22 	& 	0.53 	$\pm$	0.04 	& 	0.90 	$\pm$	0.02 	& 	23.17 	$\pm$	5.10 	& 	2.20 	$\pm$	0.02 	& 	0.20 	$\pm$	0.00 	& 	0.86 	$\pm$	0.00 	& 	28.85 	$\pm$	1.03 	\\
11702 	&	214.9794 	&	52.9032 	&	1.23	&	9.37	&	5.25 	$\pm$	0.65 	& 	0.17 	$\pm$	0.01 	& 	0.76 	$\pm$	0.03 	& 	-17.80 	$\pm$	4.47 	& 	1.98 	$\pm$	0.05 	& 	0.15 	$\pm$	0.00 	& 	0.68 	$\pm$	0.01 	& 	-22.27 	$\pm$	1.34 	\\
11843 	&	215.0390 	&	52.9471 	&	1.46	&	10.74	&	0.30 	$\pm$	0.02 	& 	0.59 	$\pm$	0.01 	& 	0.38 	$\pm$	0.00 	& 	26.42 	$\pm$	0.44 	& 		-		& 		-		& 		-		& 		-		\\
11944 	&	215.0370 	&	52.9454 	&	1.04	&	9.27	&	2.93 	$\pm$	0.18 	& 	0.44 	$\pm$	0.02 	& 	0.46 	$\pm$	0.01 	& 	71.53 	$\pm$	0.86 	& 		-		& 		-		& 		-		& 		-		\\
12015 	&	215.0323 	&	52.9432 	&	1.24	&	10.08	&	0.67 	$\pm$	0.01 	& 	0.34 	$\pm$	0.00 	& 	0.67 	$\pm$	0.00 	& 	-41.57 	$\pm$	0.58 	& 		-		& 		-		& 		-		& 		-		\\
12071 	&	215.0220 	&	52.9361 	&	1.28	&	9.47	&	1.38 	$\pm$	0.08 	& 	0.38 	$\pm$	0.01 	& 	0.30 	$\pm$	0.01 	& 	-14.81 	$\pm$	0.66 	& 		-		& 		-		& 		-		& 		-		\\
12239 	&	215.0495 	&	52.9560 	&	0.89	&	9.49	&		-		& 		-		& 		-		& 		-		& 		-		& 		-		& 		-		& 		-		\\
12253 	&	215.0442 	&	52.9521 	&	1.03	&	9.17	&	8.13 	$\pm$	1.36 	& 	0.37 	$\pm$	0.06 	& 	0.43 	$\pm$	0.02 	& 	77.17 	$\pm$	1.39 	& 		-		& 		-		& 		-		& 		-		\\
12773 	&	215.0298 	&	52.9452 	&	0.95	&	9.23	&	1.44 	$\pm$	0.07 	& 	0.43 	$\pm$	0.01 	& 	0.45 	$\pm$	0.01 	& 	-14.79 	$\pm$	0.83 	& 		-		& 		-		& 		-		& 		-		\\
13182 	&	214.9998 	&	52.9268 	&	1.54	&	10.25	&	0.71 	$\pm$	0.02 	& 	0.34 	$\pm$	0.00 	& 	0.87 	$\pm$	0.01 	& 	25.41 	$\pm$	2.91 	& 		-		& 		-		& 		-		& 		-		\\
13416 	&	215.0253 	&	52.9457 	&	1.54	&	10.76	&	1.17 	$\pm$	0.06 	& 	0.60 	$\pm$	0.02 	& 	0.52 	$\pm$	0.01 	& 	70.66 	$\pm$	1.09 	& 		-		& 		-		& 		-		& 		-		\\
18188 	&	214.9840 	&	52.9414 	&	0.82	&	9.56	&	1.04 	$\pm$	0.05 	& 	0.78 	$\pm$	0.03 	& 	0.74 	$\pm$	0.01 	& 	-79.40 	$\pm$	1.96 	& 		-		& 		-		& 		-		& 		-		\\
18586 	&	214.9712 	&	52.9338 	&	0.76	&	9.01	&	1.42 	$\pm$	0.06 	& 	0.22 	$\pm$	0.00 	& 	0.50 	$\pm$	0.01 	& 	-75.56 	$\pm$	0.74 	& 		-		& 		-		& 		-		& 		-		\\
19382 	&	214.9766 	&	52.9415 	&	1.03	&	9.38	&		-		& 		-		& 		-		& 		-		& 		-		& 		-		& 		-		& 		-		\\
29470 	&	214.9690 	&	52.9454 	&	1.04	&	9.71	&	2.48 	$\pm$	0.21 	& 	1.39 	$\pm$	0.18 	& 	0.64 	$\pm$	0.01 	& 	-47.02 	$\pm$	1.83 	& 		-		& 		-		& 		-		& 		-		\\

\hline
\end{tabular}\\

Notes:
$^a$ The stellar population properties are obtained from the 3D-HST survey catalogue \citep{Brammer2012,Skelton2014}.\\
\end{table*}

\subsection{Integral field spectra from MSA-3D}

The MSA-3D project \citep{Ivana2025} conducted its first observations on March 29-30, 2023 (JWST Cycle 1, GO-2136; PI: Jones), targeting 43 star-forming galaxies at redshifts $0.5<z<1.7$ in the Extended Groth Strip (EGS) field. The sample was selected based on spectroscopic redshift, stellar mass, and star formation rate using data from the CANDELS \citep{Koekemoer2011} and 3D-HST \citep{Skelton2014, Momcheva2016} surveys (Table~\ref{table:1}). The target selection was intentionally blind to galaxy morphology to ensure a representative population of star-forming galaxies at $z\sim1$.

Observations were carried out with JWST/NIRSpec in multi-object spectroscopy mode using the MSA. A slit-stepping strategy was used to create integral-field-like spectroscopic coverage across each galaxy. We adopted the G140H/F100LP grating/filter configuration, covering the wavelength range $\rm 0.97-1.82~\mu m$ at a spectral resolution of $\rm R\sim2700$, allowing detection of key nebular emission lines including H$\alpha$, \nii, \oiii, and H$\beta$.

The typical spatial resolution of the reconstructed data cubes is $\sim 0\farcs20 \times 0\farcs08$, corresponding to physical scales of $\sim1.6 \times 0.7$ kpc at $z\sim1$. In this work, we use datacubes resampled to a uniform grid of 0\farcs08 per pixel for structural and kinematic analysis. Further detail on observations and data processing\footnote{https://github.com/barisiciv/msa3d} is presented in the MSA-3D overview paper \cite{Ivana2025}.

We fit the observed spectra in the rest-frame wavelength range 6520 - 6630 \AA\ using a simple model consisting of a linear continuum plus three Gaussian components representing the H$\alpha +\nii\lambda\lambda$6548,6584 emission complex. All three lines are assumed to share the same velocity and velocity dispersion, The fits are performed on a spaxel-by-spaxel basis, following the methodology of \citetalias{Ju2025}. The velocity fields and velocity dispersion maps are derived from the centroid and width of the H$\alpha$ line, corrected for instrumental broadening using the wavelength-dependent line spread function.

For metallicity gradient estimation, we adopt the N2 calibration of \cite{pp04}, using the ratio of $\rm \nii\lambda6584/H\alpha$, where available. Typical metallicity gradient slopes range from -0.03 to 0.02 dex kpc$^{-1}$, with uncertainties primarily driven by systematic errors in flux calibration (\citetalias{Ju2025}). Overall, the MSA-3D sample provides a uniquely high-resolution view of the interplay between gas-phase chemical enrichment and internal dynamical structure of a representative galaxy sample at intermediate redshift. The metallicity gradient slopes used in this work are those reported by \citetalias{Ju2025} and summarized in Table~\ref{table:vs}.


\subsection{Galaxy morphologies}
\label{sec:galfit}

\begin{figure}
    \centering
    \includegraphics[width=1.0\textwidth,clip,trim={0 0 0 0}]{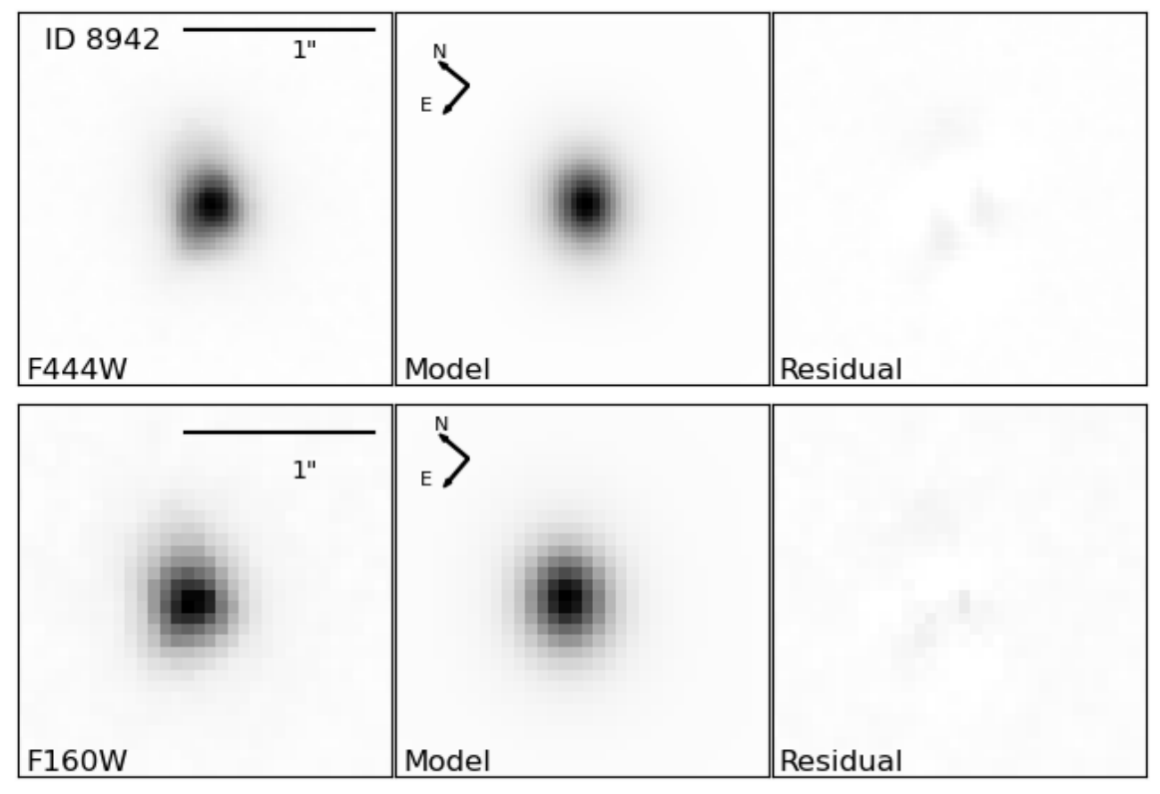} 
    \caption{Two-dimensional surface brightness modeling of galaxy ID 8942 using GALFIT. The top row shows the JWST/NIRCam F444W imaging; the bottom row shows the HST/WFC3 F160W imaging. From left to right, the panels display the observed image, the best-fit S\'ersic model, and the residual map after model subtraction. The fitting region is $\sim 2\arcsec \times 2\arcsec$ box. The same color scale is used in all panels. The photometric centroids and inclinations derived from the fits are used to extract the kinematic profiles.
}    
    \label{fig:example}
\end{figure}

We use high-resolution imaging to constrain the structural properties and determine galaxy centers for dynamical analysis. The fitting is performed on image cutouts approximately $2\arcsec\times2\arcsec$ in size, covering the central regions of each target galaxy. For galaxies with available JWST/NIRCam F444W imaging, we prioritize the longer-wavelength data
(pixel scale $0\farcs03$, cutout size $1\farcs98\times1\farcs98$) from the CEERS survey\footnote{\url{https://ceers.github.io/index.html}}. For the remaining galaxies, we use HST/WFC3 F160W imaging (pixel scale $0\farcs06$, cutout size $2\farcs16 \times 2\farcs16$) obtained from the CANDELS archive \citep{Koekemoer2011}. 
All galaxies in our sample have F160W imaging, while 21 of them additionally have F444W coverage. We present these images in Appendix~\ref{sec:appendix}, showing F444W for galaxies where available and F160W otherwise. Three galaxies (IDs 9527, 11539, and 11702) have F444W imaging but lack H$\alpha$ kinematic measurements, and are therefore not included in the appendix.
Using the WCS information, we find that the position angles (PAs) inferred from the HST images differ from those of the CEERS and MSA-3D data by 0.29$^\circ$ and 36.68$^\circ$, respectively.

A single-component S\'ersic model is fit to each galaxy using $\tt GALFIT~v3.0.5$ \citep{peng2002,peng2010}, and the best-fit central position is adopted as the galaxy center for kinematic modeling. 
We use the empirical point-spread function (PSF) results from the CANDELS and CEERS data, for the F160W and F444W images, respectively. The F444W data are close to Nyquist-sampled, and since GALFIT is used primarily to determine the galaxy center, the sampling does not affect our results.
In total, we obtain reliable structural fits for 35 galaxies in the F160W band and 18 galaxies in the F444W band, constrained by the spatial coverage of the respective imaging data, and 8 F160W and 3 F444W images could not be fitted reliably. We list the S\'ersic index (n), effective radius (r$\rm _e$), axis ratio (q) and PA in Table~\ref{table:1}.
The photometric centers derived from these models are adopted as the reference coordinates for extracting the kinematic maps and metallicity gradients throughout this work. 
For the 17 galaxies with GALFIT measurements from both F444W and F160W data, some structural parameters such as n and r$\rm _e$ differ beyond the quoted uncertainties. However, the PAs and inclinations (derived from the axis ratios assuming $q=\rm{cos}(i)$ and listed in Table~\ref{table:vs}) are largely consistent, with median differences of 2.75$^\circ$ and 2.5$^\circ$, respectively. The fitted galaxy centroids show no significant offsets.

For galaxies that cannot be well fitted with a single-component model, we instead use the spaxel with the highest H$\alpha$ flux near the galaxy center as the kinematic center.
Among the 43 galaxies in our sample, six lack GALFIT measurements from both the F444W and F160W images. All 6 galaxies have available H$\alpha$ velocity fields; however, two of them (ID~9482 and ID~10863) are excluded from kinematic modeling due to an insufficient number of spaxels and highly irregular velocity fields, respectively. For the remaining four galaxies (ID~7314, ID~9337, ID~10752, and ID~12239), the kinematic centers are determined using the peak position of the H$\alpha$ emission. Combined with the metallicity gradients reported in \citetalias{Ju2025}, two of these systems (ID~7314 and ID~12239) are included in the subsequent analysis of the $v/\sigma$-gradient relation.

\begin{table*}[]
  \scriptsize
  \caption{Kinematic measurements for the 43 galaxies in the MSA-3D project.}
  \tabcolsep=0.1cm
  \label{table:vs}
  \centering
\begin{tabular}{l|ccccccc|cc|c}
\hline\hline
ID	&	 \multicolumn{7}{c|}{Fixed Inclination} & \multicolumn{2}{c|}{Free Inclination} & gradient$^b$ \\
	&	 inc & $R_t$ &	$v_{rot}$	&	$v_{rot}(1.5R_e)$	&	$\sigma_0$ & $v/\sigma$ & $\Delta PA^a$ & inc 	&   $ v/\sigma$ &  $\rm (\nabla\, N2)$ \\
-	&	 $^\circ$	& kpc  &	km/s	&	km/s	& km/s &   - &	 $^\circ$& $^\circ$  & - &  dex/kpc  \\
\hline
2111	& 	55.94$^c$	& 	3.40 	$^{+	0.25	}_{-	0.23 	}$	& 	233.56	$^{+	22.71 	}_{-	22.33 	}$	& 	162.71	$^{+	16.35 	}_{-	16.18 	}$	& 	37.24 	$\pm$	2.22 	& 	4.37 	$\pm$	0.67 	& 	10.37$^c$	& 	75.37 	$^{+	2.19 	}_{-	2.35	}$	& 	3.89 	$\pm$	0.63 	& 		-		\\
2145$^*$	& 	65.17$^c$	& 	1.13 	$^{+	0.21	}_{-	0.19 	}$	& 	59.36	$^{+	8.36 	}_{-	8.16 	}$	& 	43.37	$^{+	6.62 	}_{-	6.45 	}$	& 	39.03 	$\pm$	4.57 	& 	1.11 	$\pm$	0.27 	& 	27.89$^c$	& 	61.62 	$^{+	9.49 	}_{-	12.36	}$	& 	1.17 	$\pm$	0.34 	& 	0.0376 	$\pm$	0.0135 	\\
2465	& 	73.74$^c$	& 	2.11 	$^{+	0.16	}_{-	0.15 	}$	& 	152.47	$^{+	14.11 	}_{-	13.85 	}$	& 	116.69	$^{+	10.96 	}_{-	10.91 	}$	& 	38.34 	$\pm$	2.75 	& 	3.04 	$\pm$	0.46 	& 	20.72$^c$	& 	75.70 	$^{+	1.12 	}_{-	0.53	}$	& 	2.94 	$\pm$	0.44 	& 		-		\\
2824$^*$	& 	62.61$^c$	& 	1.70 	$^{+	0.21	}_{-	0.18 	}$	& 	186.6	$^{+	20.20 	}_{-	19.24 	}$	& 	137.68	$^{+	15.33 	}_{-	15.16 	}$	& 	41.46 	$\pm$	3.17 	& 	3.32 	$\pm$	0.58 	& 	2.84$^c$	& 	77.44 	$^{+	2.87 	}_{-	1.72	}$	& 	3.14 	$\pm$	0.63 	& 	0.0086 	$\pm$	0.0056 	\\
3399	& 	72.54$^c$	& 	2.57 	$^{+	0.18	}_{-	0.17 	}$	& 	145.31	$^{+	13.48 	}_{-	13.34 	}$	& 	102.33	$^{+	9.81 	}_{-	9.72 	}$	& 	51.17 	$\pm$	3.05 	& 	2.00 	$\pm$	0.30 	& 	16.43$^c$	& 	80.03 	$^{+	2.09 	}_{-	2.00	}$	& 	1.93 	$\pm$	0.29 	& 		-		\\
4391$^*$	& 	25.84$^c$	& 	0.23 	$^{+	0.08	}_{-	0.07 	}$	& 	148.19	$^{+	19.58 	}_{-	18.84 	}$	& 	138.53	$^{+	18.19 	}_{-	18.16 	}$	& 	46.22 	$\pm$	4.76 	& 	3.00 	$\pm$	0.64 	& 	85.43$^c$	& 	88.73 	$^{+	0.90 	}_{-	1.49	}$	& 	1.16 	$\pm$	0.25 	& 	-0.0245 	$\pm$	0.0051 	\\
6199$^*$	& 	60.66$^c$	& 	1.77 	$^{+	0.4	}_{-	0.35 	}$	& 	40.54	$^{+	4.41 	}_{-	4.22 	}$	& 	37.62	$^{+	4.04 	}_{-	4.03 	}$	& 	37.97 	$\pm$	2.15 	& 	0.99 	$\pm$	0.16 	& 	17.31$^c$	& 	79.58 	$^{+	5.23 	}_{-	5.75	}$	& 	0.88 	$\pm$	0.14 	& 	-0.0492 	$\pm$	0.0027 	\\
6430$^*$	& 	70.12	& 	3.78 	$^{+	0.04	}_{-	0.04 	}$	& 	186.61	$^{+	16.08 	}_{-	16.08 	}$	& 	97.18	$^{+	8.40 	}_{-	8.40 	}$	& 	46.46 	$\pm$	2.50 	& 	2.09 	$\pm$	0.28 	& 	26.34	& 	64.51 	$^{+	0.41 	}_{-	0.42	}$	& 	2.09 	$\pm$	0.28 	& 	0.0007 	$\pm$	0.0017 	\\
6848	& 	48.70$^c$	& 		-					& 		-					& 		-					& 		-		& 		-		& 	-	& 		-					& 		-		& 		-		\\
7314	& 	-	& 	1.87 	$^{+	0.05	}_{-	0.05 	}$	& 	96.99 	$^{+	11.93 	}_{-	11.94 	}$	& 		-					& 	52.31 	$\pm$	5.94 	& 		-		& 	-	& 	89.88 	$^{+	0.09 	}_{-	0.13	}$	& 		-		& 	-0.0069 	$\pm$	0.0053 	\\
7561	& 	71.34$^c$	& 	2.33 	$^{+	0.23	}_{-	0.21 	}$	& 	105.13	$^{+	10.21 	}_{-	10.08 	}$	& 	75.5	$^{+	7.68 	}_{-	7.63 	}$	& 	41.30 	$\pm$	2.52 	& 	1.83 	$\pm$	0.28 	& 	27.75$^c$	& 	62.73 	$^{+	5.17 	}_{-	5.57	}$	& 	1.96 	$\pm$	0.32 	& 		-		\\
8365$^*$	& 	59.34$^c$	& 	2.20 	$^{+	0.02	}_{-	0.02 	}$	& 	183.57	$^{+	15.82 	}_{-	15.82 	}$	& 	116.06	$^{+	10.01 	}_{-	10.01 	}$	& 	38.58 	$\pm$	3.25 	& 	3.01 	$\pm$	0.45 	& 	5.74$^c$	& 	68.89 	$^{+	0.58 	}_{-	0.56	}$	& 	2.78 	$\pm$	0.41 	& 	0.0087 	$\pm$	0.0194 	\\
8512$^*$	& 	46.37$^c$	& 	1.38 	$^{+	0.01	}_{-	0.02 	}$	& 	83.13	$^{+	7.15 	}_{-	7.15 	}$	& 	79.9	$^{+	6.88 	}_{-	6.88 	}$	& 	33.37 	$\pm$	1.99 	& 	2.39 	$\pm$	0.32 	& 	87.26$^c$	& 	72.76 	$^{+	0.27 	}_{-	0.27	}$	& 	2.00 	$\pm$	0.27 	& 	-0.0300 	$\pm$	0.0020 	\\
8576$^*$	& 	31.79$^c$	& 	1.02 	$^{+	0	}_{-	0.00 	}$	& 	199.92	$^{+	17.20 	}_{-	17.20 	}$	& 	167.24	$^{+	14.39 	}_{-	14.39 	}$	& 	44.54 	$\pm$	2.73 	& 	3.75 	$\pm$	0.51 	& 	49.55$^c$	& 	89.92 	$^{+	0.06 	}_{-	0.10	}$	& 	2.02 	$\pm$	0.27 	& 	-0.0820 	$\pm$	0.0065 	\\
8942$^*$	& 	42.27$^c$	& 	0.78 	$^{+	0.06	}_{-	0.06 	}$	& 	212.88	$^{+	26.32 	}_{-	26.30 	}$	& 	175.73	$^{+	21.90 	}_{-	21.87 	}$	& 	54.42 	$\pm$	5.65 	& 	3.23 	$\pm$	0.66 	& 	7.58$^c$	& 	70.56 	$^{+	1.88 	}_{-	1.88	}$	& 	2.11 	$\pm$	0.44 	& 	0.0098 	$\pm$	0.0027 	\\
9337	& 	-	& 	4.65 	$^{+	1.45	}_{-	1.71 	}$	& 	82.54 	$^{+	24.73 	}_{-	24.50 	}$	& 		-					& 	36.39 	$\pm$	4.39 	& 		-		& 	-	& 	74.45 	$^{+	11.47	}_{-	18.93	}$	& 		-		& 		-		\\
9424$^*$	& 	66.42	& 	3.25 	$^{+	0.21	}_{-	0.20 	}$	& 	183.59	$^{+	17.00 	}_{-	16.86 	}$	& 	133.11	$^{+	12.61 	}_{-	12.56 	}$	& 	39.69 	$\pm$	2.82 	& 	3.35 	$\pm$	0.51 	& 	22.71	& 	72.62 	$^{+	2.01 	}_{-	1.93	}$	& 	3.24 	$\pm$	0.51 	& 	0.0288 	$\pm$	0.0014 	\\
9482	& 	-	& 		-					& 		-					& 		-					& 		-		& 		-		& 	-	& 		-					& 		-		& 		-		\\
9527	& 	49.46$^c$	& 		-					& 		-					& 		-					& 		-		& 		-		& 	-	& 		-					& 		-		& 		-		\\
9636$^*$	& 	75.52	& 	3.17 	$^{+	0.82	}_{-	0.58 	}$	& 	153.5	$^{+	31.00 	}_{-	23.66 	}$	& 	115.32	$^{+	22.03 	}_{-	21.00 	}$	& 	20.45 	$\pm$	2.49 	& 	5.64 	$\pm$	1.64 	& 	16.80	& 	84.09 	$^{+	4.10 	}_{-	6.52	}$	& 	5.64 	$\pm$	1.75 	& 	-0.0212 	$\pm$	0.0092 	\\
9812$^*$	& 	43.11	& 	1.35 	$^{+	0.09	}_{-	0.09 	}$	& 	171.28	$^{+	15.26 	}_{-	15.17 	}$	& 	156.3	$^{+	13.92 	}_{-	13.92 	}$	& 	28.95 	$\pm$	1.66 	& 	5.40 	$\pm$	0.75 	& 	17.89	& 	76.31 	$^{+	2.44 	}_{-	2.39	}$	& 	4.19 	$\pm$	0.61 	& 	-0.0358 	$\pm$	0.0012 	\\
9960$^*$	& 	63.26	& 	0.97 	$^{+	0.03	}_{-	0.03 	}$	& 	277.01	$^{+	23.90 	}_{-	23.90 	}$	& 	270.16	$^{+	23.31 	}_{-	23.31 	}$	& 	52.42 	$\pm$	2.98 	& 	5.15 	$\pm$	0.69 	& 	6.65	& 	61.19 	$^{+	0.00 	}_{-	0.00	}$	& 	5.15 	$\pm$	0.69 	& 	-0.0307 	$\pm$	0.0045 	\\
10107	& 	33.90$^c$	& 		-					& 		-					& 		-					& 		-		& 		-		& 	-	& 		-					& 		-		& 	-0.0123 	$\pm$	0.0019 	\\
10502	& 	53.84$^c$	& 		-					& 		-					& 		-					& 		-		& 		-		& 	-	& 		-					& 		-		& 		-		\\
10752	& 	-	& 	5.50 	$^{+	0.99	}_{-	1.44 	}$	& 	193.69 	$^{+	51.03 	}_{-	40.03 	}$	& 		-					& 	89.45 	$\pm$	10.82 	& 		-		& 	-	& 	87.39 	$^{+	1.82	}_{-	3.10	}$	& 		-		& 		-		\\
10863	& 	-	& 		-					& 		-					& 		-					& 		-		& 		-		& 	-	& 		-					& 		-		& 	-0.0133 	-	0.0119 	\\
10910$^*$	& 	39.65	& 	0.75 	$^{+	0.07	}_{-	0.07 	}$	& 	136.49	$^{+	12.04 	}_{-	12.01 	}$	& 	132.01	$^{+	11.63 	}_{-	11.62 	}$	& 	30.16 	$\pm$	1.90 	& 	4.38 	$\pm$	0.61 	& 	5.88	& 	57.24 	$^{+	2.32 	}_{-	2.42	}$	& 	3.34 	$\pm$	0.48 	& 	-0.0220 	$\pm$	0.0032 	\\
11225$^*$	& 	66.42	& 	1.77 	$^{+	0.16	}_{-	0.15 	}$	& 	147.55	$^{+	18.67 	}_{-	18.61 	}$	& 	86.94	$^{+	11.76 	}_{-	11.63 	}$	& 	55.17 	$\pm$	5.62 	& 	1.58 	$\pm$	0.34 	& 	9.54	& 	61.91 	$^{+	0.79 	}_{-	0.36	}$	& 	1.00 	$\pm$	0.22 	& 	0.0011 	$\pm$	0.0020 	\\
11539	& 	30.68$^c$	& 		-					& 		-					& 		-					& 		-		& 		-		& 	-	& 		-					& 		-		& 		-		\\
11702	& 	47.16$^c$	& 		-					& 		-					& 		-					& 		-		& 		-		& 	-	& 		-					& 		-		& 		-		\\
11843	& 	67.67	& 	2.51 	$^{+	0.11	}_{-	0.11 	}$	& 	377.73	$^{+	33.63 	}_{-	33.55 	}$	& 	301.38	$^{+	26.94 	}_{-	26.97 	}$	& 	58.52 	$\pm$	4.92 	& 	5.15 	$\pm$	0.78 	& 	10.98	& 	69.67 	$^{+	2.18 	}_{-	2.12	}$	& 	5.10 	$\pm$	0.76 	& 		-		\\
11944	& 	62.61	& 		-					& 		-					& 		-					& 		-		& 		-		& 	-	& 		-					& 		-		& 	-0.0069 	$\pm$	0.0048 	\\
12015$^*$	& 	47.93	& 	0.53 	$^{+	0.06	}_{-	0.05 	}$	& 	166.33	$^{+	14.70 	}_{-	14.66 	}$	& 	153.45	$^{+	13.60 	}_{-	13.59 	}$	& 	49.98 	$\pm$	2.75 	& 	3.07 	$\pm$	0.42 	& 	6.12	& 	75.54 	$^{+	1.68 	}_{-	1.82	}$	& 	2.28 	$\pm$	0.32 	& 	-0.0417 	$\pm$	0.0025 	\\
12071	& 	72.54	& 		-					& 		-					& 		-					& 		-		& 		-		& 	-	& 		-					& 		-		& 		-		\\
12239	& 	-	& 	3.26 	$^{+	0.7	}_{-	0.51 	}$	& 	179.31	$^{+	35.54 	}_{-	29.63 	}$	& 		-					& 	44.61 	$\pm$	4.84 	& 		-		& 	-	& 	84.45 	$^{+	3.73 	}_{-	4.30	}$	& 		-		& 	-0.0231 	$\pm$	0.0069 	\\
12253	& 	64.53	& 		-					& 		-					& 		-					& 		-		& 		-		& 	-	& 		-					& 		-		& 		-		\\
12773$^*$	& 	63.26	& 	2.86 	$^{+	0.84	}_{-	0.60 	}$	& 	61.05	$^{+	12.26 	}_{-	9.25 	}$	& 	41.39	$^{+	8.70 	}_{-	8.04 	}$	& 	39.55 	$\pm$	2.78 	& 	1.05 	$\pm$	0.31 	& 	44.28	& 	88.33 	$^{+	1.16	}_{-	1.82	}$	& 	0.99 	$\pm$	0.33 	& 	-0.0156 	$\pm$	0.0065 	\\
13182	& 	29.54	& 	0.35 	$^{+	0.09	}_{-	0.08 	}$	& 	139.82	$^{+	15.47 	}_{-	13.58 	}$	& 	132.83	$^{+	13.77 	}_{-	13.75 	}$	& 	68.78 	$\pm$	3.95 	& 	1.93 	$\pm$	0.30 	& 	19.63	& 	81.07 	$^{+	3.73	}_{-	3.42	}$	& 	0.93 	$\pm$	0.15 	& 		-		\\
13416$^*$	& 	58.67	& 	0.50 	$^{+	0.04	}_{-	0.04 	}$	& 	203.64	$^{+	17.80 	}_{-	17.80 	}$	& 	195.31	$^{+	17.08 	}_{-	17.07 	}$	& 	53.23 	$\pm$	3.29 	& 	3.67 	$\pm$	0.51 	& 	9.09	& 	60.37 	$^{+	1.97 	}_{-	2.11	}$	& 	3.61 	$\pm$	0.51 	& 	0.0044 	$\pm$	0.0023 	\\
18188$^*$	& 	42.27	& 	3.11 	$^{+	0.51	}_{-	0.45 	}$	& 	91.88	$^{+	11.18 	}_{-	10.50 	}$	& 	72.42	$^{+	9.04 	}_{-	8.92 	}$	& 	30.06 	$\pm$	2.09 	& 	2.41 	$\pm$	0.45 	& 	2.78	& 	87.75 	$^{+	1.59 	}_{-	2.25	}$	& 	1.71 	$\pm$	0.30 	& 	-0.0528 	$\pm$	0.0069 	\\
18586$^*$	& 	60.00 	& 	5.53 	$^{+	0.33	}_{-	0.61 	}$	& 	193.82	$^{+	27.09 	}_{-	30.19 	}$	& 	51.92	$^{+	8.69 	}_{-	8.43 	}$	& 	28.64 	$\pm$	3.17 	& 	1.81 	$\pm$	0.47 	& 	25.28	& 		-					& 		-		& 	0.0052 	$\pm$	0.0110 	\\
19382	& 	-	& 		-					& 		-					& 		-					& 		-		& 		-		& 	-	& 		-					& 		-		& 		-		\\
29470$^*$	& 	50.21	& 	4.70 	$^{+	0.35	}_{-	0.33 	}$	& 	255.92	$^{+	25.09 	}_{-	24.57 	}$	& 	212.29	$^{+	20.75 	}_{-	20.72 	}$	& 	38.80 	$\pm$	2.29 	& 	5.47 	$\pm$	0.82 	& 	4.22	& 	81.63 	$^{+	0.99 	}_{-	1.00	}$	& 	4.98 	$\pm$	0.68 	& 	-0.0471 	$\pm$	0.0034 	\\

\hline
\end{tabular}\\

Notes:
$^a$ $\rm \Delta PA=|PA_{morph}-PA_{kin}|$\\
$^b$ the N2 gradients are taken from \citetalias{Ju2025}. \\
$^c$ Morphological parameters are taken from F444W measurements.\\
$^*$ These 21 galaxies have both gradients and derived $v/\sigma$ measurements, and are used in the analysis of the $v/\sigma$-gradient relation.
\end{table*}

\subsection{Kinematic Fitting}
\label{sec:kinematic_fitting}

\begin{figure*}
    \centering
    \includegraphics[width=1.0\textwidth,clip,trim={0 0 0 0}]{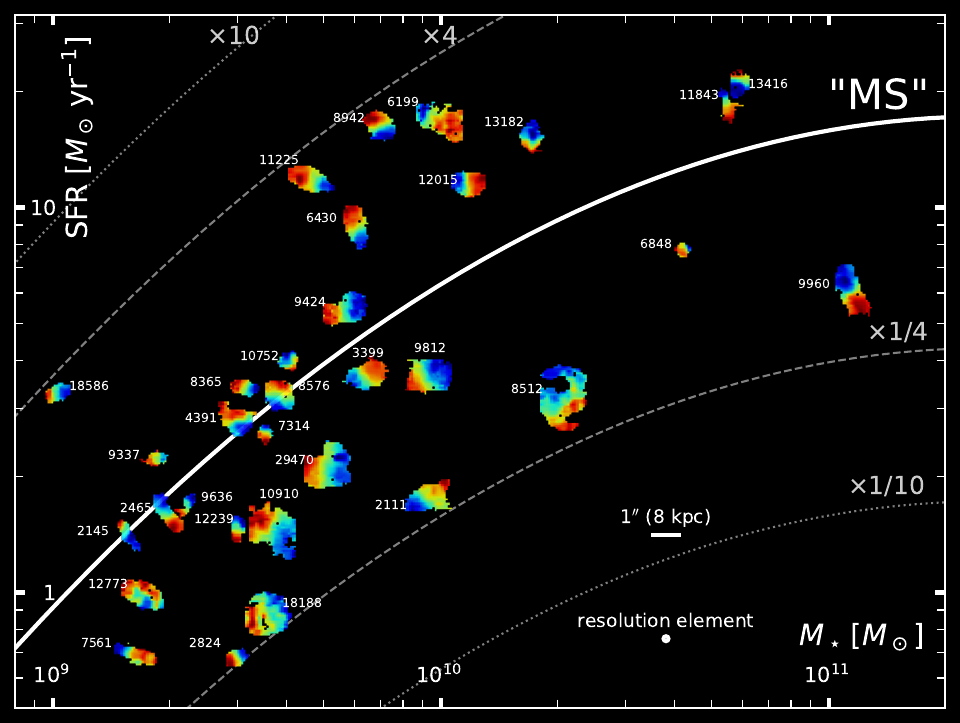} 
    \caption{H$\alpha$ velocity fields of the resolved MSA-3D galaxies at $0.5<z<1.7$ shown at their approximate locations in the SFR-M$_*$ plane. The solid line indicates the star-forming main sequence from \cite{Whitaker2014} at $z \sim 1$, while the dashed and dotted lines show offsets by factors of ×4 and ×10.}

    \label{fig:ms}
\end{figure*}

\begin{figure*}
    \centering
    \includegraphics[width=1.0\textwidth,clip,trim={0 0 0 0}]{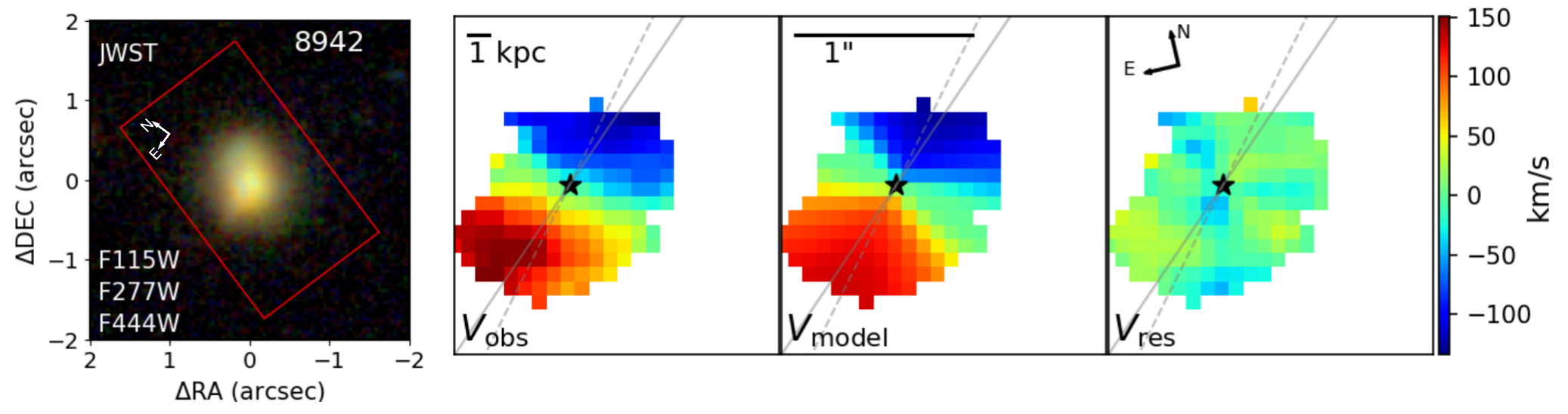} 
    \caption{Example of kinematic modeling for galaxy ID~8942. From left to right: the composite imaging cutout (used for GALFIT modeling), the observed H$\alpha$ velocity field, the best-fit velocity model assuming a rotating disk, and the residual map (data minus model). The observed velocity field exhibits a rotation-dominated morphology with a spider-like pattern. The residual map confirms the quality of the fit, with a reduced $\chi^2 = 1.03$. The gray solid lines indicate the morphological PAs, while the gray dashed lines show the PAs derived from the kinematic models, with a difference of 7.58$^\circ$.
}    
    \label{fig:example}
\end{figure*}

We perform kinematic fits for 43 galaxies in our sample. Four galaxies lack H$\alpha$ coverage and are excluded from H$\alpha$-based analysis. The H$\alpha$ velocity maps of the remaining 39 galaxies are presented in Figure~\ref{fig:ms}.
Following the method described in \citet{Ju2022}, we model the gas velocity maps of 39 galaxies using an arctangent rotation curve:

\begin{equation}
\begin{split}
&V(R) = v_0 + \frac{2}{\pi} v_{rot} \arctan(R/R_{t}),\\
&V_{obs}(R) = V(R)\cdot\sin i\cdot\cos \phi.\\
   \label{eq:vel}
\end{split}
\end{equation} 
where $v_0$ is the residual velocity offset, $v_{rot}$ is the asymptotic rotation velocity, $R_t$ is a turnover radius, 
$i$ is the inclination angle of the rotating disk, and $\phi$ is the azimuthal angle in the galaxy plane. 
We adopt a thin rotating disk model. For a given spaxel, its sky-plane coordinates are first rotated by the kinematic position angle PA and deprojected by the inclination $i$, yielding the in-plane radius $R$ and azimuthal angle $\phi$, where $\phi=0$ corresponds to the major axis.
We apply this kinematic model to the observed H$\alpha$ velocity fields, fixing the center position and inclination to the values from {\tt GALFIT} (for the 17 galaxies with both F444W and F160W measurements, the F444W values are adopted), and use the nested sampling code \texttt{nautilus} \citep{Lange2023} to explore the posterior distributions. For galaxies not well fitted by a single-component model, we use the H$\alpha$ peak as the kinematic center and leave the inclination as a free parameter.
As an example of our kinematic fitting procedure, Figures~\ref{fig:example} show the results for galaxy ID~8942. In Figure~\ref{fig:example}, from left to right, the panels display the false-color composite image (blue: F115W, green: F277W, red: F444W), the observed H$\alpha$ velocity field, the best-fit rotating disk model, and the residual map. The small residuals indicate an excellent fit, with a reduced $\chi^2$ of 1.03. The morphological PAs (gray solid lines) and kinematic model PAs (gray dashed lines) differ by 7.58$^\circ$. The kinematic parameters are listed in Table~\ref{table:vs}.

For the remaining 39 galaxies, two (IDs 6848 and 9482) contain too few spaxels for a reliable constraint on $v_{rot}$, and six exhibit highly irregular velocity fields that cannot be modeled with a simple arctangent function. The remaining 31 galaxies yield reliable rotational velocity measurements, these results are summarized in Table~\ref{table:vs}.

Among the six galaxies whose velocity fields cannot be reliably modeled due to irregular kinematics, two galaxies (IDs 10107 and 11944) have measured metallicity gradients. Both exhibit gradients of approximately $-0.02$ dex/kpc, which are slightly lower than the median metallicity gradient of the MSA-3D sample.
For the 31 galaxies with successful kinematic modeling, 27 have both morphological and kinematic position angles. We computed $\rm \Delta PA = |PA_{morph} - PA_{kin}|$, which are listed in Table~\ref{table:vs}, with a median difference of $17^\circ$. When both F444W and F160W morphological measurements are available, $\rm PA_{morph}$ from F444W is adopted preferentially. In the 27 galaxies, we selected 17 galaxies with axis ratios $q < 0.6$ for which the morphological measurements are more reliable, as galaxies that are nearly face-on can show less reliable morphological PA measurements \citep{Wisnioski2015, Schreiber2018}. The median $\rm \Delta PA$ for this subset of galaxies is also $17^\circ$.

We compute the intrinsic velocity dispersion of each galaxy as
\begin{equation}
\sigma_{0} = \sqrt{\sigma_{\rm obs}^2-\sigma_{\rm inst}^2},
\end{equation}
where $\sigma_{\rm obs}$ is the observed velocity dispersion measured from the emission lines and $\sigma_{\rm inst}$ represents the instrumental line-spread function (LSF). Among them, 12 galaxies (39\%) satisfy $\sigma_{\rm obs}/\sigma_{\rm inst} < 1.3$, indicating that a subset of the sample lies relatively close to the instrumental resolution limit \citep{Nidever2024, Shajib2025, Isobe2023}. The differences between the pre-launch resolution estimates and updated IFS-based models are modest for the G140H configuration over the wavelength range relevant to our sample \citep{Shajib2025}. For the NIRSpec G140H grating ($R \approx 2700$), we derive the instrumental FWHM at the observed H$\alpha$ wavelength of each galaxy by interpolating the wavelength-dependent pre-launch resolution curve.
Uncertainties on $\sigma_{\rm obs}$ are estimated using 1000 bootstrap realizations and are propagated to $\sigma_0$ through standard error propagation.
Specifically, we define the intrinsic dispersion $\sigma_0$ as the median value of the velocity dispersion measured from spaxels located at galactocentric radii beyond 1.5~kpc, where the rotation curves are typically flat and the impact of beam smearing is minimized \citep[e.g.,][]{Wisnioski2015}. Following \citet{Wisnioski2015, Burkert2016, Rizzo2024}, any remaining effects of finite spatial resolution and pixel sampling are incorporated into the error budget as additional systematic uncertainties in both $v$ and $\sigma$.
We also note that the effective LSF can in principle be influenced by the detailed morphology of the source within the slit \citep[e.g.,][]{Graaff2024}. A fully self-consistent treatment would require forward modeling of the spatial light distribution and its coupling to the spectrograph optics, which is beyond the scope of the present work. We therefore treat the adopted instrumental resolution as an effective, aperture-averaged approximation.

We note that in some cases the best-fit rotation curve continues to rise beyond the radial extent of the available kinematic data. In such cases, defining a characteristic rotation velocity based on extrapolated models may introduce systematic uncertainties. To mitigate this effect, we adopt the rotation velocity evaluated at the effective radius, ($v_{rot}(1.5R_e)$), where the kinematics are directly constrained by the observations. This definition follows the approach of \citet{Lee_2025} and ensures a more robust and uniform comparison of rotation velocities across the sample. The corresponding $v_{rot}(1.5R_e)$ values are used to compute $v/\sigma$, which are listed in Table~\ref{table:vs} for the 27 galaxies with reliable measurements of $R_e$.

We present the H$\alpha$ kinematic information for these 39 galaxies in Appendix~\ref{sec:appendix}. For the 31 galaxies with successful kinematic fits, the first row shows the galaxy image, the observed H$\alpha$ velocity map, the best-fit model, and the residual map; the second row displays the intrinsic velocity dispersion map and the one-dimensional radial profiles of $v$ and $\sigma$. For the remaining eight galaxies, which cannot be reliably modeled, we show only the galaxy image, the observed velocity map, and the intrinsic velocity dispersion map.

\begin{figure}
    \centering
    \includegraphics[width=1.0\textwidth,clip,trim={0 0 0 0}]{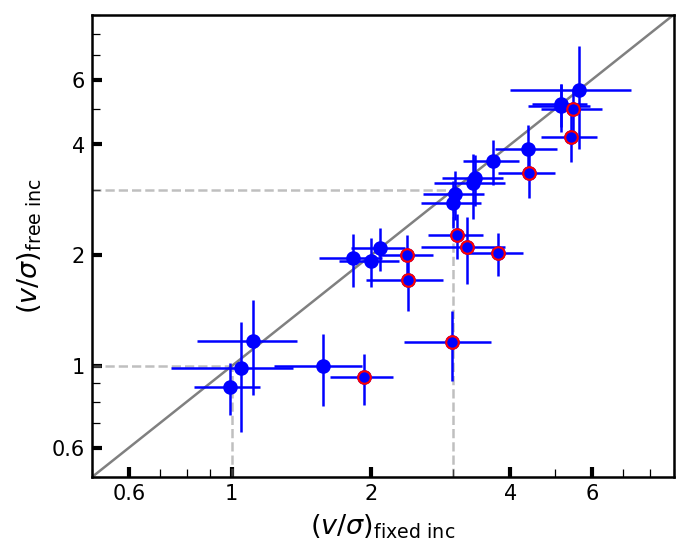} 
    \caption{
Comparison of the $v/\sigma$ values obtained by fixing the inclination and by leaving it as a free parameter. The two measurements show good agreement, with most points lying close to the one-to-one relation. Galaxies with $q>0.6$ ($i<53^\circ$) are marked with red circles (10 objects) and show the largest discrepancies between the two fitting approaches, resulting in strong deviations from the one-to-one relation.}
    \label{fig:b2}
\end{figure}

For comparison, we also performed a fit without fixing the inclination angle, allowing it to be a free parameter. However, for ID 18586, the fit did not succeed when the inclination was treated as a free parameter. The results from this comparison are also shown in Table~\ref{table:vs}.
We also compared the $v/\sigma$ values derived from the two methods, as shown in Figure~\ref{fig:b2}. The x-axis represents the results obtained by fixing the inclination angle, while the y-axis shows the results with the inclination angle treated as a free parameter. It can be seen that most of the galaxies fall near the one-to-one line (shown in gray). The $v/\sigma$ values obtained from the two methods are generally consistent, with a median difference of approximately 0.26, indicating that fixing the inclination has a minimal effect on the derived $v/\sigma$. Galaxies with relatively round morphologies ($q>0.6$, corresponding to $i<53^\circ$) are highlighted with red circles (10 objects in total). 
These systems show the largest discrepancies between the two fitting approaches.

For galaxies with high axis ratios ($q>0.6$), we find that the inclination inferred from free-inclination fits is systematically larger than the photometrically fixed value, with a mean difference of $\sim40^\circ$. This behavior reflects the strong degeneracy between inclination and rotation velocity in nearly face-on systems. Adopting a larger inclination leads to a lower deprojected rotation velocity and hence a reduced $v/\sigma$, explaining the systematic offset observed between the two measurements. We therefore rely on the fixed-inclination results in our main analysis, which provide more stable estimates for these galaxies.

We perform forward-modelling tests using {\tt DysmalPy} on several galaxies \citep{Davies_2004a, Wuyts_2016, Lee_2025}. The resulting rotation curve and $v/\sigma$ agree well with those from our arctangent-based analysis, indicating that our simplified method reliably captures the intrinsic kinematics. For example, in the case of galaxy ID 11225, the forward modeling yields $v/\sigma = 1.45 \pm 0.02$ at $1\ R_e$, while our arctangent-based analysis gives $\sigma_0 = 55.17$ km s$^{-1}$ and $v/\sigma = 1.24$ at the same radius. This close consistency indicates that our simplified method reliably captures the intrinsic kinematics. A detailed description of the modeling procedure and a full analysis of rotation curve shapes and dark matter content will be presented in Espejo Salcedo et al. (in prep.).

\section{Results}\label{sec:method}

\begin{figure}
    \centering
    \includegraphics[width=1.0\textwidth,clip,trim={0 0 0 0}]{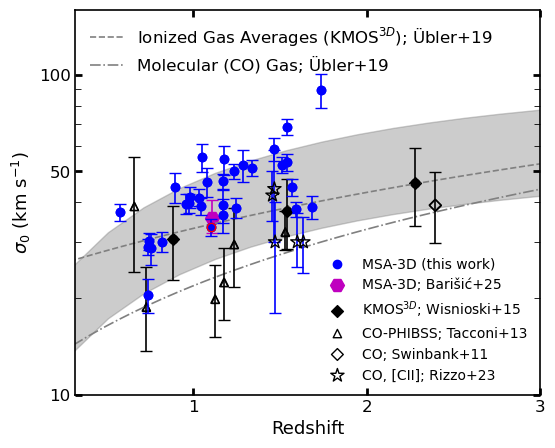} 
    \caption{Intrinsic velocity dispersion $\sigma_0$ as a function of redshift. Blue filled circles represent the MSA-3D sample (this work). Galaxy 8512 is shown as a blue point with a red edge in this work, while the purple hexagon indicates the result from \citet{Ivana2025}. Other samples shown for comparison are 
    KMOS$^{\rm 3D}$ \citep[diamonds;][]{Wisnioski2015}, 
    and CO-based measurements from CO-PHIBSS \citep[open triangles;][]{Tacconi2013} and \citet[][open diamonds]{Swinbank2011}. The gray dashed and solid curves indicate average trends for ionized gas and molecular gas, respectively, from \citet{Ubler2019}. 
    The gray band shows predictions from a simplified Toomre stability model \citep{Romeo2010, Wisnioski2025}, assuming marginally stable disks with $\log(M_*/M_\odot)=9.5-$10 and constant rotational velocity. The upper and lower bounds correspond to $v_{\rm obs}=110-$202 km/s.
    Our sample generally lie slightly above the dashed line, which likely reflects systematic differences in measurement methodology. 
    }
    \label{fig:sigma}
\end{figure}

\begin{figure*}
    \centering
    \includegraphics[width=1.0\textwidth,clip,trim={0 0 0 0}]{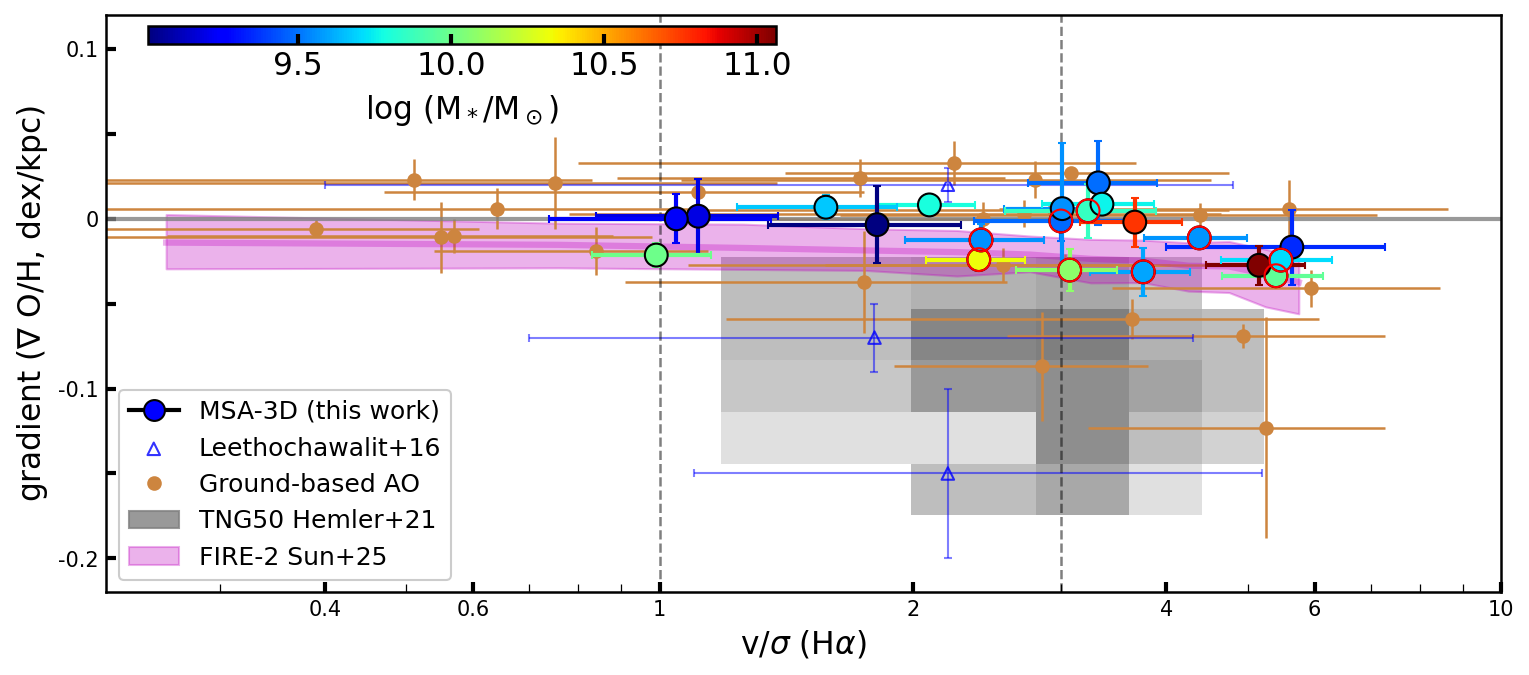} 
    \caption{Metallicity gradient ($\rm \nabla\, O/H$, dex/kpc) as a function of the dynamical support parameter $v/\sigma$ (from H$\alpha$) for our MSA-3D galaxies, shown as filled circles color-coded by stellar mass. Comparison samples are from \citet[][open triangles]{Leethochawalit2016}, \citet[][small light brown]{Sharda2021b}, 
    and the TNG50 simulations density histogram \citep[grey shaded;][]{Hemler2021} and the median trends from FIRE-2 \citep[magenta band;][]{Sun2025} simulations. \cite{Sharda2021b} summarized ground-based AO-assisted observations.
    The MSA-3D galaxies show a moderate anti-correlation, with a Spearman rank coefficient of $\rho(v/\sigma, \nabla \mathrm{O/H}) = -0.30^{+0.16}_{-0.15}$ ($n=21$), and a best-fit slope of $\sim$0.005 dex per dex, consistent with the FIRE-2 simulations.
    Galaxies with $v/\sigma < 1$ exhibit flat metallicity gradients, while those with $v/\sigma > 1$ show a wider range of gradient slopes. }    
    \label{fig:thiswork}
\end{figure*}

\begin{figure}
    \centering
    \includegraphics[width=1.0\textwidth,clip,trim={0 0 0 0}]{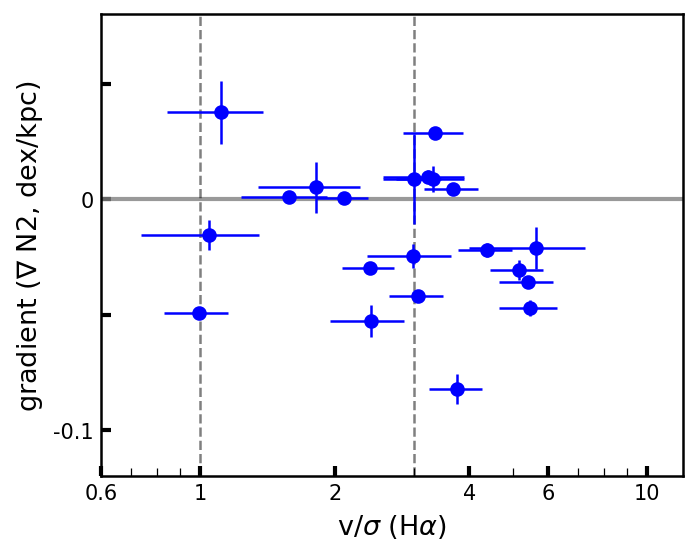} 
    \caption{N2 gradient ($\rm \nabla\, N2$, dex/kpc) as a function of the dynamical support parameter $v/\sigma$ (from H$\alpha$) for our MSA-3D galaxies. The Spearman rank coefficient is $\rho(v/\sigma, \nabla \mathrm{N2}) = -0.19 \pm 0.11$ ($n=21$), where the uncertainty is estimated from 1000 Monte Carlo realizations that perturb the data within their measurement errors. The anti-correlation remains present when expressed directly in terms of N2 gradients.}   
    \label{fig:thiswork-n2}
\end{figure}

\begin{figure}
    \centering
    \includegraphics[width=1.0\textwidth,clip,trim={0 0 0 0}]{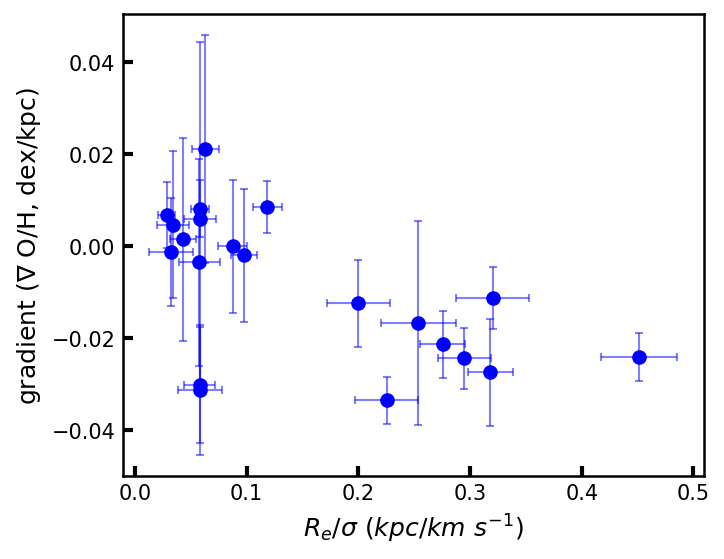} 
    \caption{Metallicity gradient versus $R_e/\sigma$ for the MSA-3D sample. Galaxies with smaller $R_e/\sigma$ show systematically flatter metallicity gradients, whereas systems with larger $R_e/\sigma$ tend to exhibit more negative gradients. The Spearman rank correlation coefficient is $\rho = -0.43^{+0.14}_{-0.13}$.} 
    \label{fig:thiswork2}
\end{figure}

In this work, we fit the H$\alpha$ velocity fields from the MSA-3D survey using an arctangent function. The morphological center and inclination angle were fixed as parameters in the fit. A total of 31 galaxy velocity fields are successfully fitted, yielding their rotation velocities ($v_{rot}$) and intrinsic velocity dispersions ($\sigma_0$); requiring a reliable measurement of $R_e$ to evaluate $v/\sigma$ at $1.5R_e$, we obtain robust $v/\sigma$ estimates for 27 galaxies.

Figure~\ref{fig:sigma} presents the intrinsic velocity dispersion $\sigma_0$ measured for our MSA-3D galaxies (blue filled circles) as a function of redshift, in comparison with previous observational studies. 
Our measurement for ID~8512 is consistent with the value reported in \citet{Ivana2025}. 
Most of our 31 galaxies exhibit $\sigma_0$ values that lie above the average trend for ionized gas in KMOS$^{\rm 3D}$ (gray dashed line). 
This offset likely arises from systematic differences in measurement methodology. 
Although the absolute values of $\sigma_0$ and $v/\sigma$ may therefore differ from other surveys, the relative trends, such as the correlation between $v/\sigma$ and metallicity gradient, remain robust.

The measured $v/\sigma$ values, based on $v_{rot}(1.5R_e)$, are listed in Table~\ref{table:vs} for 27 galaxies with reliable kinematic fits, excluding four objects for which no reliable effective radius could be obtained.
According to the classification criteria proposed by \citet{Kassin2012} and \citet{Girard2020}, galaxies with $v/\sigma > 1$ are classified as rotation-dominated systems, while those with $v/\sigma < 1$ are considered turbulence-dominated systems. Systems with $v/\sigma > 3$ are generally interpreted as more regularly rotating, kinematically settled disks.
Our measured $v/\sigma$ ratios are generally $>1$ (with the smallest value being 0.99), supporting classification as rotationally supported disks.
This conclusion is further supported by observed morphologies. In particular we examine the S\'ersic indices $n$, where values of $n=1$ and $n=4$ correspond to exponential disks and de Vaucouleurs bulges, respectively.
From Table~\ref{table:1}, most galaxies indeed have S\'ersic indices around $n\sim1-2$, consistent with disk-like structures.
We have examined the dependence of $v/\sigma$ on the adopted measurement radius by evaluating the ratio at $R_{e}$, $1.5\,R_{e}$, $2\,R_{e}$, and in the asymptotic limit. While the rotational velocity increases gradually with radius. Consequently, the values of $v/\sigma$ measured at $R_{e}$, $1.5\,R_{e}$, and $2\,R_{e}$ differ only marginally, and most galaxies have $v/\sigma \gtrsim 1$ even at $R_{\rm e}$. In contrast, the asymptotic values of $v/\sigma$ are systematically higher, with the distribution primarily concentrated at $v/\sigma \gtrsim 3$.

We have previously presented spatially resolved gas-phase metallicity (derived using the N2 indicator with the \cite{pp04} calibration, hereafter PP04 N2) for 25 galaxies from MSA-3D project, finding mostly flat or negative metallicity gradients ($\rm \nabla\, O/H$) that correlate with stellar mass, consistent with disk-like structures and predictions from cosmological simulations (\citetalias{Ju2025}). These trends are also consistent with the expected transition from dynamically hot, thick disks to colder, younger thin disks, where gas-phase metallicity patterns trace the chemical imprint of this evolution \citep{Tsukui2025}.
Figure~\ref{fig:thiswork} presents the correlation between metallicity gradients and the dynamical ratio $v/\sigma$ for the 21 galaxies in our sample with both kinematic fits (Section~\ref{sec:kinematic_fitting}) and metallicity gradients from \citetalias{Ju2025}. Our MSA-3D sample is shown as filled circles color-coded by stellar mass. These galaxies span a wide range in $v/\sigma$ (from $\sim$1 to $\sim$6), and most exhibit negative metallicity gradients consistent with inside-out growth. The horizontal error bars represent uncertainties in $v/\sigma$ propagated from both rotation velocity and intrinsic dispersion measurements. 
We find a moderate anti-correlation between the$\rm \nabla\, O/H$ and $v/\sigma$ in the MSA-3D sample, with a Spearman rank correlation coefficient of $\rho(v/\sigma, \nabla \mathrm{O/H}) = -0.30^{+0.16}_{-0.15}$ ($n=21$) where the uncertainty is estimated from 1000 Monte Carlo realizations that perturb each data point within its measurement errors;
a linear fit yields a slope of $\sim$0.005 dex change in the metallicity gradient per dex change in $v/\sigma$, which is weaker than the 0.014 dex per dex dependence on stellar mass reported in \citetalias{Ju2025}.

For comparison, we include two literature samples with spatially resolved metallicity and kinematic measurements. The comparison sample is restricted to galaxies matched in redshift ($0.5<z<1.7$) and stellar mass ($\rm 10^{9.5}<M_*/M_\odot<10^{11.5}$), and further limited to systems with $v/\sigma<10$. We additionally require gas-phase metallicities to be derived using the PP04 N2 calibration to ensure consistency.
From the sample of 15 gravitationally lensed galaxies at $z\sim2$ observed with Keck/LGSAO and OSIRIS by \citet{Leethochawalit2016}, three galaxies satisfy these criteria. From the compilation of five published surveys of non-lensed, star-forming galaxies observed with ground-based IFUs, presented by \citet{Sharda2021b}, a total of 28 galaxies meet our selection, which are restricted to those observed with adaptive optics (AO) to ensure sufficient spatial resolution.
We also show the density histogram from the TNG50 simulations \citep{Hemler2021} and the median trends from the FIRE-2 simulations \citep{Sun2025} for comparison.

Turbulence-dominated galaxies ($v/\sigma < 1$) exhibit nearly flat gradients with little evidence of chemical stratification, whereas rotation-dominated systems ($\rm v/\sigma > 1$) show increasingly negative gradients as ordered motion becomes more prominent. 
This trend is broadly consistent with the predictions from the FIRE-2 simulations \citep[shown as the magenta band;][]{Sun2025}.
A similar pattern is seen in the FIRE-2 galaxies themselves, where high-$v/\sigma$ systems with stronger feedback-driven outflows (e.g., m12c) exhibit flatter metallicity gradients than high-$v/\sigma$ disks with weaker feedback and limited radial mixing (e.g., m12b).
The transition from dynamically hot to cold disks is expected to leave chemical signatures because turbulence regulates radial metal redistribution. In dynamically hot systems (low $v/\sigma$), strong turbulence and gas inflows efficiently mix metals, flattening metallicity gradients. As disks settle and become dynamically colder (higher $v/\sigma$), turbulence weakens and radial mixing becomes less efficient. Inside-out growth and centrally concentrated star formation can then re-establish or steepen negative gradients. The observed anti-correlation between $v/\sigma$ and chemical gradients is therefore consistent with disk settling modulating radial metal mixing. A more detailed discussion can be found in \cite{Sun2025}.

While $v/\sigma$ shows only a weak and statistically insignificant correlation with stellar mass (r = 0.33, p = 0.15), the rotation velocity $v_{rot}$ itself is strongly correlated with stellar mass. We find a significant positive correlation between $v$ and stellar mass, with a Pearson coefficient of r = 0.63 and a p-value of 0.002, such that more massive galaxies rotate faster. 
This trend reflects the well-known Tully-Fisher relation at $z\sim1$ \citep{Tully1977,Miller2011,Kassin2007}.
The lack of a corresponding trend in $v/\sigma$ therefore suggests that the increase in rotation velocity with stellar mass is largely accompanied by a similar increase in intrinsic velocity dispersion $\sigma_0$, effectively compensating the mass dependence in the ratio.

Gas-phase metallicities at intermediate redshift are commonly inferred from strong-line diagnostics, most frequently using the PP04 N2. While this calibration has been widely adopted, strong-line methods are known to carry systematic uncertainties, and different calibrations can yield different slopes in the N2-O/H relation \citep[e.g.,][]{Sanders2025arXiv,Montero2021}. In this work, we therefore examine the robustness of our results to the adopted strong-line calibration by also reporting the directly measured N2 gradients, enabling comparison independent of any particular N2-O/H conversion, which are obtained from \citetalias{Ju2025}. In Figure~\ref{fig:thiswork-n2}, we show the relation between $v/\sigma$ and N2 gradients. We compute the Spearman rank correlation coefficient between $v/\sigma$ and $\nabla\, \mathrm{N2}$ and obtain $\rm \rho(v/\sigma, \nabla\, N2) = -0.23^{+0.07}_{-0.08}$. For comparison, the correlation derived using metallicity gradients yields $\rho(v/\sigma, \nabla\, \mathrm{O/H}) = -0.30^{+0.16}_{-0.15}$. The two measurements are fully consistent within their uncertainties. Although different strong-line calibrations can rescale the absolute amplitude of metallicity gradients, the anti-correlation with $v/\sigma$ remains present when expressed directly in terms of the observed emission-line ratio gradients. This indicates that the qualitative trend does not depend sensitively on the adopted N2-O/H calibration, but is already encoded in the underlying radial N2 measurements. We list the N2 gradients in Table~\ref{table:vs} and present the corresponding maps and 1D gradients in Appendix~\ref{sec:appendix}.

We find a moderate anti-correlation between metallicity gradients and $v/\sigma$ in the MSA-3D sample, indicating that galaxies with higher relative rotational support tend to exhibit more negative metallicity gradients. 
The quantity $R_{\rm e}/\sigma$ can be interpreted as an order-of-magnitude proxy for the radial mixing timescale within a simple turbulent diffusion framework. If radial metal transport is approximated as a diffusive process with diffusion coefficient $D \sim \sigma\, l_{\rm drive}$, where $l_{\rm drive}$ denotes the characteristic driving scale of turbulence, the mixing timescale over a radial distance $R$ is
\begin{equation}
t_{\rm mix} \sim \frac{R^2}{D}.
\end{equation}
If the dominant driving scale is comparable to the disk scale ($l_{\rm drive} \sim R$), as expected for gravitational instabilities and large-scale feedback-driven motions in high-redshift disks, this reduces to
\begin{equation}
t_{\rm mix} \sim \frac{R}{\sigma}.
\end{equation}
Adopting $R_{\rm e}$ as the characteristic radial scale, $R_{\rm e}/\sigma$ therefore provides a dimensional estimate of the turbulent mixing timescale. Systems with larger velocity dispersions are expected to redistribute metals more efficiently and thus develop flatter metallicity gradients, while more extended disks require longer mixing times. This interpretation is broadly consistent with theoretical models and numerical simulations in which turbulence, gravitational instability, clump migration, and feedback-driven flows regulate the evolution of abundance gradients \citep[e.g.,][]{Wisnioski2015, Burkert2016,Ma2017, Hopkins2018}.
In this case, we find a statistically significant anti-correlation, with a Spearman rank coefficient of $\rho = -0.43^{+0.14}_{-0.13}$ (Figure~\ref{fig:thiswork2}).
This comparison suggests that while $v/\sigma$ captures the balance between ordered and random motions, $R_e/\sigma$ more directly reflects the efficiency of radial mixing, thereby providing a more physically motivated predictor of metallicity gradient strength.

From MSA-3D sample, the scatter in metallicity gradients increases with $v/\sigma$. Galaxies with low $v/\sigma$ ($<3$) mostly show flat or weak gradients, with a standard deviation of $\sim0.01$ dex/kpc, whereas rotation-dominated systems ($v/\sigma>3$) span a wider range of slopes, with $\mathrm{std} \sim0.02$ dex/kpc. 
This suggests that strong rotational support, characteristic of dynamically cold disks, is necessary but not sufficient for establishing significant radial metallicity gradients. Additional processes, such as gas inflows or feedback, likely influence how metals are distributed. Dynamically hotter disks with lower $v/\sigma$ may experience enhanced turbulence, which effectively erases or flattens their metallicity gradients.


\section{Conclusion and Discussion} \label{sec:summary}

The MSA-3D survey takes advantage of the powerful spectroscopic capabilities of JWST/NIRSpec to spatially resolve the internal structure of star-forming galaxies. The full sample comprises 43 galaxies, of which 39 have resolved H$\alpha$ kinematic maps and 31 yield robust dynamical fits. 
For an axisymmetric, oblate rotating disk, the kinematic line of nodes is generally expected to align with the projected morphological major axis. In our sample, 27 galaxies have both morphological and kinematic position angles, with a median $\rm \Delta PA \sim 17^\circ$. Most galaxies (22/27) show $\rm \Delta PA < 30^\circ$, indicating broadly consistent orientations. A strong agreement is found in the subset of 17 galaxies with axis ratios $q < 0.6$, where the measurements are most reliable.
Larger misalignments occur in three compact systems (IDs 2145, 12773, 8576) and two nearly face-on galaxies (IDs 8512 and 4391), consistent with \cite{Schreiber2018}, who found that smaller or more face-on galaxies tend to exhibit higher $\rm \Delta PA$. The agreement in the $q < 0.6$ subset further supports the interpretation of these galaxies as rotating disks.

In this work, the ratio $v/\sigma$ is evaluated at $1.5\,R_e$. We find that the majority of galaxies in our sample have $v/\sigma > 1$, indicating rotation-dominated kinematics, with a substantial fraction reaching $v/\sigma > 3$, characteristic of dynamically cold, regularly rotating disks.

Combined with the 25 galaxies for which metallicity gradients were measured in \citetalias{Ju2025}, this results in a final subsample of 21 galaxies for which the connection between gas kinematics and metallicity gradients can be examined. We identify a moderate anti-correlation between $\rm \nabla\, O/H$ and $v/\sigma$, with a Spearman rank correlation coefficient of $\rho(v/\sigma, \nabla \mathrm{O/H}) = -0.30^{+0.16}_{-0.15}$ ($n=21$), such that galaxies with higher relative rotational support tend to exhibit more negative metallicity gradients. However, this trend is accompanied by considerable scatter, particularly among rotation-dominated systems. A linear fit yields a slope of $\sim 0.005$ dex change in the metallicity gradient per dex change in $v/\sigma$, which is weaker than the 0.014 dex per dex dependence on stellar mass reported in \citetalias{Ju2025}. We compute the Spearman rank correlation coefficient between $v/\sigma$ and $\nabla \mathrm{N2}$ and obtain $\rho(v/\sigma, \nabla \mathrm{N2}) = -0.23^{+0.07}_{-0.08}$. This indicates that the qualitative trend does not depend sensitively on the adopted N2-O/H calibration, but is already reflected in the underlying radial N2 measurements.

The overall scatter in metallicity gradients within the MSA-3D sample is relatively small, and no galaxies exhibit extremely steep gradients. This limited dynamic range naturally weakens correlations with global galaxy properties, including $v/\sigma$, which characterizes the relative kinematic support from ordered versus random motions at disk scales. A clearer and statistically stronger trend emerges when considering the ratio $R_e/\sigma$. This quantity can be interpreted as a proxy for the effective radial mixing timescale. We find a statistically significant anti-correlation between $v/\sigma$ and the metallicity gradients, with a Spearman rank coefficient of $\rho = -0.43^{+0.14}_{-0.13}$,
indicating that galaxies with shorter characteristic timescales (implying more efficient mixing) tend to have flatter gradients, while systems with larger $R_e/\sigma$ are able to sustain more pronounced chemical stratification.

Finally, we find that the scatter in metallicity gradients increases toward higher $v/\sigma$. Galaxies with $v/\sigma < 3$ predominantly show flat or weak gradients, whereas systems with $v/\sigma > 3$ span a wider range of slopes. This indicates that strong rotational support is a necessary but not sufficient condition for establishing significant metallicity gradients. Additional processes, such as gas inflows, feedback-driven outflows, and internal radial mixing, likely play a key role in regulating the chemical structure of dynamically cold disks.

To date, three other JWST programs have adopted a similar slit-stepping approach (i.e., GO-3426 (PI: Jones), GO-2123 (PI: Kassin) and GO-4291 (PI: Kassin)).
The application of {\tt DysmalPy} demonstrates the potential of forward modeling to recover intrinsic kinematic properties, providing physically motivated estimates of rotational velocity and velocity dispersion. Building on this framework, future work will extend the analysis to the full MSA-3D sample, enabling a comprehensive study of rotation curves, velocity dispersions, and dark matter contributions (Espejo Salcedo et al., in prep.). These efforts will facilitate more detailed investigations of the relationship between gas kinematics and chemical structure, offering stronger constraints on the physical processes that drive galaxy evolution.

\section*{Acknowledgements}

We thank the anonymous referee for the constructive comments, which significantly improved the manuscript.
This work is supported by the National Key R\&D Program of China No.2025YFF0510603, the National Natural Science Foundation of China (grant 12373009), the CAS Project for Young Scientists in Basic Research Grant No. YSBR-062, the China Manned Space Program with grant no. CMS-CSST-2025-A06, and the Fundamental Research Funds for the Central Universities. XW acknowledges the support by the Xiaomi Young Talents Program, and the work carried out, in part, at the Swinburne University of Technology, sponsored by the ACAMAR visiting fellowship.
TJ acknowledges support from a Chancellor's Fellowship and a Dean's Faculty Fellowship, and from NASA through grant 80NSSC23K1132. 
TT is supported by the JSPS Grant-in-Aid for Research Activity Start-up (25K23392) and the JSPS Core-to-Core Program (JPJSCCA20210003).
This work is based on observations made with the NASA/ESA/CSA James Webb Space Telescope. The data were obtained from the Mikulski Archive for Space Telescopes at the Space Telescope Science Institute, which is operated by the Association of Universities for Research in Astronomy, Inc., under NASA contract NAS 5-03127 for JWST.
These observations are associated with program JWST-GO-2136. The specific observations analyzed can be accessed via \dataset[doi:10.17909/s8wp-5w10]{https://doi.org/10.17909/s8wp-5w10}.
We acknowledge financial support from NASA through grant JWST-GO-2136. 

\facilities{JWST (NIRSpec MSA)}
\software{nautilus \citep{Lange2023}}

\appendix

\section{The spatially resolved 2D maps of our sample galaxies}
\label{sec:appendix}

Here we present the spatially resolved 2D maps of our sample galaxies in Figure~\ref{fig:1}1.

\begin{figure*}
    \centering
    \includegraphics[width=0.8\textwidth,clip,trim={0 0 0 0}]{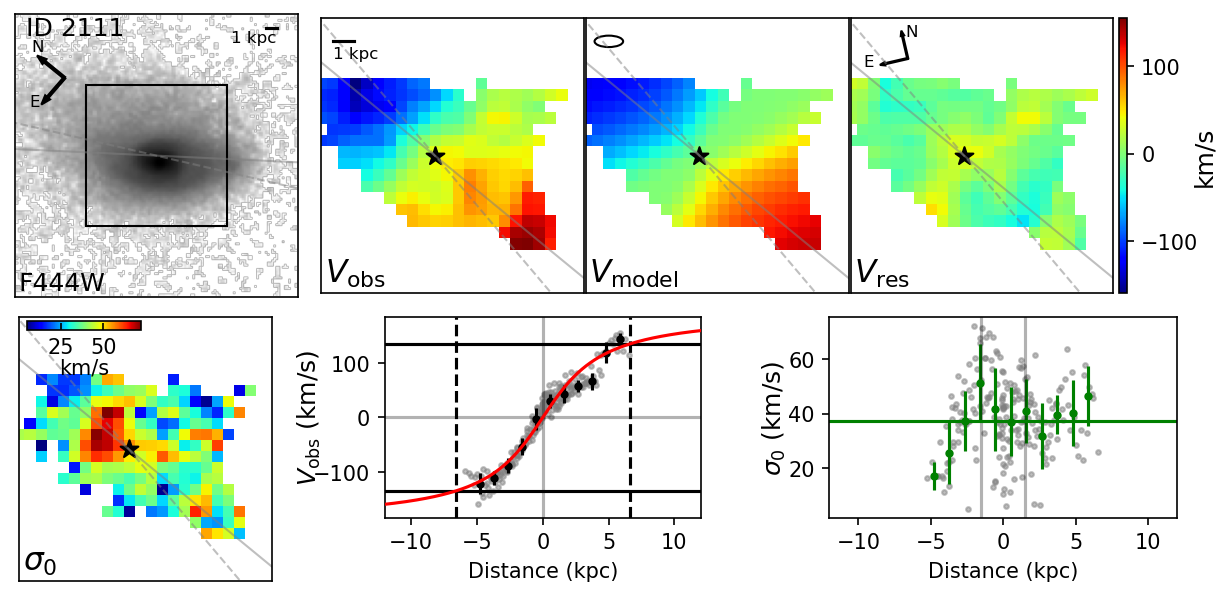}\\
    \vspace{1.2em}
    \includegraphics[width=0.8\textwidth,clip,trim={0 0 0 0}]{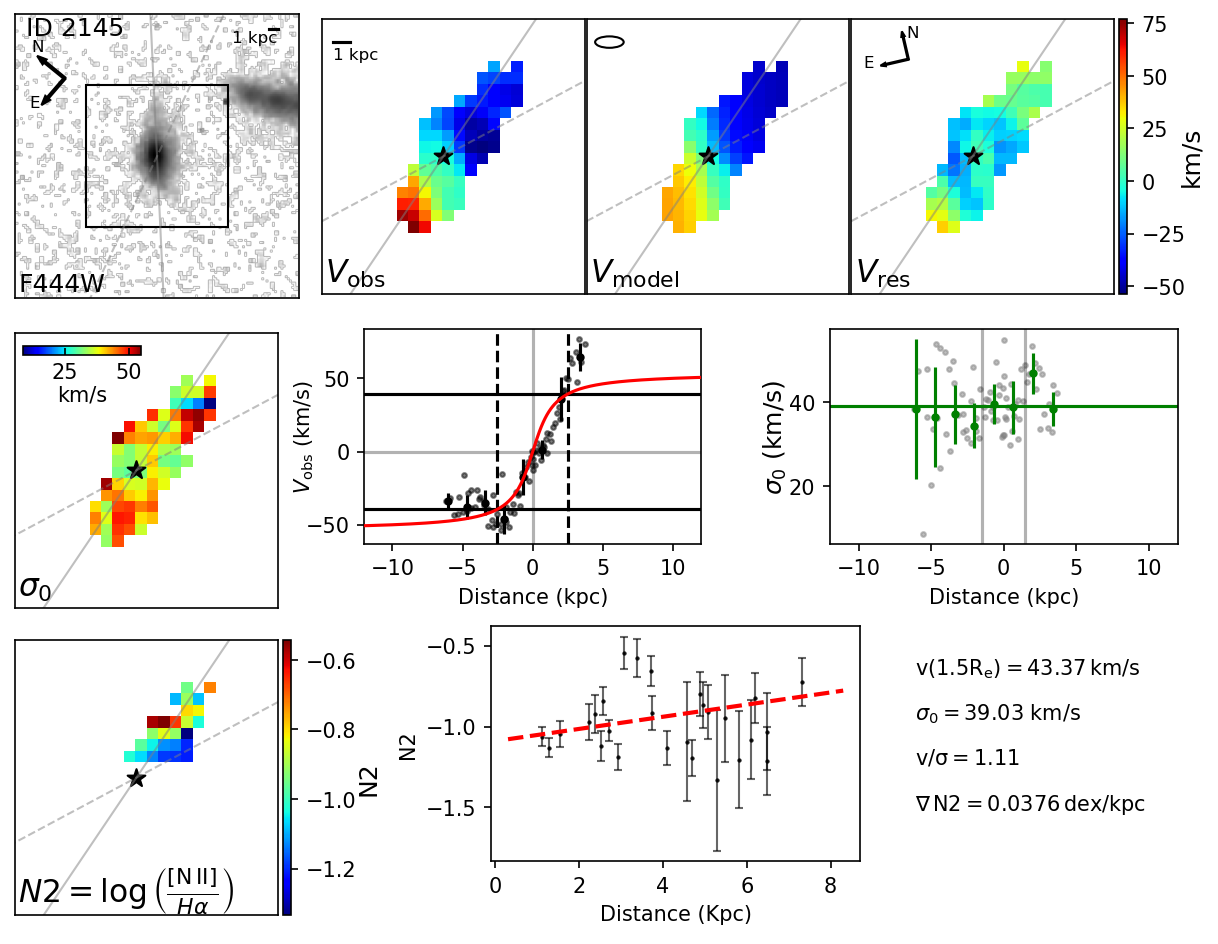}\\
    \flushleft
    {\bf Figure A1.} 
    A comprehensive spatially-resolved view of our galaxy sample with gas kinematic information. For galaxies that have converged kinematic fits, the top row shows (from left to right): the F444W/F160W image, the observed H$\alpha$ velocity field, the best-fit kinematic model, and the residual velocity map. The gray solid lines indicate the morphological PAs, while the gray dashed lines show the PAs derived from the kinematic models. The bottom row presents: the intrinsic velocity dispersion map, 1D radial curves of $v$ and $\sigma_0$ derived from kinematic extractions. The red curve in the 1D radial profiles of $v$ shows the best-fit 1D kinematic model. The black dashed lines mark the effective radius ($\rm 1.5 R_{e}$). For galaxies without converged kinematic fits, we show the F444W/F160W image, the observed H$\alpha$ velocity field, and the intrinsic velocity dispersion map. In all panels, the black star marks the adopted kinematic center, the ellipse indicates the \msasd\ resolution element, and the scale bar corresponds to 1 kpc. Grey lines in the 1D $\sigma_0$ panels indicate the $\pm 1.5$ kpc radial range. The F444W/F160W images have a field of view of $4\arcsec \times 4\arcsec$, with the boxes indicating the {\tt GALFIT} fitting region of $2\arcsec \times 2\arcsec$. The N2 maps and corresponding gradients, taken from \citet{Ju2025}, for the 21 galaxies used in the kinematic–metallicity analysis are shown in the third row.\\

    \label{fig:1}

\end{figure*}

\begin{figure*}
\ContinuedFloat
\centering
    \includegraphics[width=0.8\textwidth,clip,trim={0 0 0 0}]{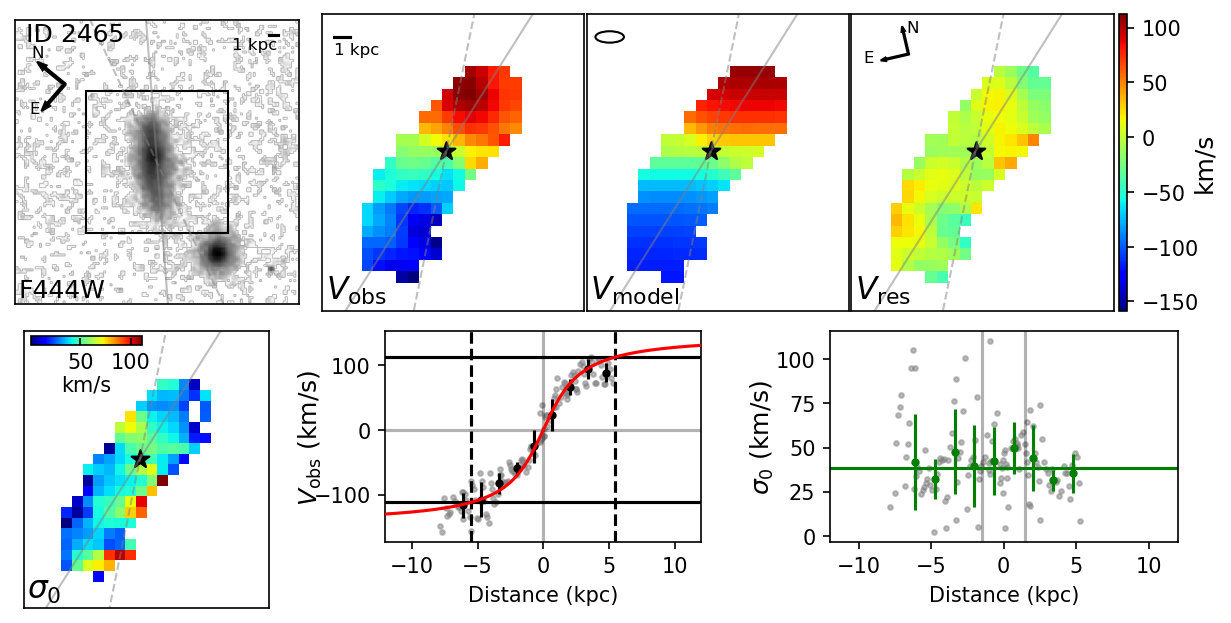}\\
    \vspace{1.2em}
    \includegraphics[width=0.8\textwidth,clip,trim={0 0 0 0}]{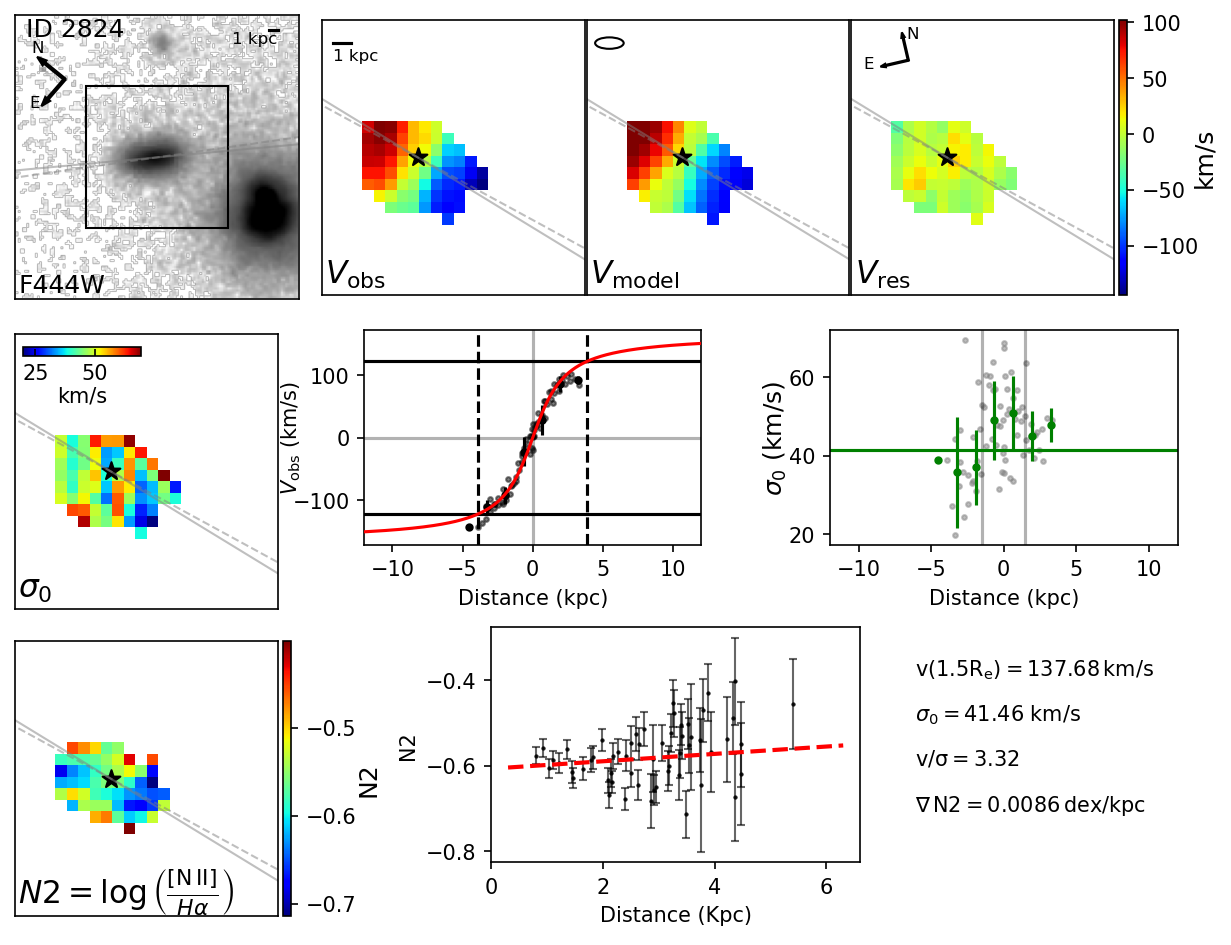}\\
    {\bf Figure A1.}  continued \\
\end{figure*}

\begin{figure*}
\centering
    \ContinuedFloat
    \includegraphics[width=0.8\textwidth,clip,trim={0 0 0 0}]{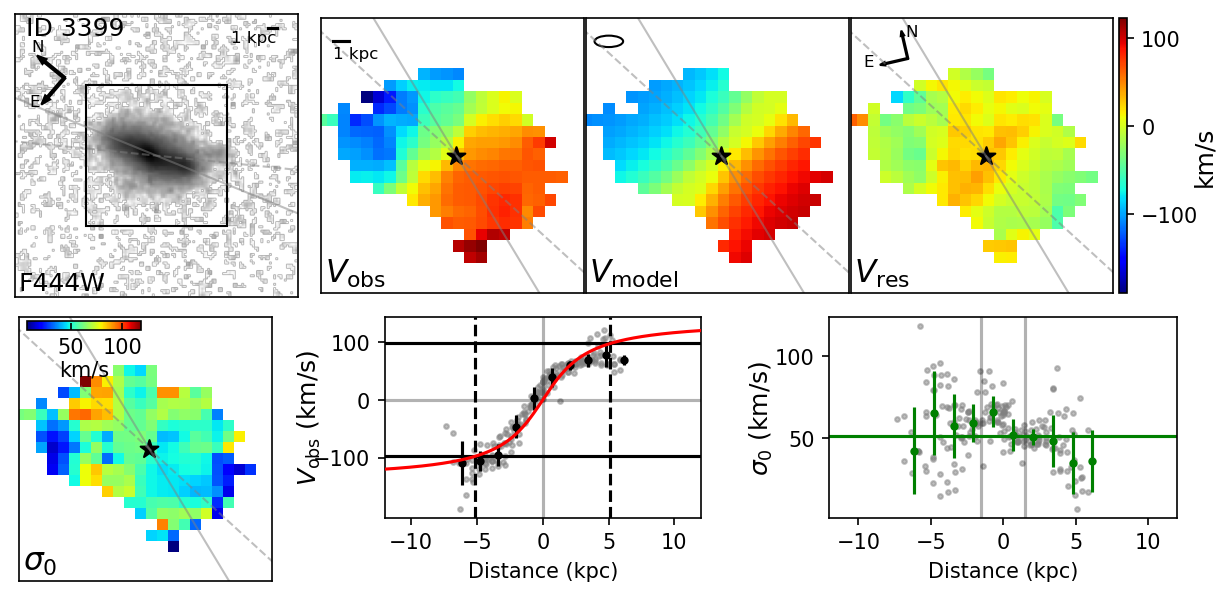}\\
    \vspace{1.2em}
    \includegraphics[width=0.8\textwidth,clip,trim={0 0 0 0}]{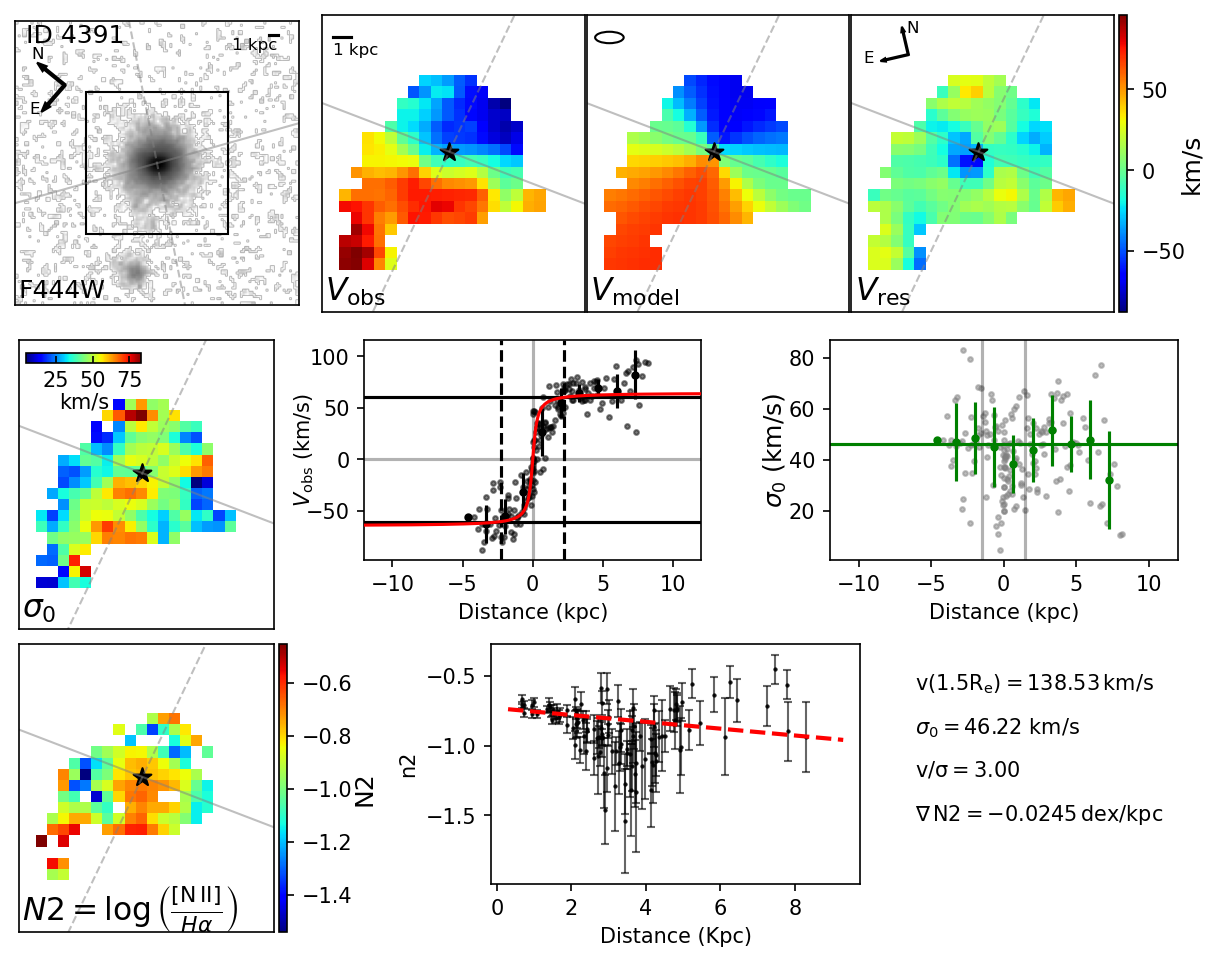}\\
    {\bf Figure A1.} continued\\
\end{figure*}

\begin{figure*}
\centering
    \ContinuedFloat
    \includegraphics[width=0.8\textwidth,clip,trim={0 0 0 0}]{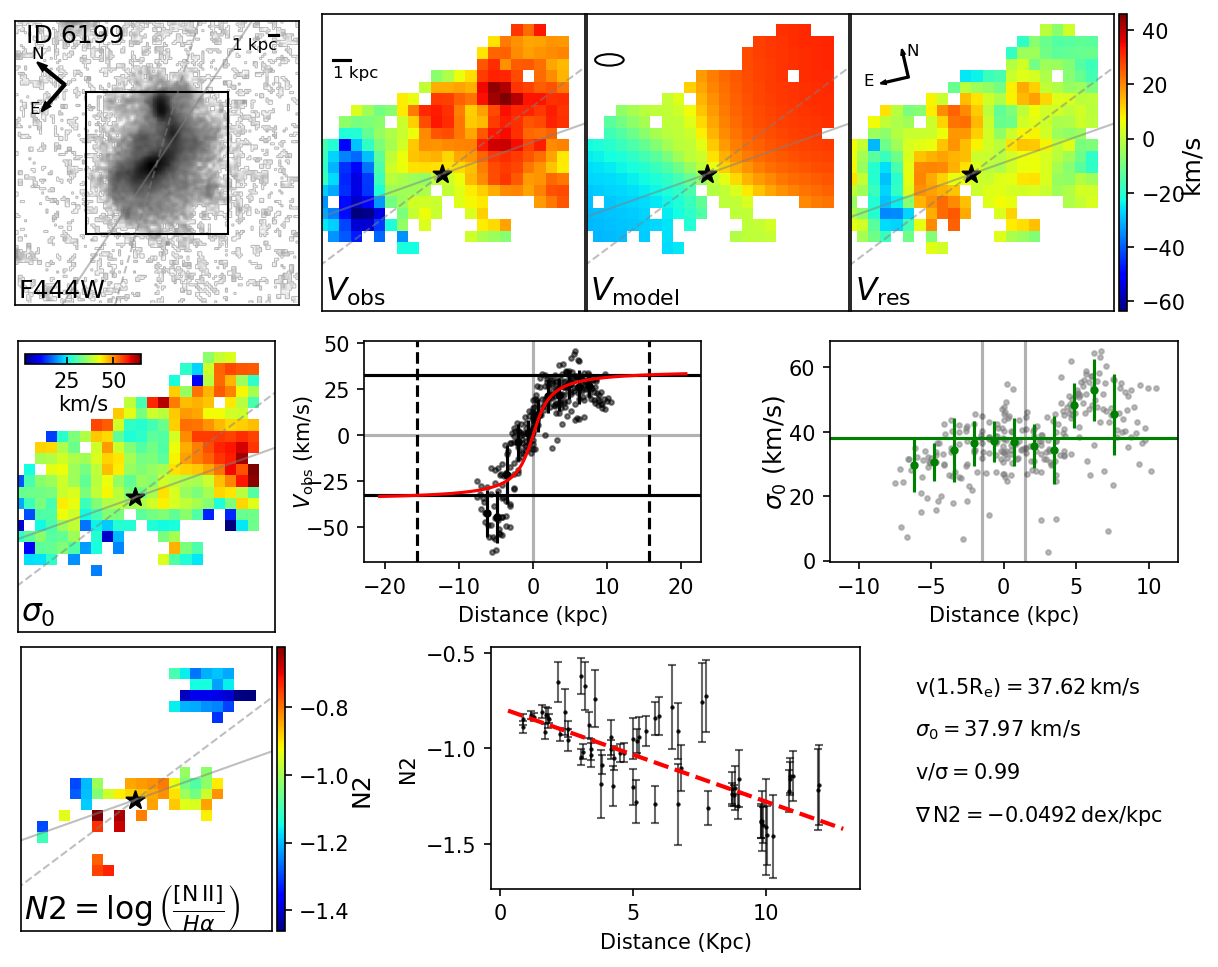}\\
    \vspace{1.2em}
    \includegraphics[width=0.8\textwidth,clip,trim={0 0 0 0}]{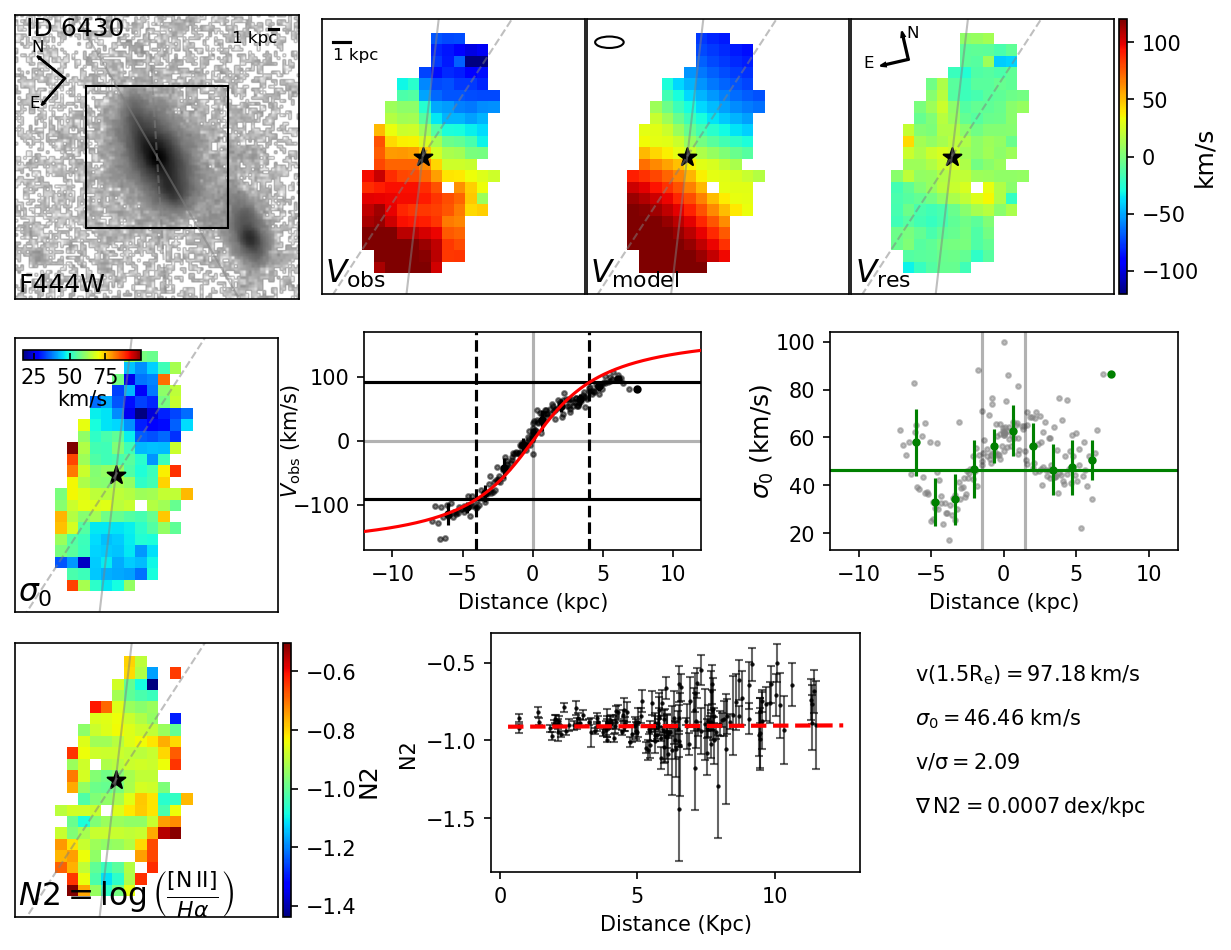}\\
    \vspace{1.2em}
    {\bf Figure A1.} continued\\
\end{figure*}

\begin{figure*}
\centering
    \ContinuedFloat
    \includegraphics[width=0.8\textwidth,clip,trim={0 0 0 0}]{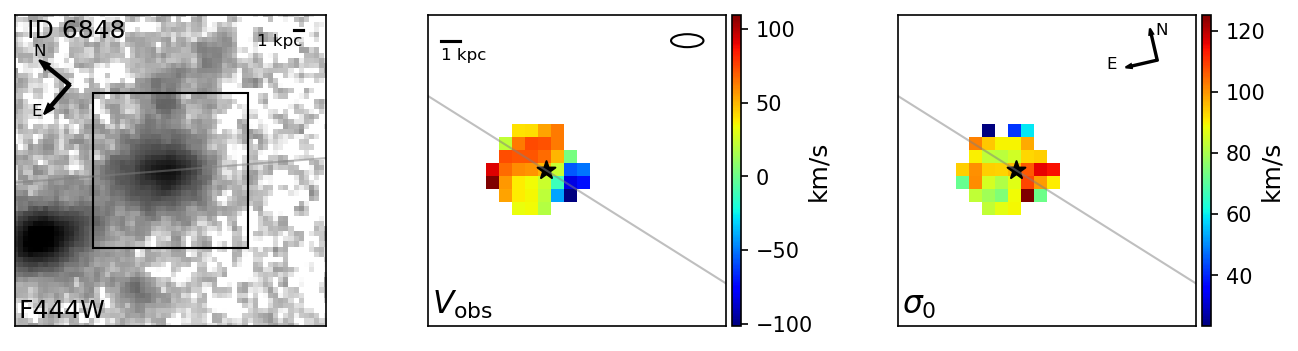}\\
    \vspace{1.2em}
    \includegraphics[width=0.8\textwidth,clip,trim={0 0 0 0}]{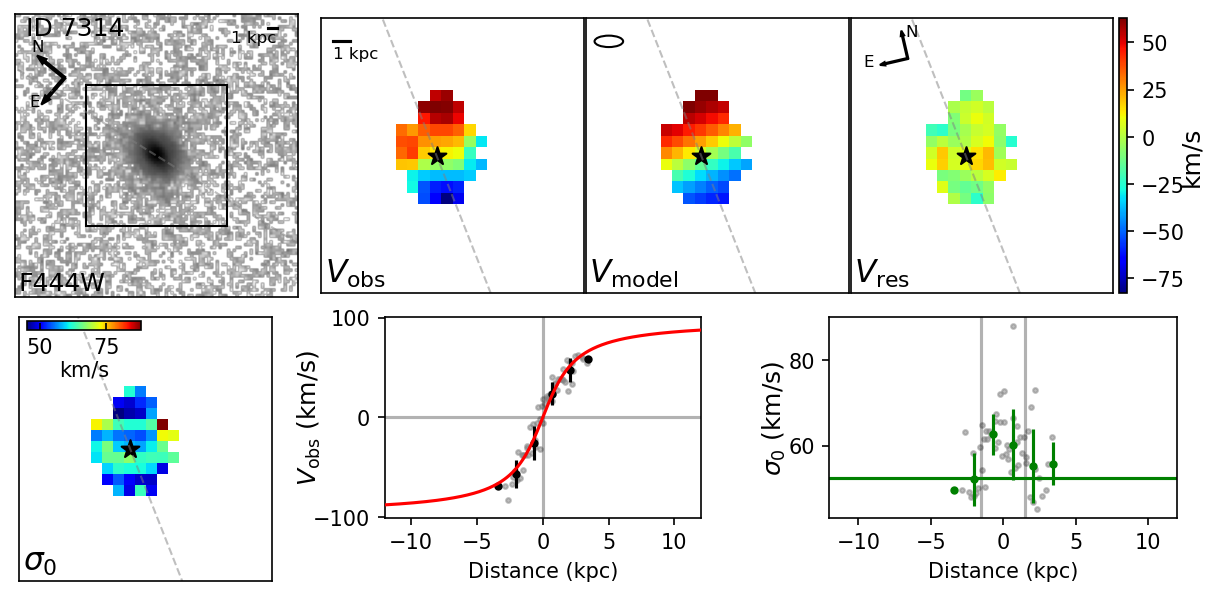}\\
    \vspace{1.2em}
    \includegraphics[width=0.8\textwidth,clip,trim={0 0 0 0}]{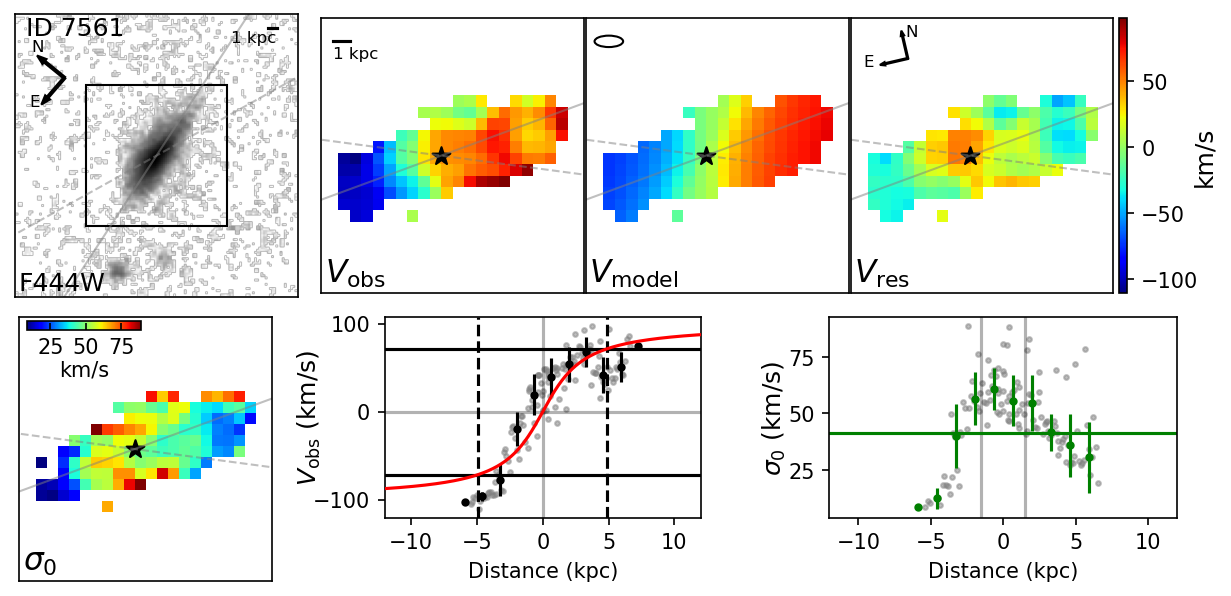}\\
    \vspace{1.2em}

    {\bf Figure A1.} continued\\
\end{figure*}

\begin{figure*}
\centering
    \ContinuedFloat
    \includegraphics[width=0.8\textwidth,clip,trim={0 0 0 0}]{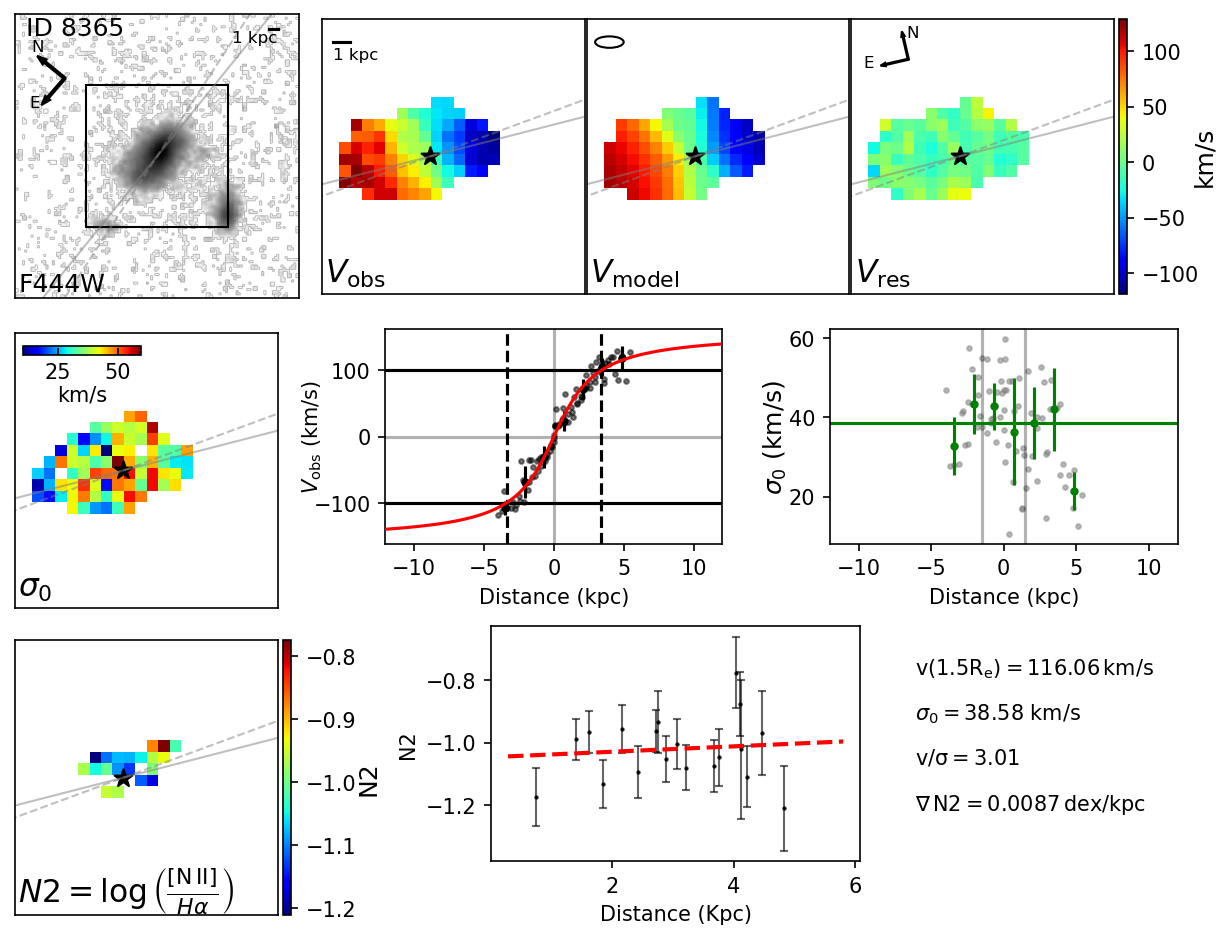}\\
    \vspace{1.em}
    \includegraphics[width=0.8\textwidth,clip,trim={0 0 0 0}]{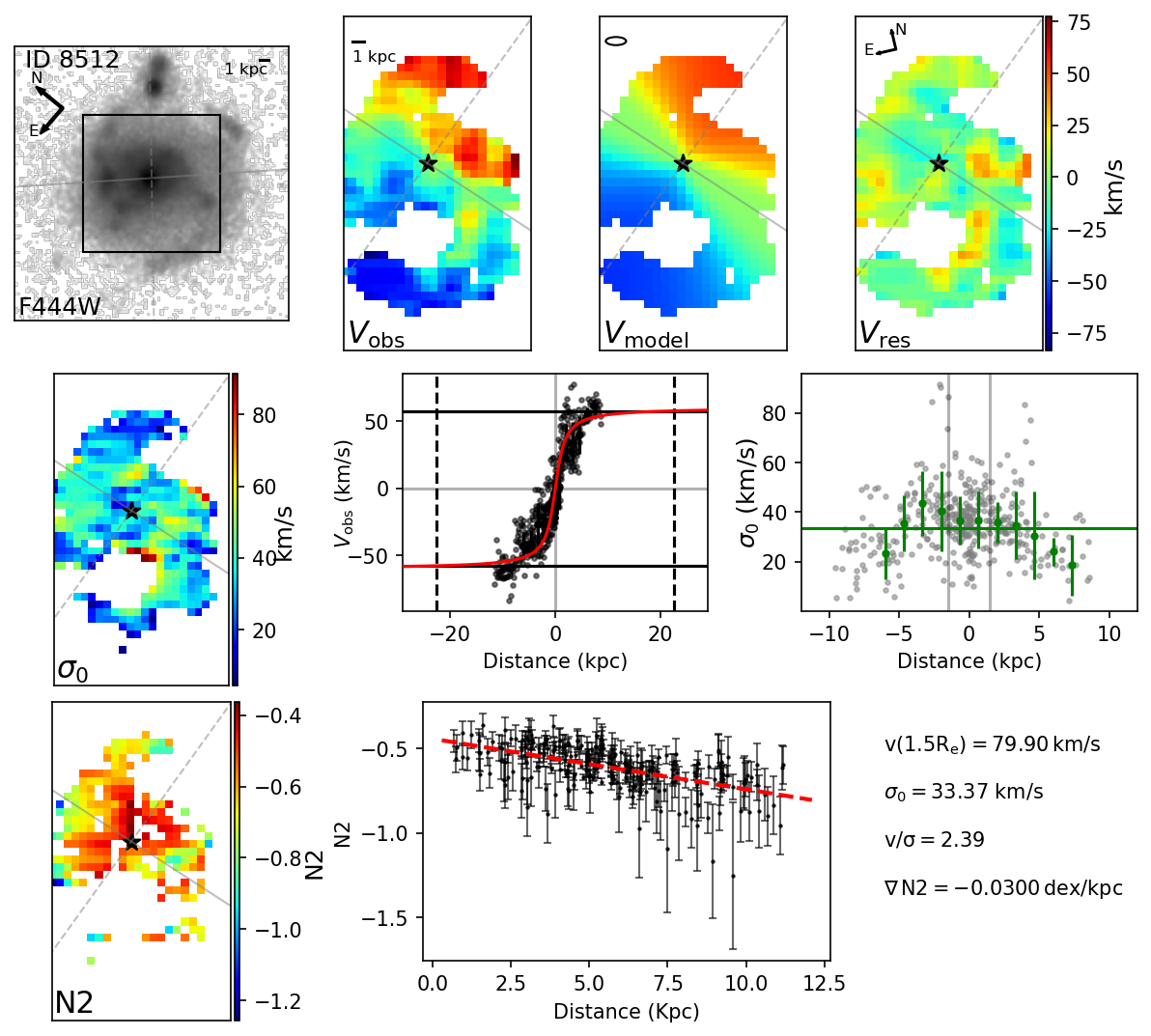}\\
    {\bf Figure A1.} continued\\
\end{figure*}

\begin{figure*}
\centering
    \ContinuedFloat
    \includegraphics[width=0.8\textwidth,clip,trim={0 0 0 0}]{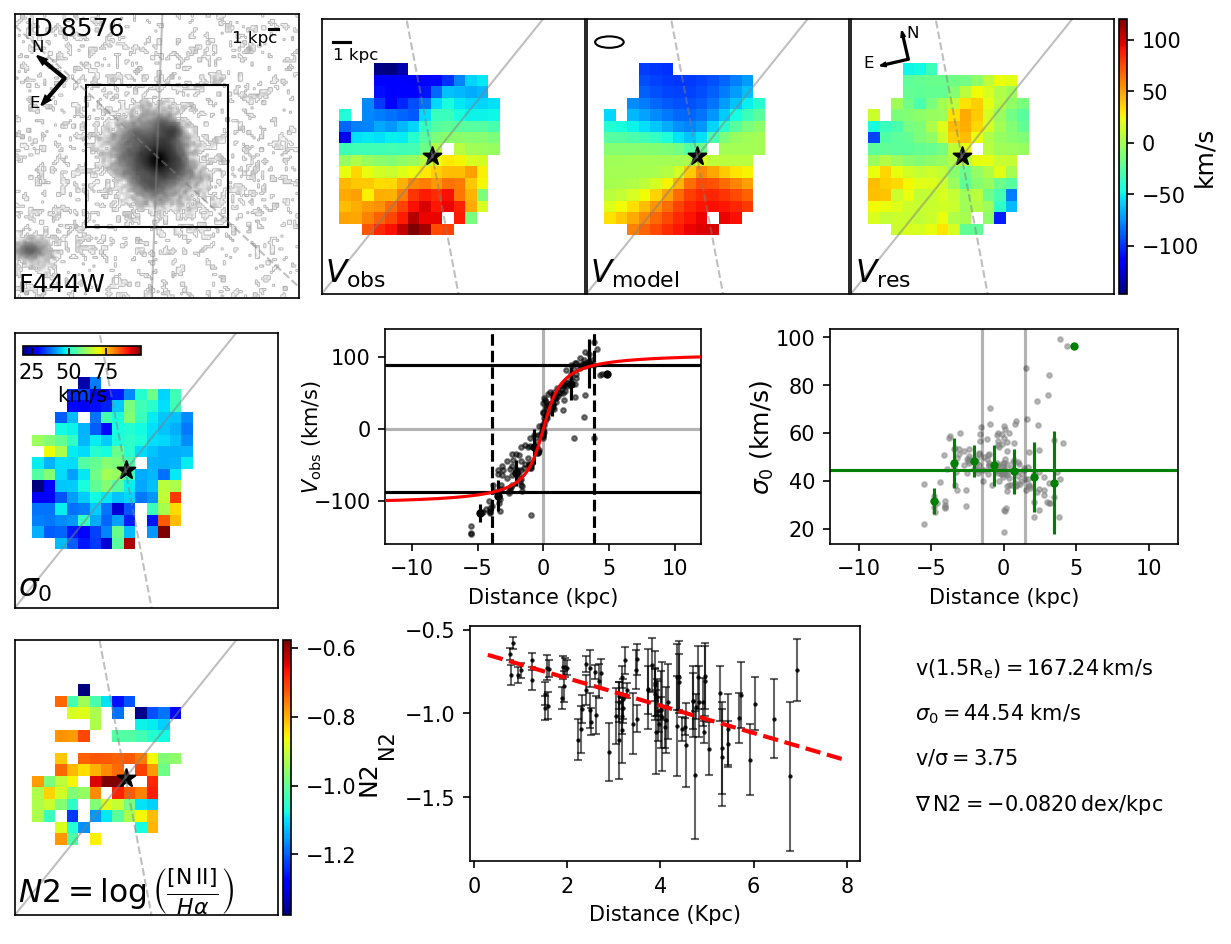}\\
    \vspace{1.2em}
    \includegraphics[width=0.8\textwidth,clip,trim={0 0 0 0}]{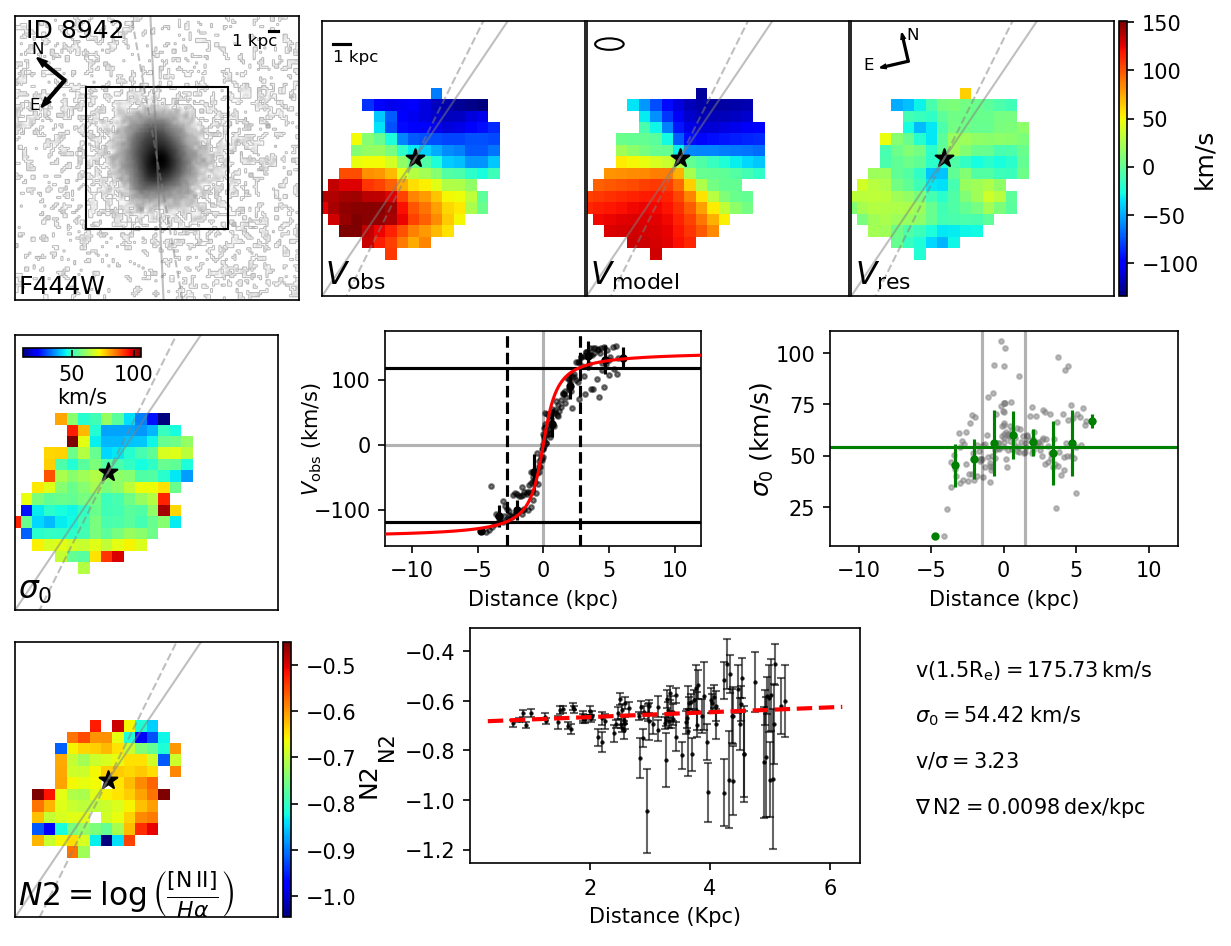}\\
    {\bf Figure A1.} continued\\
\end{figure*}

\begin{figure*}
\centering
    \ContinuedFloat
    \includegraphics[width=0.8\textwidth,clip,trim={0 0 0 0}]{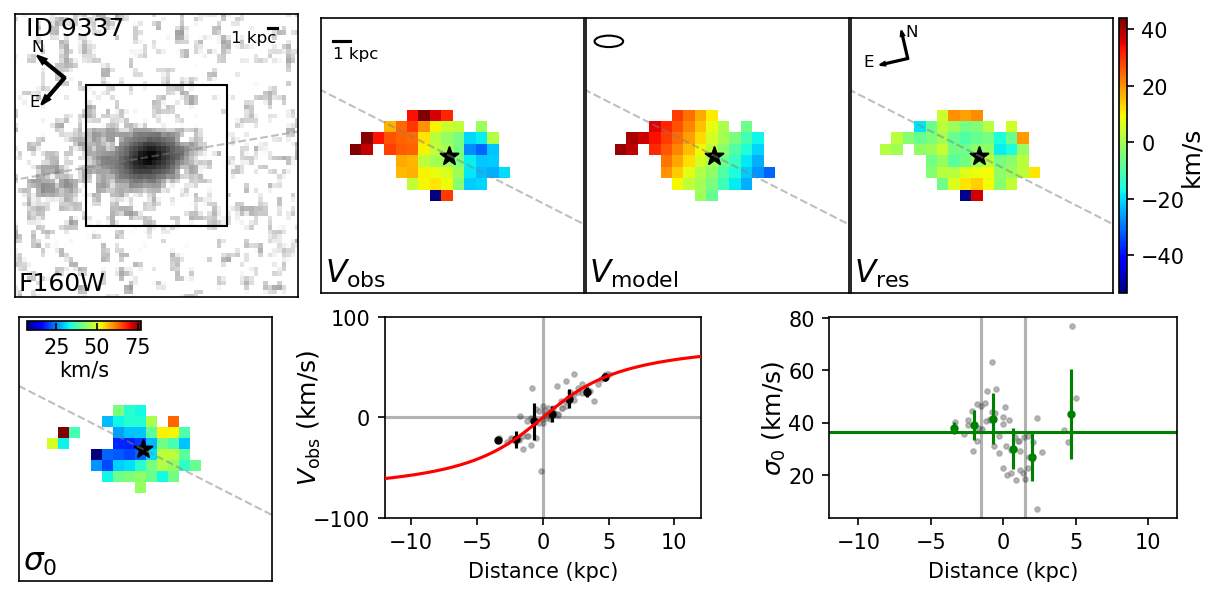}\\
    \vspace{1.2em}
    \includegraphics[width=0.8\textwidth,clip,trim={0 0 0 0}]{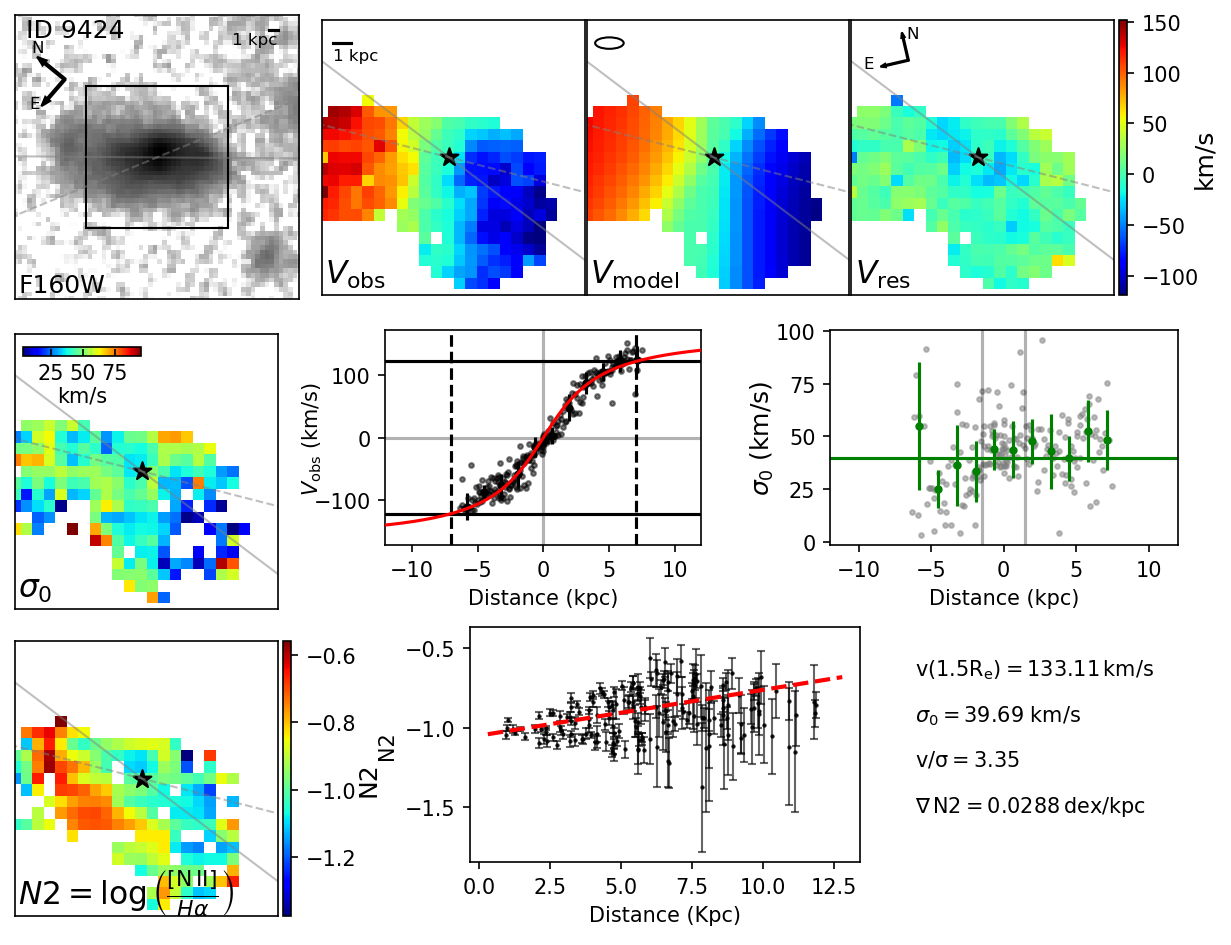}\\
    \vspace{1.2em}
    \includegraphics[width=0.8\textwidth,clip,trim={0 0 0 0}]{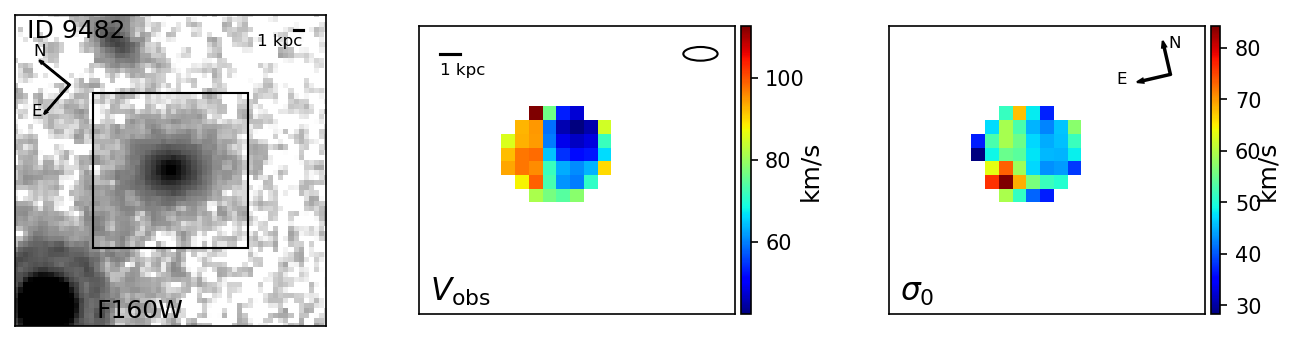}\\
    {\bf Figure A1.} continued\\
\end{figure*}

\begin{figure*}
\centering
    \ContinuedFloat
    \includegraphics[width=0.8\textwidth,clip,trim={0 0 0 0}]{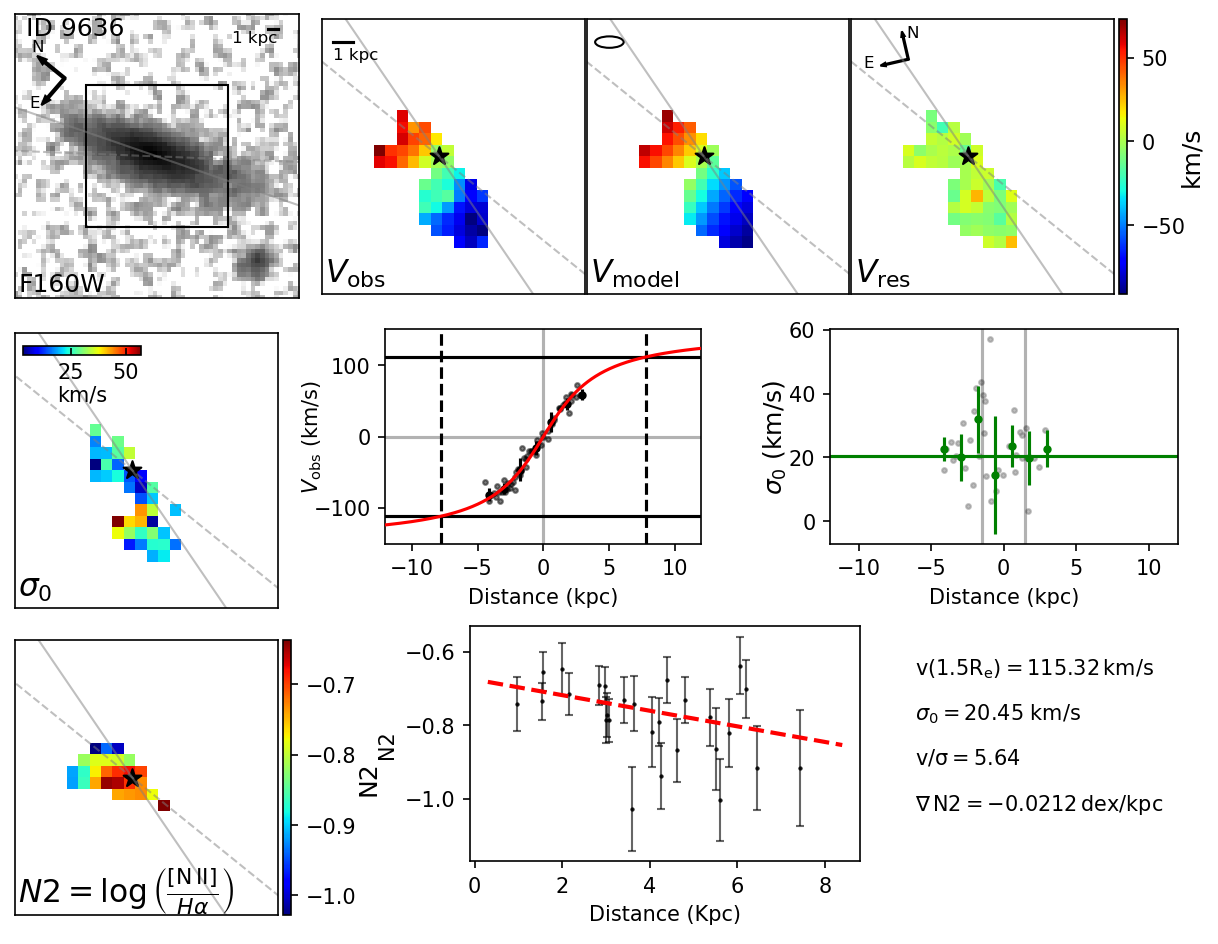}\\
    \vspace{1.2em}
    \includegraphics[width=0.8\textwidth,clip,trim={0 0 0 0}]{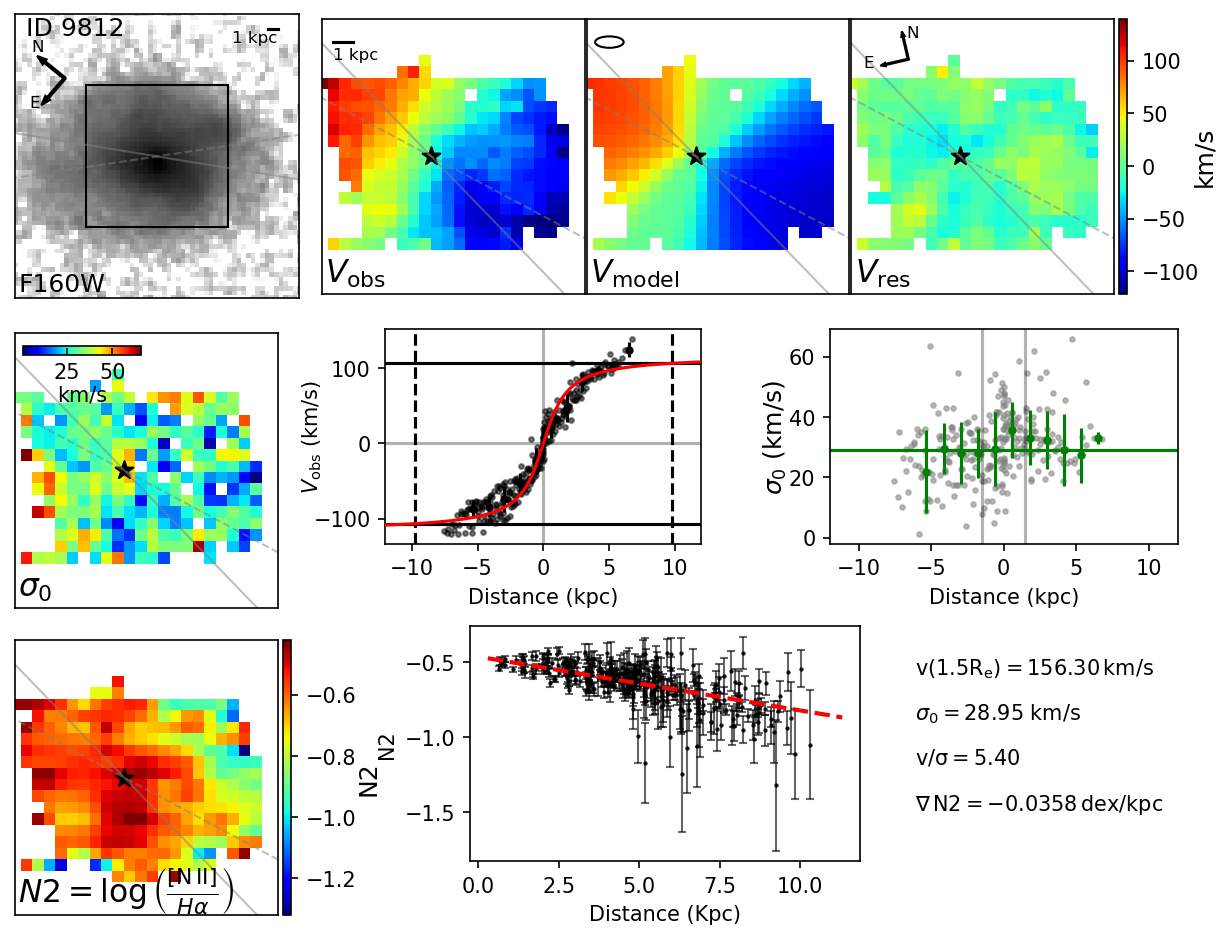}\\
    {\bf Figure A1.} continued\\
\end{figure*}

\begin{figure*}
\centering
    \ContinuedFloat
    \includegraphics[width=0.8\textwidth,clip,trim={0 0 0 0}]{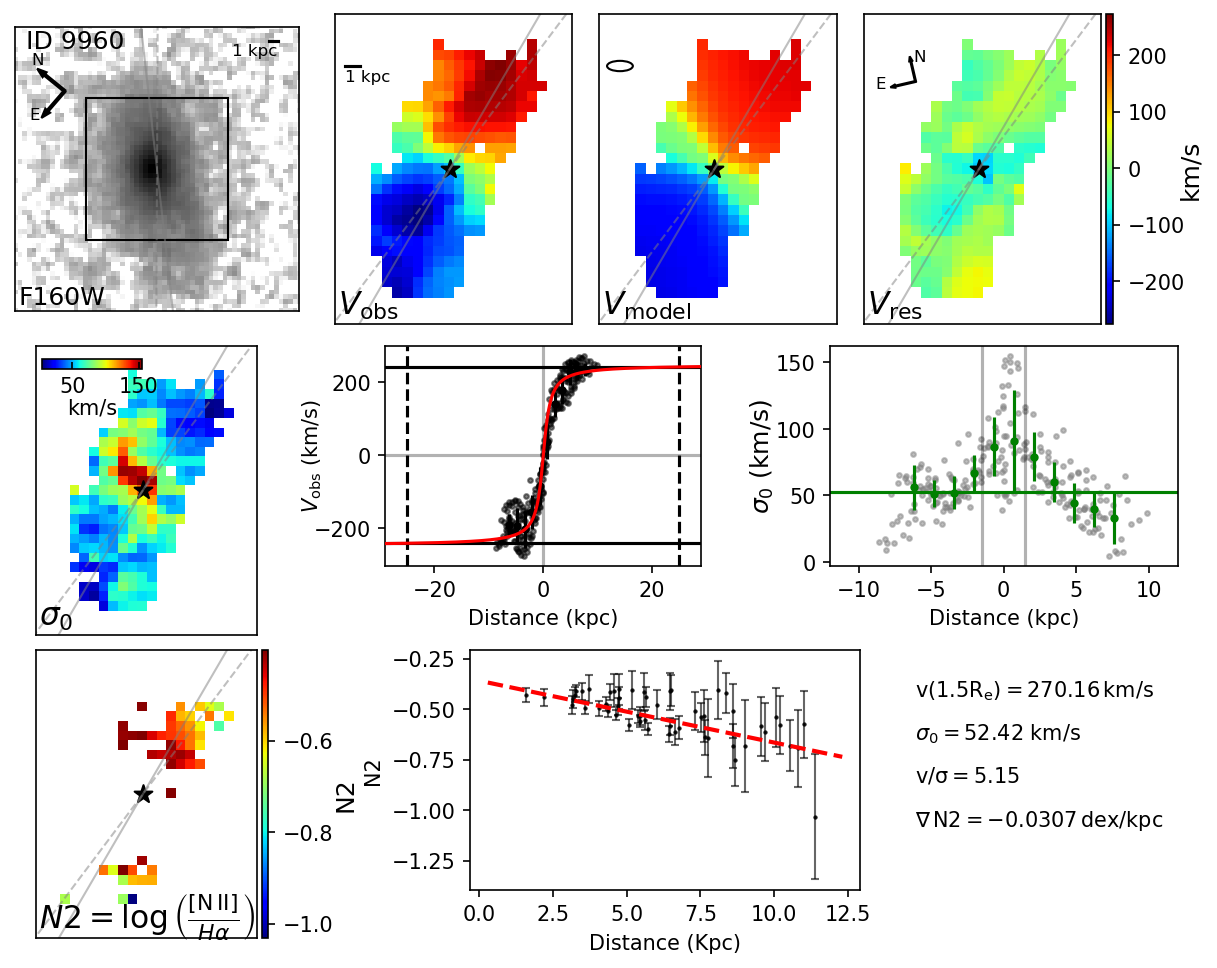}\\
    \vspace{1.2em}
    \includegraphics[width=0.8\textwidth,clip,trim={0 0 0 0}]{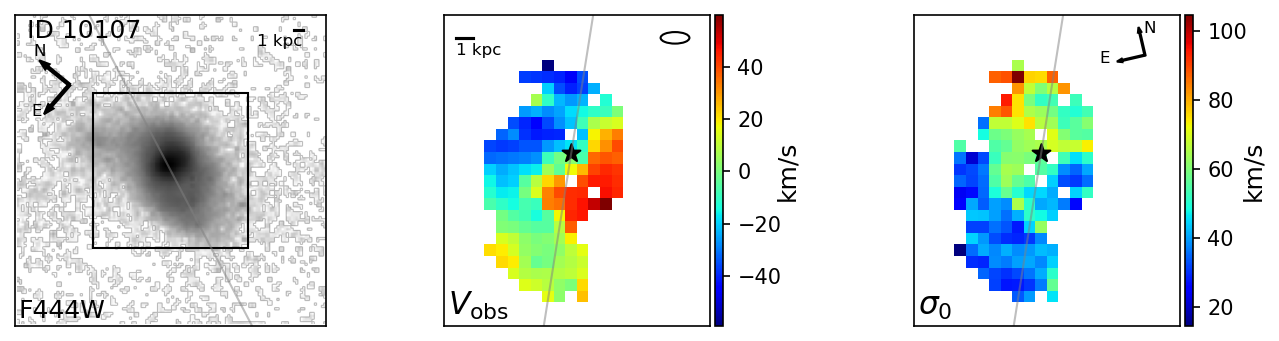}\\ 
    \vspace{1.2em}
    \includegraphics[width=0.8\textwidth,clip,trim={0 0 0 0}]{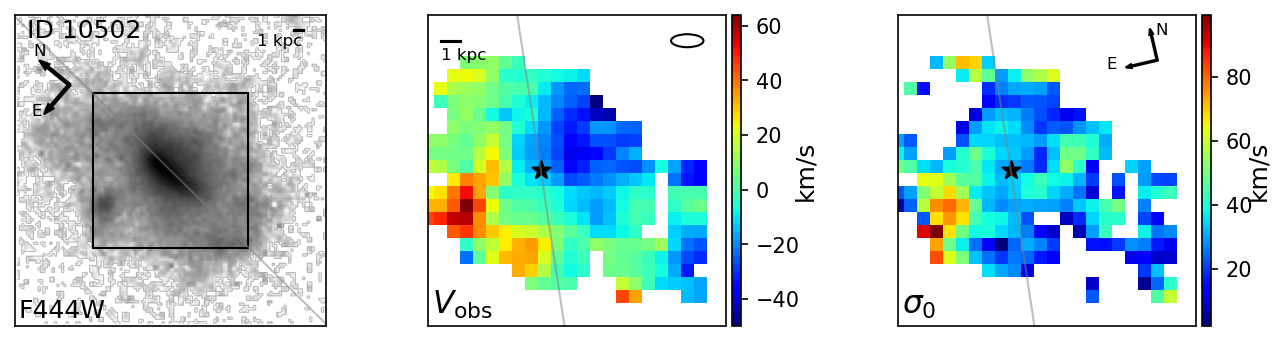}\\    
    {\bf Figure A1.} continued\\
\end{figure*}

\begin{figure*}
\centering
    \ContinuedFloat
    \includegraphics[width=0.8\textwidth,clip,trim={0 0 0 0}]{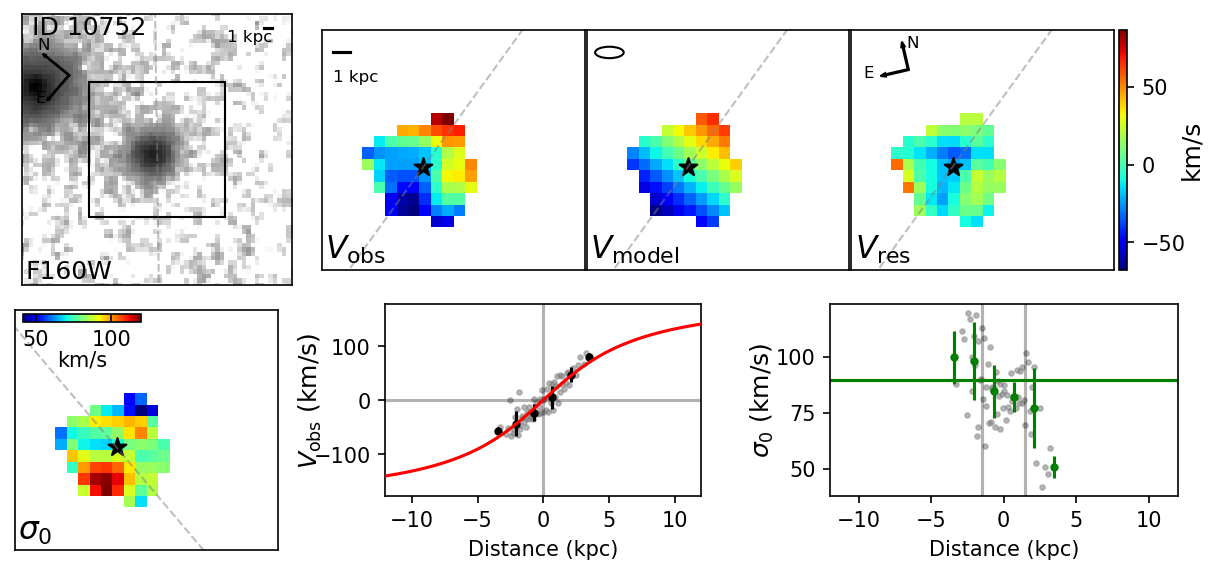}\\
    \vspace{1.2em}
    \includegraphics[width=0.8\textwidth,clip,trim={0 0 0 0}]{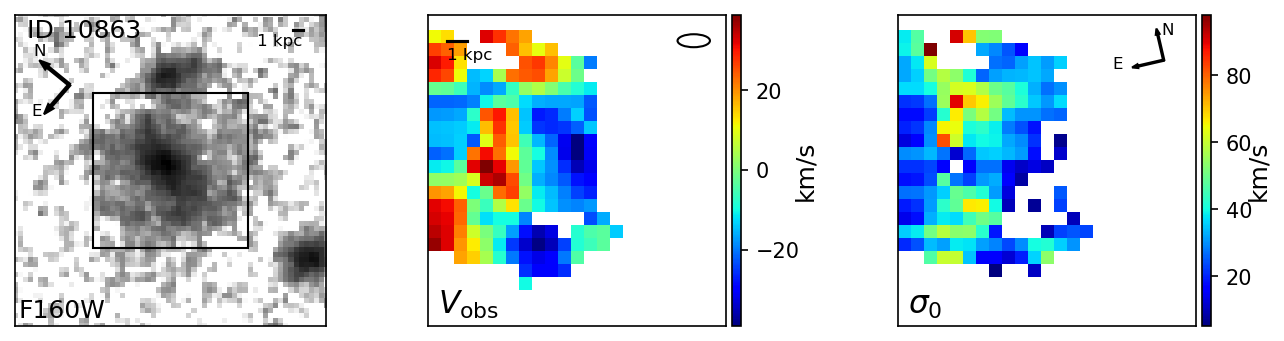}\\
    \vspace{1.2em}
    \includegraphics[width=0.8\textwidth,clip,trim={0 0 0 0}]{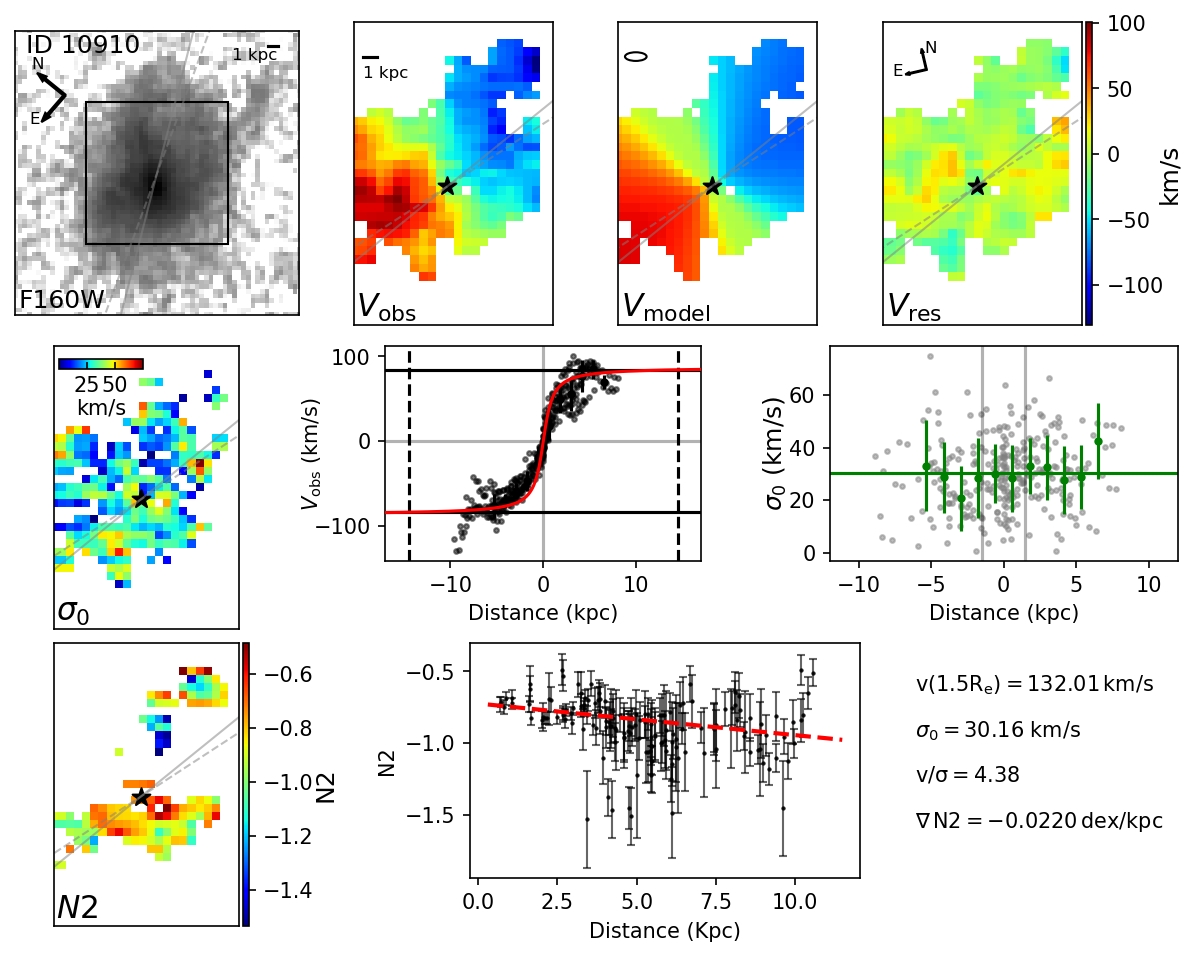}\\ 
    {\bf Figure A1.} continued\\
\end{figure*}

\begin{figure*}
\centering
    \ContinuedFloat
    \includegraphics[width=0.8\textwidth,clip,trim={0 0 0 0}]{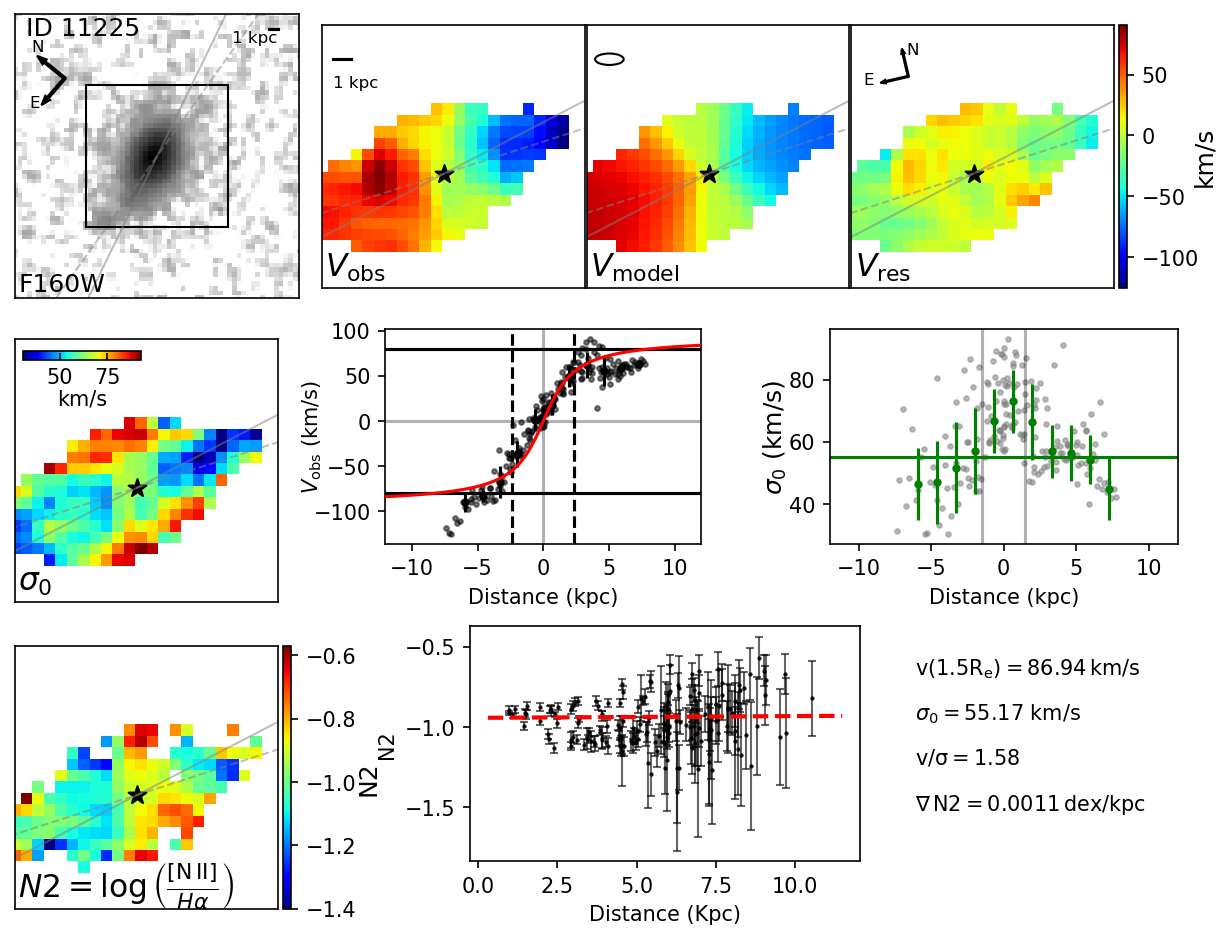}\\
    \vspace{1.2em}
    \includegraphics[width=0.8\textwidth,clip,trim={0 0 0 0}]{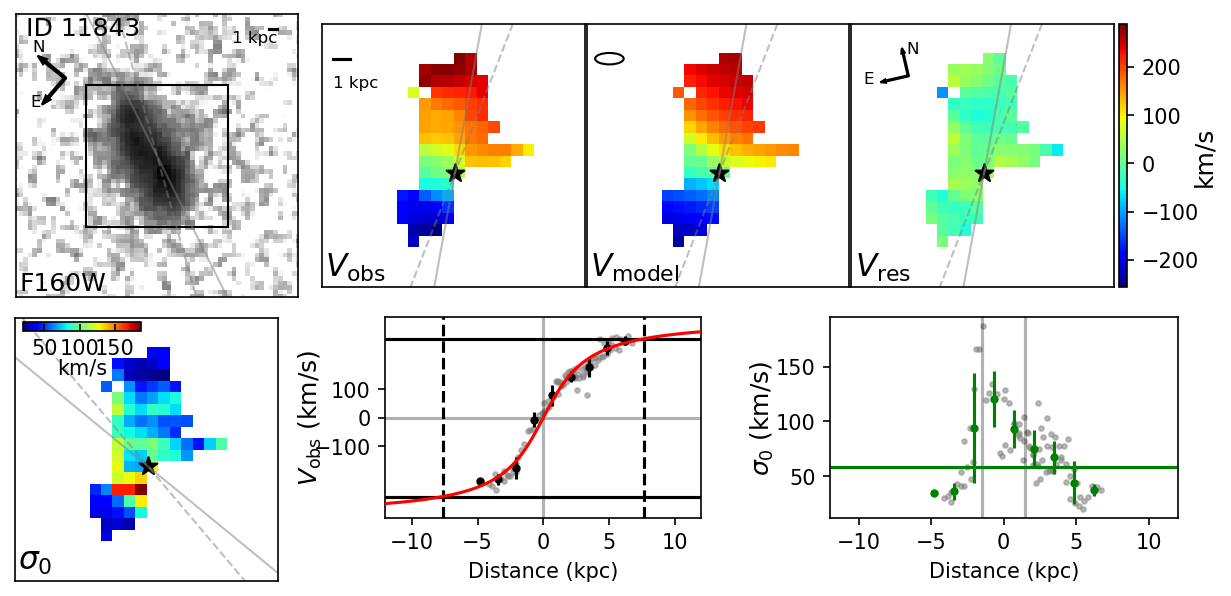}\\
    \vspace{1.2em}
    \includegraphics[width=0.8\textwidth,clip,trim={0 0 0 0}]{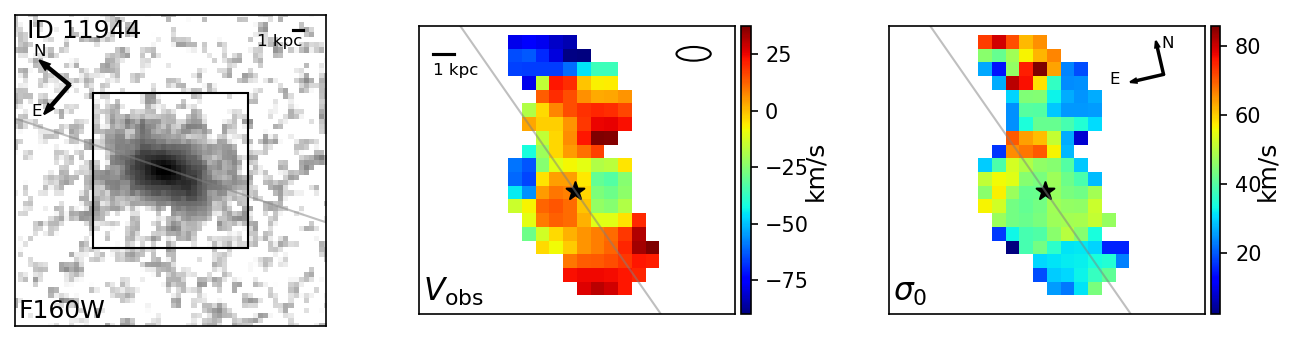}\\   
    {\bf Figure A1.} continued\\
\end{figure*}

\begin{figure*}
\centering
    \ContinuedFloat
    \includegraphics[width=0.8\textwidth,clip,trim={0 0 0 0}]{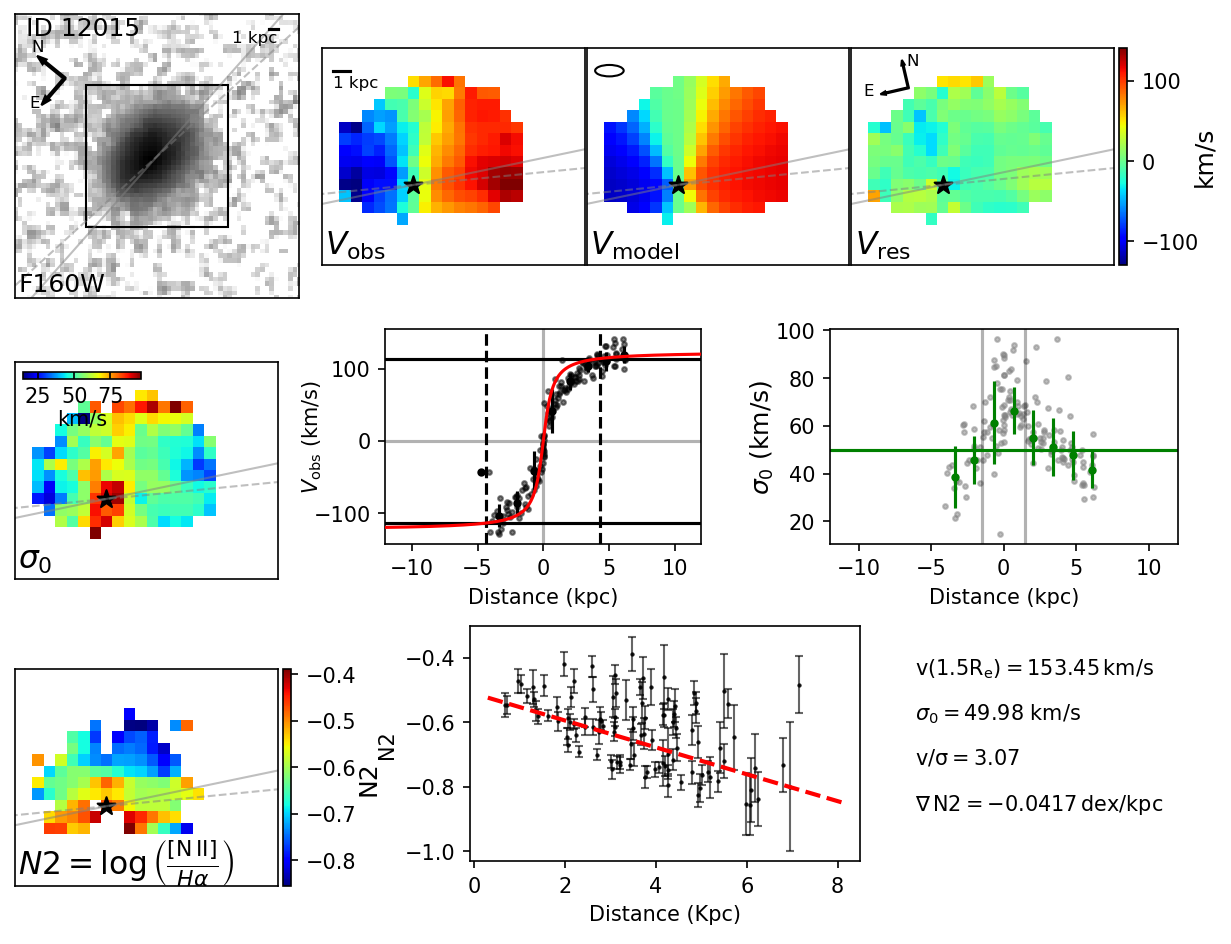}\\
    \vspace{1.2em}
    \includegraphics[width=0.8\textwidth,clip,trim={0 0 0 0}]{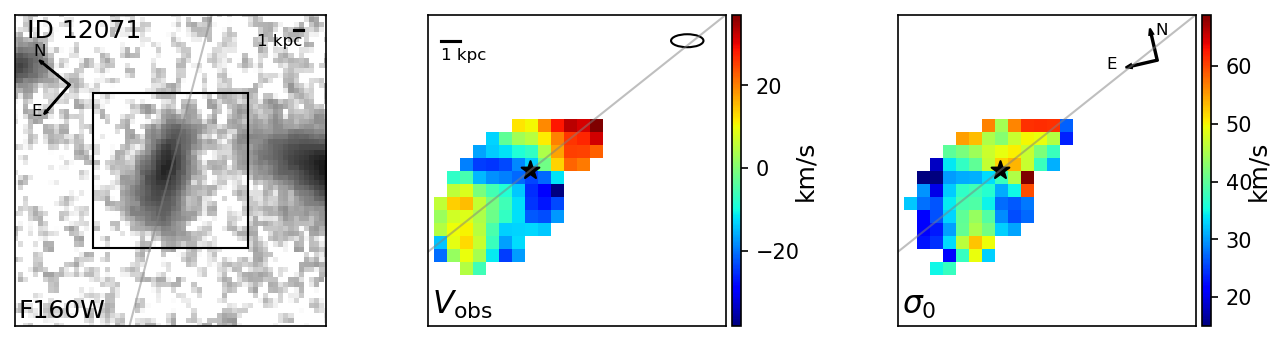}\\
    \vspace{1.2em}
    \includegraphics[width=0.8\textwidth,clip,trim={0 0 0 0}]{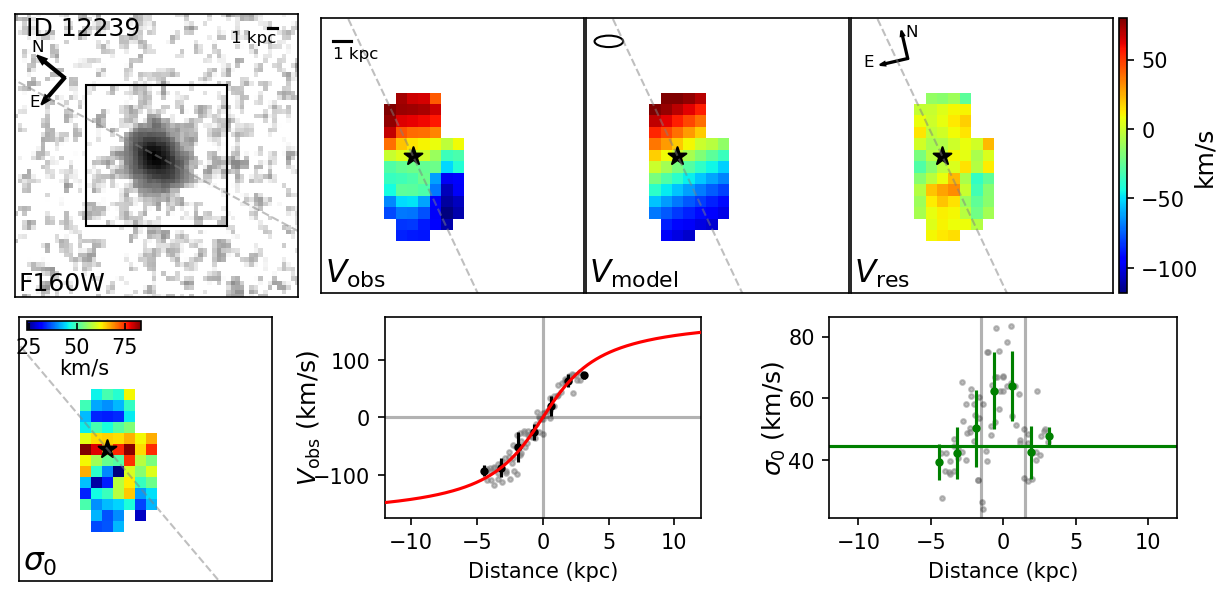}\\
    {\bf Figure A1.} continued\\
\end{figure*}

\begin{figure*}
\centering
    \ContinuedFloat
    \includegraphics[width=0.8\textwidth,clip,trim={0 0 0 0}]{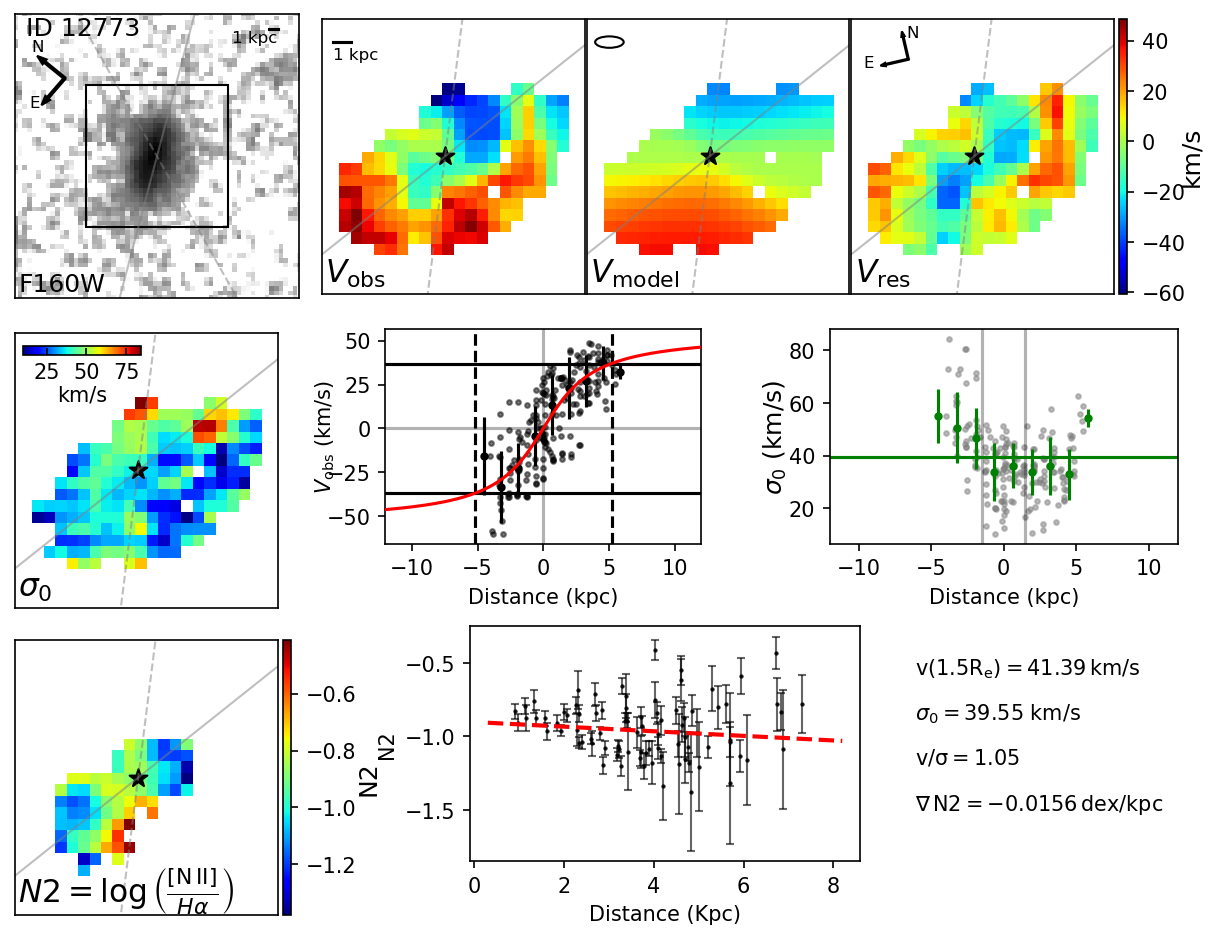}\\
    \vspace{1.2em}
    \includegraphics[width=0.8\textwidth,clip,trim={0 0 0 0}]{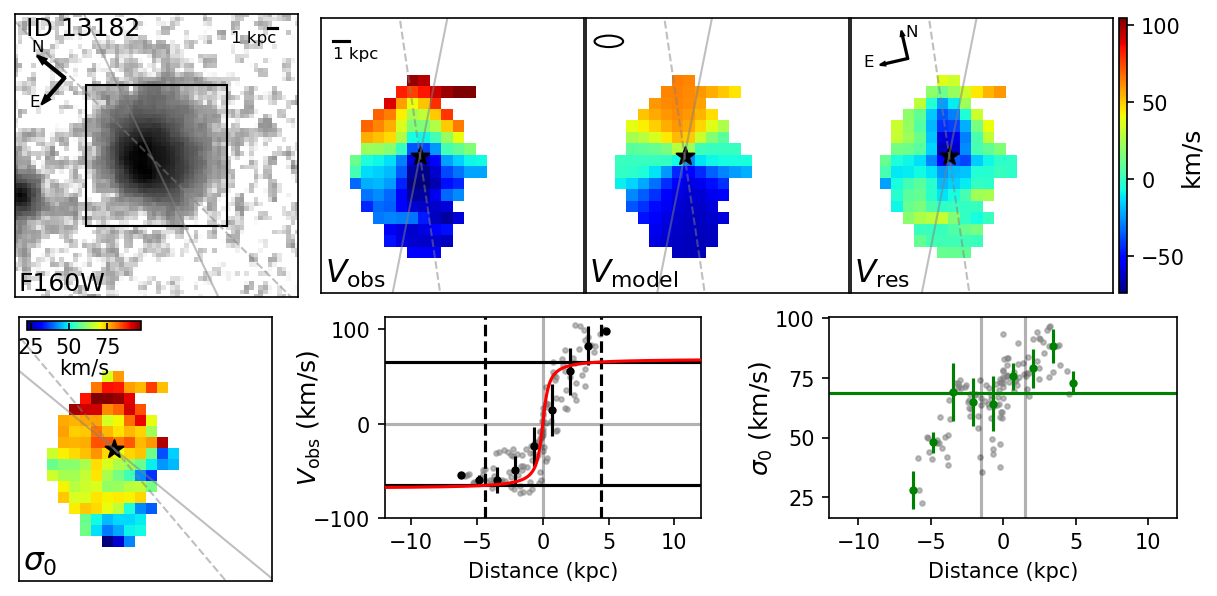}\\
    {\bf Figure A1.} continued\\
\end{figure*}

\begin{figure*}
\centering
    \ContinuedFloat
    \includegraphics[width=0.8\textwidth,clip,trim={0 0 0 0}]{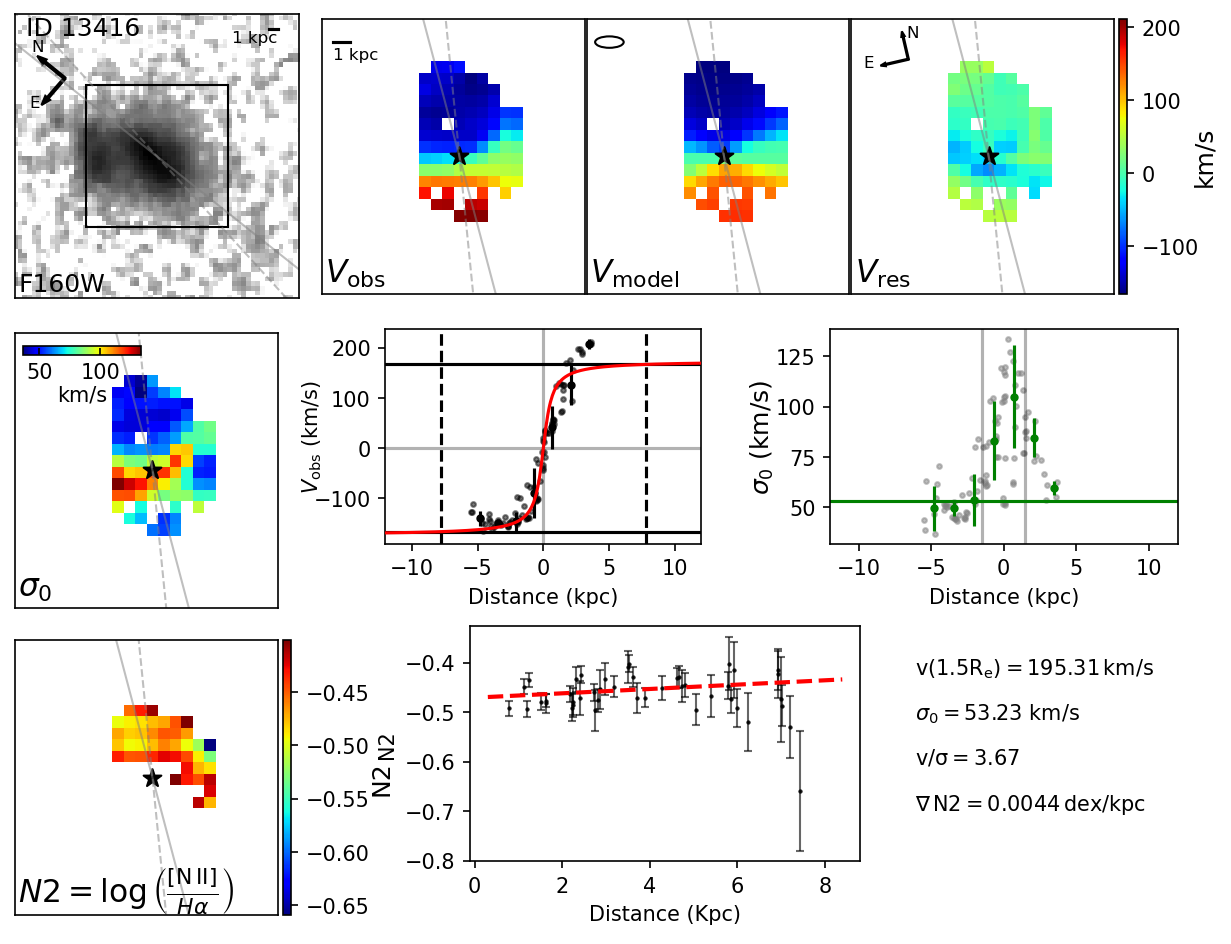}\\
    \vspace{1.2em}
    \includegraphics[width=0.8\textwidth,clip,trim={0 0 0 0}]{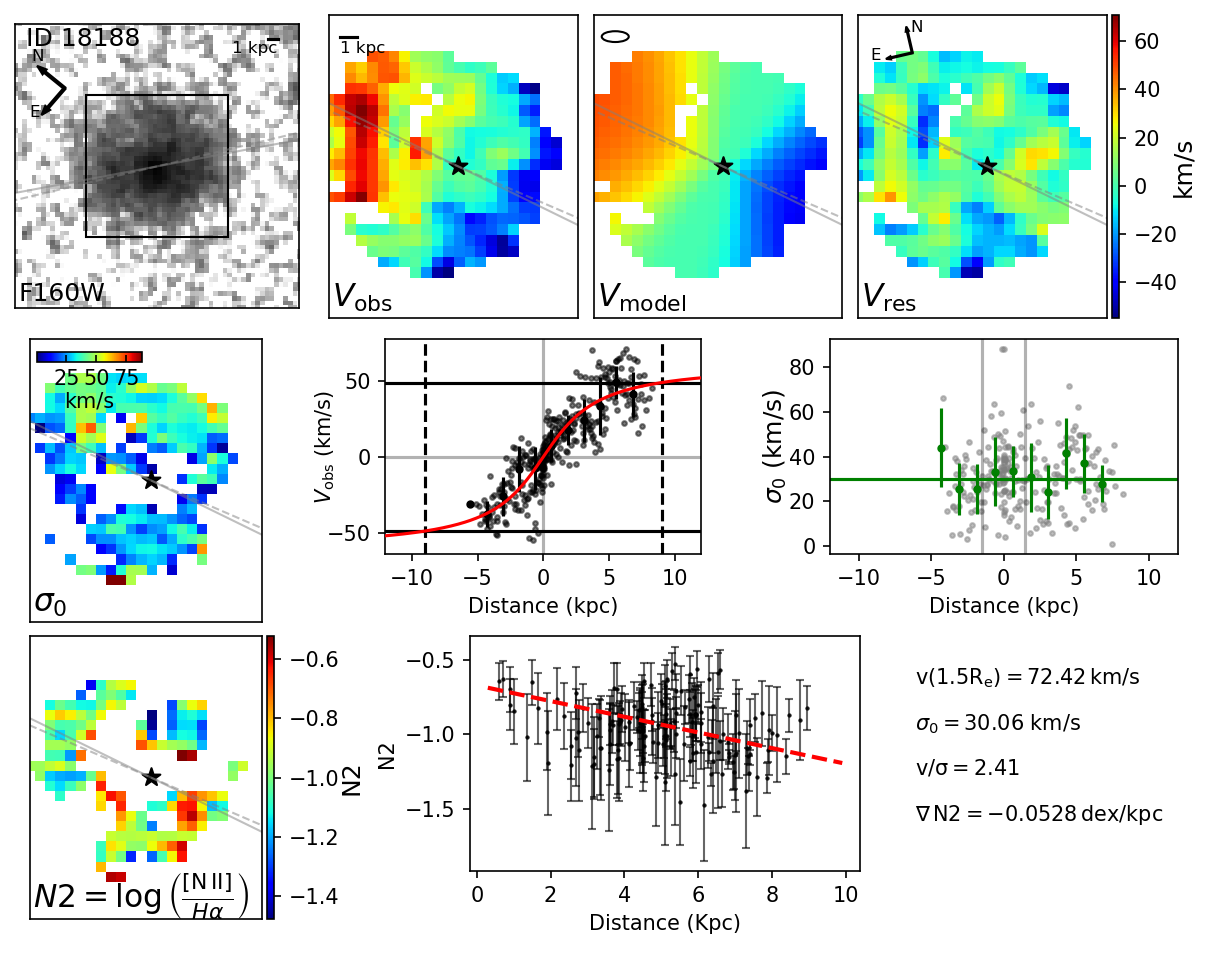}\\
    {\bf Figure A1.} continued\\
\end{figure*}

\begin{figure*}
\centering
    \ContinuedFloat
    \includegraphics[width=0.8\textwidth,clip,trim={0 0 0 0}]{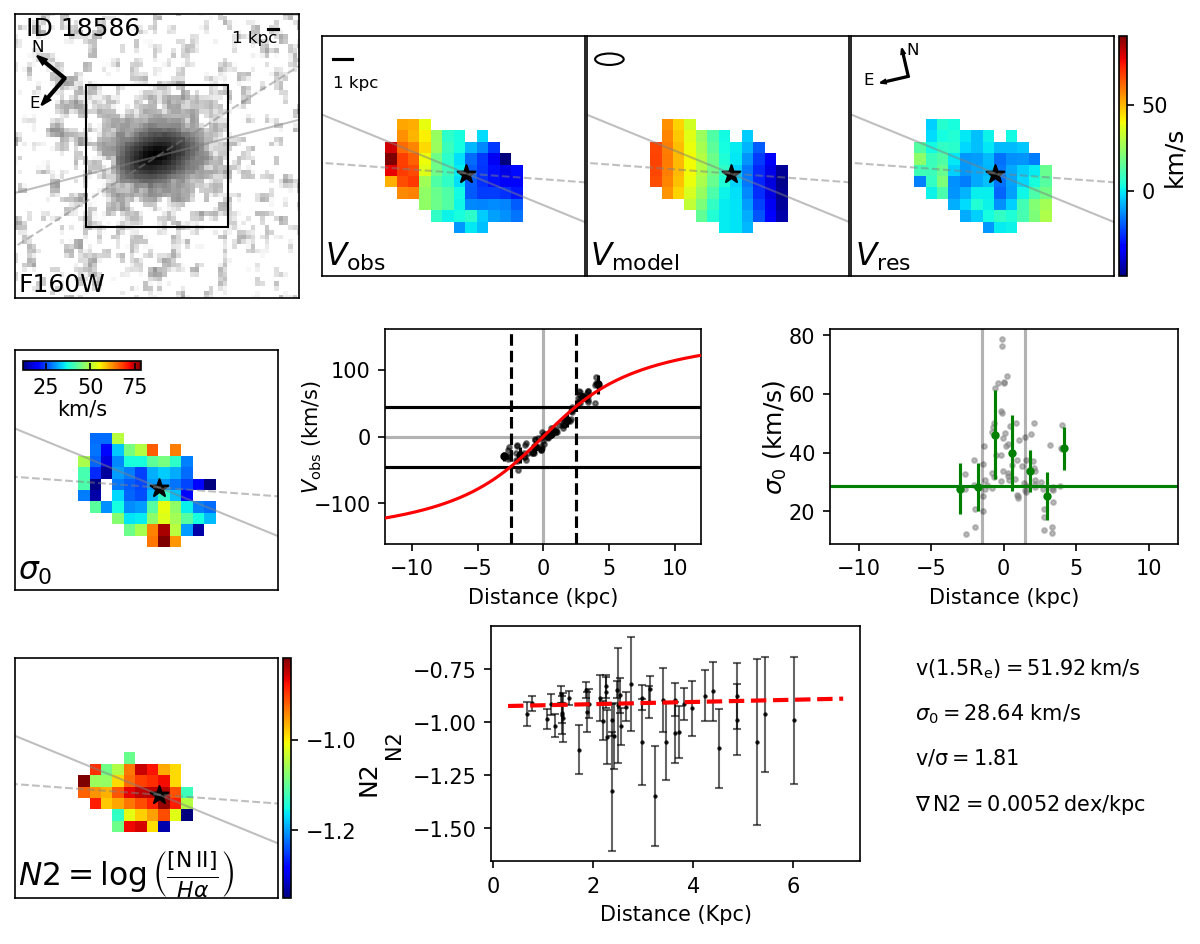}\\
    \vspace{1.2em}
    \includegraphics[width=0.8\textwidth,clip,trim={0 0 0 0}]{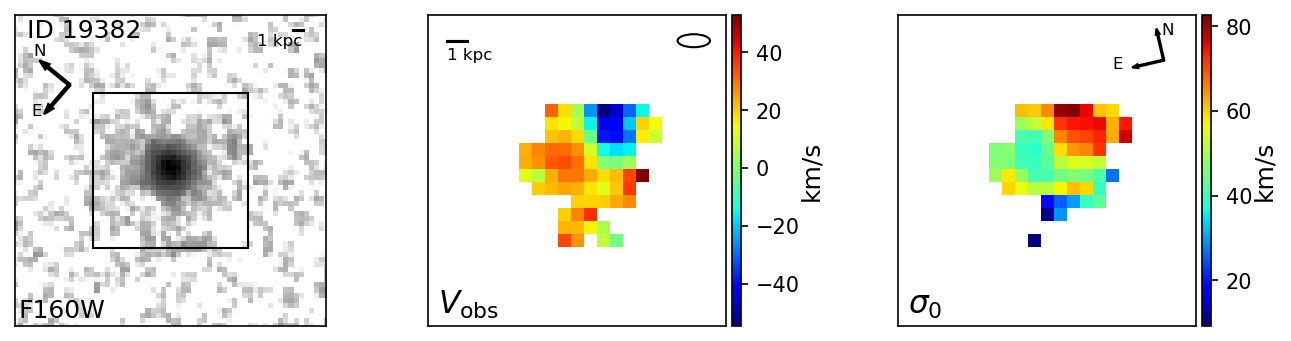}\\
    {\bf Figure A1.} continued\\
\end{figure*}

\begin{figure*}
\centering
    \ContinuedFloat
    \includegraphics[width=0.8\textwidth,clip,trim={0 0 0 0}]{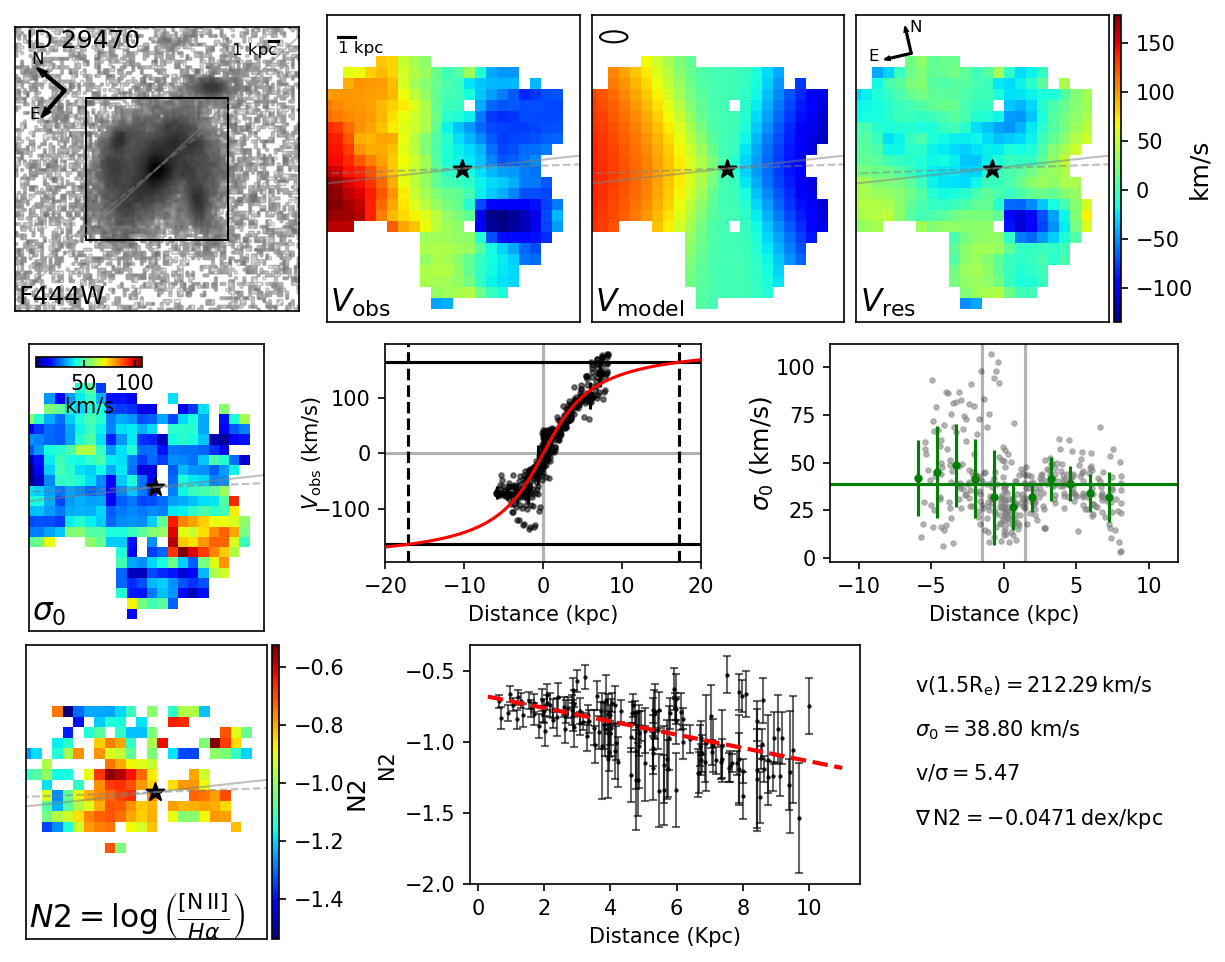}\\
    {\bf Figure A1.} continued\\
\end{figure*}

\bibliographystyle{aasjournal}
\bibliography{cita.bib}

\end{CJK*}
\end{document}